\documentclass[twocolumn,superscriptaddress,aps,prx,amssymb,amsfonts,amsmath,floatfix,longbibliography]{revtex4-1}
\usepackage{bm}
\usepackage{xcolor}
\usepackage{hyperref}
\usepackage{graphicx}
\usepackage[normalem]{ulem}
\usepackage[english]{babel}

\newcommand{\beginsupplement}{
        \clearpage
        \setcounter{table}{0}
        \renewcommand{\thetable}{S\arabic{table}}
        \setcounter{figure}{0}
        \renewcommand{\thefigure}{S\arabic{figure}}
        \setcounter{equation}{0}
        \renewcommand{\theequation}{S\arabic{equation}}
        \setcounter{section}{0}
        \renewcommand{\thesection}{S\arabic{section}}
     }

\def \chidynA{\chi^{A_0}_{\rm dyn}}
\def \chidyn{\chi_{\rm dyn}}
\def \chiback{\chi_{\rm bg}}
\def \chitot{\chi_{\rm tot}}

\begin{document}

\title{Fluctuations and the limit of predictability in protein evolution}

\author{Saverio Rossi}%
\affiliation{Dipartimento di Fisica, Sapienza Universit\`a di Roma, Piazzale Aldo Moro 5, 00185 Rome, Italy}
\affiliation{Laboratoire de Physique de l'Ecole Normale Sup\'erieure, ENS, Universit\'e PSL, CNRS, Sorbonne Universit\'e, Universit\'e Paris Cit\'e, 75005 Paris, France}

\author{Leonardo Di Bari}%
\affiliation{DISAT, Politecnico di Torino, Corso Duca degli Abruzzi, 24, I-10129, Torino, Italy}
\affiliation{Sorbonne Universit\'e, CNRS, Laboratory of Computational and Quantitative Biology, 75005 Paris, France}

\author{Martin Weigt}%
\affiliation{Sorbonne Universit\'e, CNRS, Laboratory of Computational and Quantitative Biology, 75005 Paris, France}

\author{Francesco Zamponi}%
\email{E-mail: francesco.zamponi@uniroma1.it}
\affiliation{Dipartimento di Fisica, Sapienza Universit\`a di Roma, Piazzale Aldo Moro 5, 00185 Rome, Italy}
\affiliation{Laboratoire de Physique de l'Ecole Normale Sup\'erieure, ENS, Universit\'e PSL, CNRS, Sorbonne Universit\'e, Universit\'e Paris Cit\'e, 75005 Paris, France}

\date{\today}%

\begin{abstract}
Protein evolution involves mutations occurring across a wide range of time scales. In analogy with disordered systems in statistical physics, this dynamical heterogeneity suggests strong correlations between mutations happening at distinct sites and times. To quantify these correlations, we examine the role of various fluctuation sources in protein evolution, simulated using a data-driven energy landscape as a proxy for protein fitness. By applying spatio-temporal correlation functions developed in the context of disordered physical systems, we disentangle fluctuations originating from the initial condition, i.e. the ancestral sequence from which the evolutionary process originated, from those driven by stochastic mutations along independent evolutionary paths. Our analysis shows that, in diverse protein families, fluctuations from the ancestral sequence predominate at shorter time scales. This allows us to identify a time scale over which ancestral sequence information persists, enabling its reconstruction. We link this persistence to the strength of epistatic interactions: ancestral sequences with stronger epistatic signatures impact evolutionary trajectories over extended periods. At longer time scales, however, ancestral influence fades as epistatically constrained sites evolve collectively. To confirm this idea, we apply a standard ancestral sequence reconstruction algorithm and verify that the time-dependent recovery error is influenced by the properties of the ancestor itself. 
Overall, our results reveal that the properties of ancestral sequences—particularly their epistatic constraints—influence the initial evolutionary dynamics and the performance of standard ancestral sequence reconstruction algorithms.
\end{abstract}

\maketitle

\section{Introduction}

Proteins play a key role in many biological processes that are essential for life.
At the same time, they display a huge evolutionary flexibility, in that a large diversity of protein sequences can fold into the same structure and perform the same biological function.
Such proteins are called `homologous' and grouped into the same `protein family' \cite{blum2024interpro,paysan2024pfam, Uniprot2023, PDB2022_2}.
During evolution, a single ancestral protein belonging to a given family can diversify its amino acid sequence and, given enough time, explore a portion of the `neutral space' of equivalently fit protein sequences~\cite{kimura1985neutral}. Nevertheless, a random mutation in the amino acid sequence has a high probability of negatively impacting the structure and functionality of the protein \cite{fowler2014deep,sarkisyan2016local,notin2024proteingym}.
Such kind of deleterious mutations are eliminated by natural selection, by which certain protein variants are favored over others due to their functional advantages: neutral (or even beneficial) mutations that maintain (or enhance) protein structure, stability, or function are more likely to persist in a population \cite{hartl1997principles}. 

The picture gets more complex as one takes into account the concept of epistasis~\cite{harms2013evolutionary,weinreich2013should,de2014empirical,starr2016epistasis,poelwijk2016context,cocco2018inverse,domingo2019causes,johnson2023epistasis,buda2023pervasive}: the effect of a mutation on the sequence fitness changes with the `background' in which the mutation takes place, i.e. the amino acids that are present in the other sites of the protein.
As a consequence, a mutation that would be deleterious in a particular  background sequence can be instead beneficial in another (also called sign epistasis), allowing evolution to explore different pathways. 

Understanding and characterizing the impact of epistasis in evolution requires careful experiments and modeling.
In particular, recent developments have substantially increased the sequence divergence that can be reached by laboratory evolution experiments~\cite{fantini2020protein,stiffler2020protein,erdougan2023neutral,park2022epistatic,rix2023continuous}. A large amount of data is therefore now becoming available with more expected to come soon.
Yet, the sequence diversity of natural evolution still remains out of reach of such experiments, thus leaving an unexplored gap in evolutionary time scales.
In order to fill this gap, one can simulate the evolution of protein sequences \textit{in silico} \cite{de2020epistatic,bisardi2022modeling,alvarez2022novel,alvarez2024vivo,biswas2024kinetic,DiBari2024}, relying on the data-driven approach that goes under the name of Direct Coupling Analysis  (DCA)~\cite{Morcos2011,cocco2018inverse}. DCA infers a fitness landscape (analogous to an energy function in the statistical physics vocabulary) starting from a Multiple Sequence Alignment (MSA) of natural homologs constituting a given protein family \cite{ferguson2013translating,figliuzzi2016coevolutionary,levy2017potts,couce2017mutator,vigue2023predicting}. 
The energy function that results from this inference procedure can then be used to assign a probability to each sequence.  
The resulting landscape is explored by means of a biologically motivated Monte Carlo Markov-Chain (MCMC) algorithm, providing in silico evolutionary trajectories that quantitatively mimic experimental results~\cite{Biswas2019,de2020epistatic,bisardi2022modeling,alvarez2024vivo,alvarez2022novel,biswas2024kinetic,DiBari2024}. 

Within this framework, it has been recently shown~\cite{DiBari2024} that different sites evolve with widely different time scales, which also depend on the background sequence, due to epistatic interactions. More precisely, sites that are more epistatically constrained need much longer times (or number of generations) to evolve (i.e. accumulate mutations) compared to less constrained sites~\cite{DiBari2024}. 
Furthermore, whether a site is epistatically constrained or not depends on the rest of the sequence (the background)~\cite{lyons2020idiosyncratic,vigue2022deciphering,chen2023understanding}, 
which adds sequence-to-sequence heterogeneity on top of the site-to-site heterogeneity.
This ``heterogeneity of time scales'' is strongly reminiscent of similar phenomena observed in disordered physical systems~\cite{kirkpatrick1988comparison,franz1999dynamical,franz2000non,bouchaud2005nonlinear,Berthier2005,Berthier2007,franz2011field,berthier2011dynamical,franz2013static,Seoane2018,Folena2022}. There, it has been shown that
a large heterogeneity in time scales implies the presence of strong and highly collective space-time correlations between sites that slow down the dynamical evolution.
Furthermore, it has been shown that the structural disorder in the initial configuration partially encodes future correlations, and can be used (possibly by machine learning tools) to predict the future dynamics~\cite{widmer2008irreversible,schoenholz2016structural,bapst2020unveiling,jung2024dynamic}, see~\cite{jung2025roadmap} for a recent review.
These are the main observations that motivated this work. 

In this paper, we measure space-time correlations to explore how the heterogeneity of evolutionary time scales relates to the epistatic interactions within the ancestral sequence. 
We follow the dynamics of the Hamming distance, i.e., the number of accepted mutations with respect to the ancestral sequence. We characterize how this quantity fluctuates (i) between different evolutionary trajectories originating from the same ancestor, and (ii) between different ancestors.
We observe a strong dependence of the dynamics on the ancestral sequence for short enough times, over which the first source of fluctuations is found to be subdominant with respect to the second.
On the other hand, at long times the stochasticity of evolutionary trajectories takes over, and any memory of the ancestor is lost.
This has important consequences, as it allows us to properly quantify the time scale over which it should be possible to reconstruct the ancestral sequence of a certain set of evolutionary trajectories, and how this time scale depends on the epistatic interactions in the ancestor itself.
In fact, we find that the amount of epistatically constrained sites in the ancestral sequence determines this time scale: more epistatic sequences leave their trace on evolutionary dynamics for longer times.
We then measure the correlations between the evolution of all pairs of sites at the time scale at which such correlations are stronger, finding a different pattern for each ancestral sequence, which is then reflected in the evolutionary dynamics.
Finally, we show that the magnitude of epistatic interactions, which controls the stochasticity of the evolutionary trajectories, is also related to the linear response of the evolutionary dynamics to a change in selective pressure (controlled by the temperature in our Monte Carlo simulations). Hence, we show and quantify how more epistatically constrained ancestors lead to a more complex evolutionary dynamics over longer time scales, which is also more sensitive to perturbations of the environment.

\section{Epistasis in biological data}

Epistasis plays a central role in shaping protein evolution, influencing both the structure of fitness landscapes and the fate of evolutionary trajectories, and as such it has been the subject of theoretical, computational and experimental studies, see Refs.~\cite{harms2013evolutionary,weinreich2013should,de2014empirical,starr2016epistasis,poelwijk2016context,cocco2018inverse,domingo2019causes,johnson2023epistasis,buda2023pervasive} for a few recent reviews.
It has been shown that some epistatic effects can be accounted for by a global non-linearity of the phenotype-fitness relation~\cite{otwinowski2018biophysical,otwinowski2018inferring,reddy2021global,schulte2025functional}, but epistasis is also due to a network of more specific pairwise and higher-order interactions.
While some of these interactions are sparse~\cite{sailer2017detecting,sailer2017high,domingo2018pairwise,poelwijk2019learning,ballal2020sparse,phillips2021binding,buda2023pervasive,miton2021epistasis},
multiple studies suggest that the most relevant effects tend to emerge collectively, from the accumulation of numerous weak interactions between a single site and multiple other residues across the protein sequence~\cite{lunzer2010pervasive,miton2016mutational,figliuzzi2016coevolutionary,rivoire2016evolution,poelwijk2016context,cocco2018inverse,starr2018pervasive,vigue2022deciphering,chen2023understanding}. This form of distributed epistasis suggests that the evolutionary constraints on a given mutation are shaped by the broader sequence context, rather than by a few dominant interactions, highlighting the complexity of protein fitness landscapes. 

Experiments have provided valuable insights into how mutations interact, but each methodology comes with inherent limitations. While some studies focus on measuring epistasis in a restricted mutational space, others provide broader datasets but lack information on evolutionary dynamics. Here, we summarize the main categories of experimental data available and discuss their relevance to our theoretical framework.
\begin{itemize}
\item Combinatorial mutagenesis~\cite{poelwijk2019learning, phillips2021binding, bakerlee2022idiosyncratic, buda2023pervasive, Papkou2023, Somermeyer2022,  Schulz2025} experiments  involving a small number of residues have demonstrated epistatic effects. However, these studies typically do not allow the accumulation of sufficient mutational effects to observe the large-scale evolutionary patterns we investigate in this work. 
\item Deep mutational scanning (DMS) experiments across homologous wild-type proteins~\cite{olson2014comprehensive,park2022epistatic, chen2023understanding, Romanowicz2025}, reveal epistatic interactions and can be used to explore the concept of site variability within a small region of the landscape. However, these experiments lack direct observations of evolutionary dynamics over multiple generations. 
\item Experimental studies tracking evolutionary dynamics in vitro are available~\cite{fantini2020protein,stiffler2020protein,erdougan2023neutral,rix2023continuous}, but only a few cases, such as TEM-1~\cite{fantini2020protein} and PSE-1~\cite{stiffler2020protein} $\beta$-lactamases, involve multiple wild-type homologs belonging to the same protein family, whose evolutionary dynamics can be directly compared. Moreover, these experiments do not extend far enough in sequence space to capture the long-term epistatic effects central to our study. 
\end{itemize} Given these limitations, we anticipate that future advancements in experimental techniques will provide richer datasets capable of testing the predictions outlined in our study. In the meantime, our theoretical framework serves as a guide for interpreting existing data and shaping expectations for future experimental work.

\section{Methods}

\subsection{Modeling evolution in silico}

In order to mimic the evolution of protein sequences, we use the DCA model energy as a proxy for the fitness landscape. We start by building an MSA of naturally occurring sequences for each protein family we want to study. 
From the MSA we infer the parameters (fields and couplings) of the DCA model via Boltzmann Machine learning (bmDCA)~\cite{Figliuzzi2018,muntoni2021adabmdca}. 
The resulting model assigns a probability 
$P(A) = \exp[-\mathcal{H}(A)]/Z$ to each sequence $A=(a_1, \dots, a_L)$, with $L$ the common length of aligned sequences in the MSA and $a_i$ being a symbol that takes 21 possible values corresponding to the 20 natural amino acids plus the gap symbol needed for alignment. According to the statistical physics language, low energy 
\begin{equation}
  \mathcal{H}(A)  = - \sum_{i<j} J_{ij}(a_i,a_j) - \sum_i h_i(a_i)
\end{equation}
corresponds to high probability, hence high fitness. 
In this expression, epistatic interactions between different amino acids are represented via the pairwise couplings $J_{ij}(a_i,a_j)$ that have been found to be crucial in data-driven statistical models of biological sequences, cf.~\cite{cocco2018inverse,russ2020evolution}.
Because training of these models has become a standard and well-documented procedure~\cite{Figliuzzi2018,muntoni2021adabmdca}, we do not give further details here.

Following~\cite{DiBari2024, de2020epistatic, bisardi2022modeling}, we consider 
a fixed initial sequence $A_0 = (a^0_1, \dots, a^0_L)$ as the ancestor, and
we let many trajectories evolve from it in parallel, hence restricting ourselves to a star phylogeny describing an ensemble of independent evolutionary trajectories of common initialization. 
Our goal is to characterize the statistical properties of the specific ancestral sequence and the sequence space accessible from it in a given evolutionary time.
The sequence evolution is simulated by Monte Carlo dynamics.
In this work, we use the Metropolis algorithm acting on amino acids for simplicity, and we verified that more refined sampling strategies taking into account amino-acid accessibility via the genetic code, insertions and deletions~\cite{DiBari2024, de2020epistatic, bisardi2022modeling} produce qualitatively the same results, see the Supplemental Material~(SM). 
As it was previously shown~\cite{russ2020evolution}, at large times the generated sequences accurately reproduces many statistical features of the natural sequences used for the training, and have the same probability of being biologically functional in a given experimental platform; hence, the model is generative.
Here, we are concerned with what happens at short and intermediate times, where the influence of the ancestral sequence is still important.

The results we report mainly concern the DNA-binding domain (DBD) protein family already studied in a similar setting in Ref.~\cite{DiBari2024} and experimentally in Ref.~\cite{park2022epistatic}, but for some results we also generalize to other protein families, in particular the WW domain (WW), Chorismate Mutase (CM), Aminoglycoside 6-N-acetyltransferase (AAC6), Dihydrofolate reductase (DHFR), Beta Lactamase (BL), and Serine Protease (SP) families. These families have been chosen because they span several chain lengths, and experimental data obtained either from Deep Mutational Scanning or by \textit{in vitro} evolution are available. The procedures to construct the natural MSAs for these families are given in the~SM.

\subsection{Measures of epistasis}
\label{sec:CDECIE}

By using the MSA of a given protein family as input data, and inferring the fitness landscape parameters via the DCA model, the authors of~\cite{vigue2022deciphering, DiBari2024} have been able to roughly classify the sites 
of any specific protein sequence belonging to the family 
in three categories: conserved, mutable, and epistatically constrained.
In order to do so, they defined two site-mutability metrics.
\begin{itemize}
\item The Context-Independent Entropy (CIE) is obtained by computing the empirical frequency $f_i(a)$ of occurrence of amino acid $a$ on site $i$, for every $a$ and $i$, in the MSA obtained from the natural sequences. This is then used to compute a standard Shannon entropy as 
\begin{equation}
\textrm{CIE}_i = -\sum_{a=1}^{21} f_i(a) \log_2 f_i(a)  \ .
\end{equation}
This quantity measures to what extent 
a site is variable or conserved across the input MSA.
\item
The Context-Dependent Entropy (CDE) is defined for site $i$ in sequence $A$ as 
\begin{equation}
\textrm{CDE}^{A}_i = -\sum_{a=1}^{21} P_i(a | A_{\setminus i}) \log_2 P_i( a | A_{\setminus i}) \ ,
\end{equation}
using the conditional probability $P_i(a | A_{\setminus i}) = P(a_i = a | A_{\setminus i})$ of having amino acid $a$ on site $i$ given the rest of the sequence $A_{\setminus i} = (a_1,...,a_{i-1},a_{i+1},...,a_L)$. This quantity cannot be extracted directly from the input MSA and needs to be obtained from the model parameters; due to the epistatic couplings in the energy $\mathcal{H}(A)$ it actually differs from the CIE. 
As indicated explicitly, the CDE$^A_i$ depends on the site $i$ and the sequence $A$, hence on the context in which site $i$ finds itself. As a matter of fact, this metric quantifies the local mutability of a site, i.e. its mutability within a certain reference-sequence context $A_{\setminus i}$.
\end{itemize}
A large value of CIE or CDE means that many mutations are allowed, while a small value means that only one or a few amino acids are tolerated.

In terms of these quantities, a site $i$ in a given background sequence $A$ can thus be classified as follows:
\begin{itemize}
\item
{\it Mutable sites} have a large CDE and a large CIE. These sites can tolerate many mutations, both in the natural MSA and in the considered background.
\item
{\it Conserved sites} have a small CDE and a small CIE. These sites do not tolerate mutations, neither in the natural MSA nor in the considered background. 
\item
{\it Epistatically constrained sites} have a small CDE and a large CIE.
These sites display a large variability in the natural alignment, but only because the rest of the sequence is mutating at the same time. In fact, in the considered background, only one or a few amino acids are tolerated.
\end{itemize}
Note that it is very rare that a site has a CDE larger than the CIE, because generally speaking, fixing the background reduces the number of mutations that can be tolerated \cite{vigue2022deciphering}.
We also stress that this classification depends on the background $A$, and Ref.~\cite{DiBari2024} has shown that epistatically constrained sites in a background can be mutable in another background and viceversa (while conserved sites tend to remain so in all backgrounds).

Furthermore, and most importantly for the present work, Ref.~\cite{DiBari2024} has considered a given background $A_0$ as the ancestral sequence, and starting from it has performed many parallel evolutions in silico, looking at how each site diversifies in the library of mutants obtained after a certain evolutionary time.
It was found that mutable sites evolve very rapidly and quickly reach their asymptotic mutability, i.e.~the CIE.
Conserved sites remain so at any time during evolution, hence do not display interesting dynamics.
Epistatically constrained sites, instead, are conserved at short evolutionary times, due to the epistatic constraints in the background of the ancestor, which cause a small CDE. But, as soon as the background sequence mutates enough, they can tolerate more mutations, asymptotically reaching the large CIE that they display in the natural alignment. 
Their mutation, however, is contingent on several mutations happening in the rest of the sequence, which often require a rather long evolutionary time to take place~\cite{DiBari2024}.

Along with the classification of sites discussed above, we also introduce a global measure of the level of epistasis in a given sequence $A$.
Because each sequence has a different set of variable and epistatically constrained sites, we measure its overall level of epistatic constraints by averaging the CDE over sites,
\begin{equation}\label{eq:globalCDE}
\textrm{CDE}^{A} = \frac1L \sum_{i=1}^L \textrm{CDE}^{A}_i \ .    
\end{equation}
Hence, a more epistatically constrained sequence will display a lower $\textrm{CDE}^A$.

\begin{figure*}[t]
\centering
\includegraphics[width=0.8\linewidth]{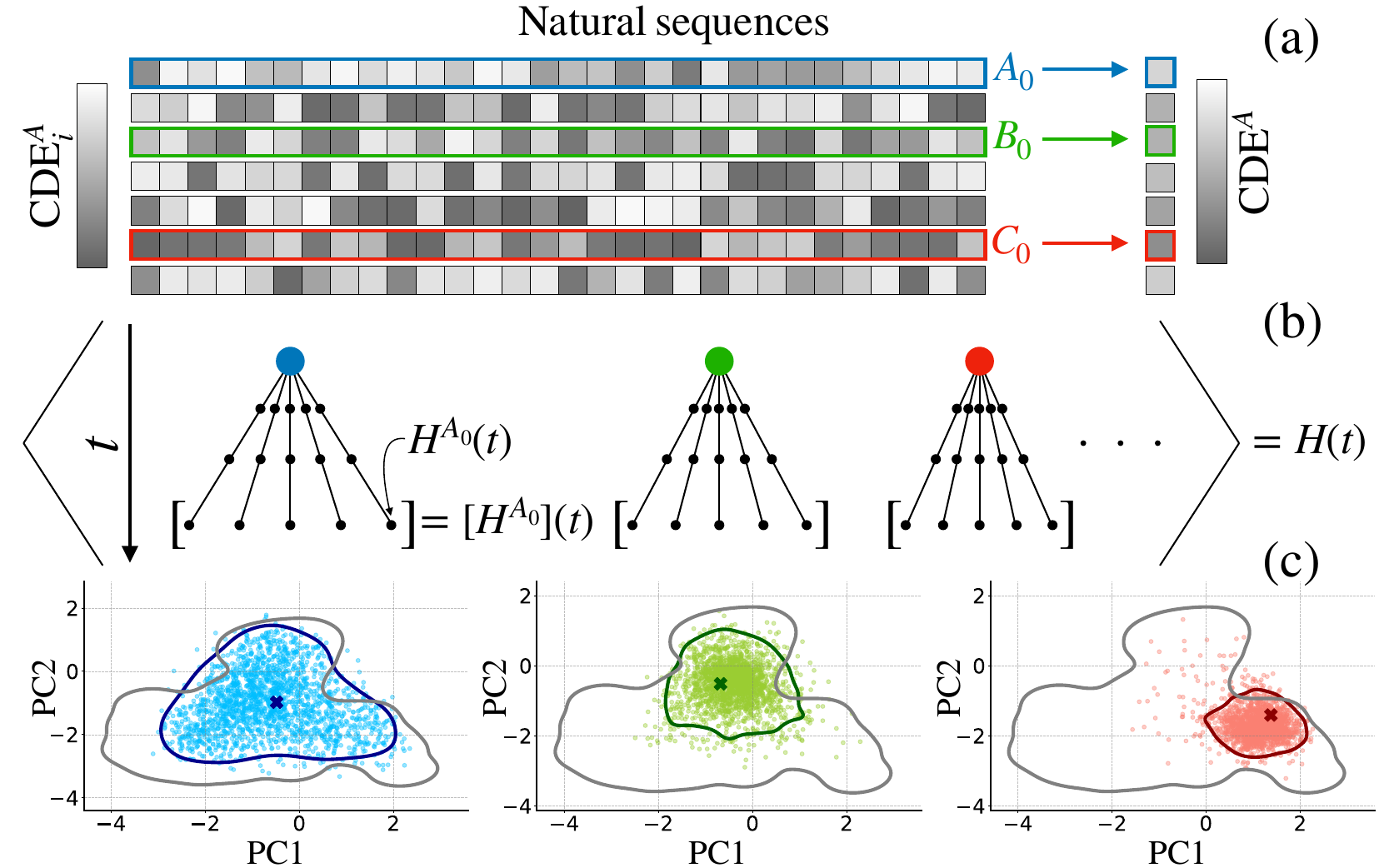}
\caption{Schematic representation of (a) an MSA in which each site is colored based on its mutability as measured by the $\textrm{CDE}_i^A$ and (b) the protocol of our evolutionary dynamics with the relevant averages.
(c) Depending on the epistatic constraints acting on the ancestral sequence and for equal evolutionary times, we observe a different level of diversity of the evolved sequences (colored), corresponding to a different exploration of the functional sequence space (grey).
}
\label{fig:Scheme}
\end{figure*}

\subsection{Dynamical fluctuations}

Building on the previous work described above, and 
inspired by works on disordered systems~\cite{kirkpatrick1988comparison,franz1999dynamical,franz2000non,bouchaud2005nonlinear,Berthier2005,franz2011field,berthier2011dynamical,franz2013static,Seoane2018,Berthier2007}, we
devised a protocol (sketched in Fig.~\ref{fig:Scheme}) to quantify how the amount of epistasis in the ancestral sequence impacts the subsequent evolution.
We consider a series of ancestral sequences, taken from the natural MSA, let us call them $A_0$, $B_0$, $C_0$, etc. Each ancestor has a different level of epistatic constraints, as measured by the average CDE introduced in Eq.~\eqref{eq:globalCDE},
as illustrated in Fig.~\ref{fig:Scheme}a.

Next, we consider a star phylogeny of independent MCMC evolutionary trajectories, all starting from the same ancestor and evolving in parallel (Fig.~\ref{fig:Scheme}b). 
Hence, for each ancestor, we construct, as a function of evolutionary time, an MSA of descendants that mimic those obtained through in vitro evolution experiments.
As a measure of diversity, we focus on the evolution of the Hamming distance,
between the ancestor $A_0$ 
and the MSA of evolving sequences at time $t$, $A_t = (a^t_1, \cdots, a^t_L)$,
defined as
\begin{equation}
\label{eq:HA0def}
H^{A_0}(t) = H(A_t,A_0) = 
\frac1L \sum_{i=1}^L (1- \delta_{a^t_i,a^0_i}) \ ,
\end{equation}
introducing the Kronecker delta symbol $\delta_{a^t_i,a^0_i}$, 
which is zero if site $i$ is mutated between the two sequences $A_t$ and $A_0$, 
and one otherwise.
This quantity
corresponds to the `overlap' in the disordered systems literature and to the number of accepted mutations in evolution. Other measures could be considered as well, but we focus on this one for simplicity.
For a given ancestor, simulating many parallel evolutionary trajectories,
we thus obtain a set of realizations of the random variable $H^{A_0}(t)$.

As illustrated through a projection in PCA space in Fig.~\ref{fig:Scheme}c, we observe that the level of epistasis in the ancestor determines the size of the portion of the fitness landscape explored by evolution. For comparable evolutionary time, ancestors with more epistatically constrained sites (e.g., sequence $C_0$ in Fig.~\ref{fig:Scheme}) lead to a less diverse set of evolved sequences. 
Conversely, ancestors with less epistatically constrained sites (e.g., $A_0$ in Fig.~\ref{fig:Scheme}) lead to a widely diverse set of evolved sequences.

To make this observation quantitative, 
we characterize the
statistical properties of the resulting MSAs, and how they depend on the ancestor, by introducing two distinct averages and corresponding fluctuations, inspired from the disordered systems literature~\cite{franz2011field,franz2013static,Seoane2018,Berthier2007,Folena2022}
and illustrated in Fig.~\ref{fig:Scheme}b.
Recall that the number of accepted mutations $H^{A_0}(t)$ at evolutionary time $t$ for fixed ancestor $A_0$ is a random variabile, whose realizations depend on the 
stochasticity of the evolutionary trajectory.
\begin{itemize}
\item
First, we consider
averaging over many evolutionary trajectories for fixed ancestor. We denote this average as $[ \cdots ]$. 
The variance of $H^{A_0}(t)$ can then be defined as 
\begin{equation}\label{eq:chiA0dyndef}
\chidynA(t) = [H^2] - [H]^2 \ ,
\end{equation}
where the dependence of $H^{A_0}(t)$ on $A_0$ and $t$ is omitted to simplify the notation.
The quantity $\chidynA(t)$ depends on the ancestor $A_0$ and on time~$t$.
\item
Second, we consider the average over different ancestors, which we define as $\langle \cdots \rangle$.
In particular, we consider the average number of mutations $[H^{A_0}(t)]$ for a given ancestor, that measures the average diversity of the evolved MSA,
and we measure how this quantity fluctuates from ancestor to ancestor via the variance
\begin{equation}
    \chiback(t) = \langle [H]^2
 \rangle - \langle [H] \rangle^2 \ ,
\end{equation}
where again the dependence of $[H] =[H^{A_0}(t)]$ on $A_0$ and $t$ is omitted to simplify the notation.
This quantity is the variance of $[H]$ associated to the fluctuations in the background of the ancestral sequence $A_0$, hence the suffix `bg'.
\end{itemize}
Note that the total variance of $H^{A_0}(t)$ over both sources of randomness, i.e. the random choice of ancestor and the stochasticity of mutations along the evolutionary trajectory, can be decomposed as
\begin{equation}
\begin{split}
  \chitot(t) &= \langle [ H^2 ] \rangle - \langle [ H ] \rangle^2 \\
  &= 
  \langle [ H^2 ] \rangle - \langle [ H ]^2 \rangle
  + \langle [ H ]^2 \rangle- \langle [ H ] \rangle^2 \\
&=  \chiback(t) + \chidyn(t)  \ ,
  \end{split}
\end{equation}
where
\begin{equation}
\chidyn(t) = \langle \chidynA(t) \rangle 
\end{equation}
is the average of the ancestor-dependent $\chidynA(t)$ over $A_0$.
Using these quantities, which are called `dynamical susceptibilities' in the physics of disordered systems, we can thus quantify the relative importance of the ancestor and the evolutionary noise in determining the diversity of the resulting MSAs at a fixed evolutionary time.
In the SM we also study some alternative definitions of susceptibilities, based for example on the Hamming distance between two chains starting from the same ancestor. However, because our MCMC algorithm is time-reversible, these two quantities are related.

\section{Results}

\subsection{Mutational dynamics and its fluctuations}
\label{sec:resA}

\begin{figure}[b]
\centering
\includegraphics[width=\linewidth]{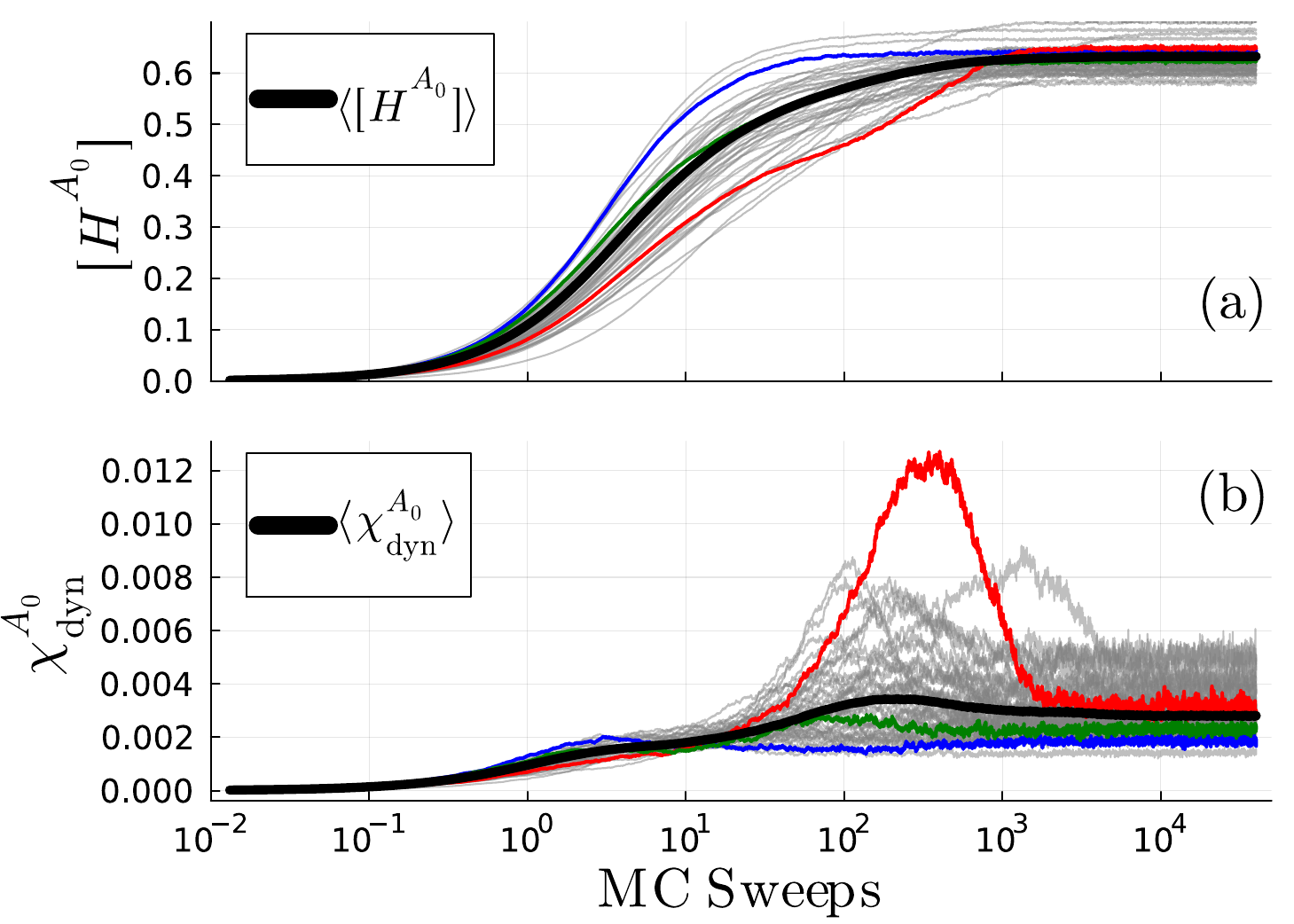}
\caption{Average (a) and variance (b) of the Hamming distance between the evolving sequence and the ancestor estimated using $10^3$ independent trajectories, for many different ancestors (grey lines). 
The thick black line represents the average over ancestors, i.e. $\langle [H^{A_0}(t)] \rangle$ and $\chidyn(t) = \langle \chidynA(t) \rangle$, over 200 ancestors. 
The green, blue and red lines highlights the same specific choices of ancestor used in
Fig.~\ref{fig:Scheme}.
}
\label{fig:HDistanceEvolution}
\end{figure}

The dynamical evolution of the Hamming distance $H^{A_0}(t)$ (number of accepted mutations with respect to the ancestor) is displayed as a function of the number of Monte Carlo sweeps in Fig.~\ref{fig:HDistanceEvolution}, with a sweep corresponding, on average, to one attempted mutation per site.
More specifically, Fig.~\ref{fig:HDistanceEvolution}a shows the average $[H^{A_0}(t)]$ over a star phylogeny of $10^3$ parallel evolutions, for 200 choices of the ancestor $A_0$ taken at random (with weights, see SM) from the DBD protein family.
Fig.~\ref{fig:HDistanceEvolution}b reports the variance $\chidynA(t) = [H^2] - [H]^2$ defined in Eq.~\eqref{eq:chiA0dyndef} over the same phylogeny, and for the same initial sequences as in Fig.~\ref{fig:HDistanceEvolution}a.

Before describing three interesting cases that we highlighted with colors, we discuss the general traits of such dynamics.
The average Hamming distance and its variance are both initially null because all chains start from the same ancestral sequence. 
Because the DCA model is generative and the evolutionary dynamics respects detailed balance, at large times we expect the simulated sequences to be independent samples from the DCA model, which then reproduce statistical features of the natural ones used for training. 
This means that the average of $H^{A_0}(t)$ and its variance converge, respectively, to the average and the variance of the Hamming distance between the chosen ancestral sequence and the rest of the natural ones.  These values can vary significantly (the final average Hamming distance in Fig.~\ref{fig:HDistanceEvolution}a varies from roughly $0.55$ to $0.65$) depending on how close $A_0$ is to the other sequences in the natural MSA.  
The different curves, corresponding to distinct ancestors $A_0$, show a wide range of behaviors and time scales: some trajectories reach the steady state rapidly while others take much longer, displaying intermediate plateaus as a hallmark of epistasis.
In some cases $\chidynA(t)$ displays a peak and then decreases, while in others the equilibrium value is reached in a monotonic way.
The large value reached by $\chidynA(t)$ for some initial sequences implies strong fluctuations between distinct evolutionary trajectories, 
making it hard to predict the dynamics. 

In both panels of Fig.~\ref{fig:HDistanceEvolution}, we highlighted with colors some representative curves.
In particular, the green curve corresponds to an ancestor that behaves in a rather `typical' way, close to the average.
The blue curve corresponds to an ancestor that has less epistatically constrained sites, hence the dynamics is faster and less heterogeneous.
Finally, the red curve corresponds to a highly epistatically constrained ancestor, which leads to a slower dynamics with an intermediate plateau in $[H^{A_0}(t)]$, and
a large peak in $\chidynA(t)$.
The same three ancestors have been used to construct the PCA plots in Fig.~\ref{fig:Scheme} that correspond to time $t=50$ MC Sweeps.

\subsection{Dynamical heterogeneity across families}
\label{sec:resB}

\begin{figure}[t]
\centering
\includegraphics[width=\linewidth]{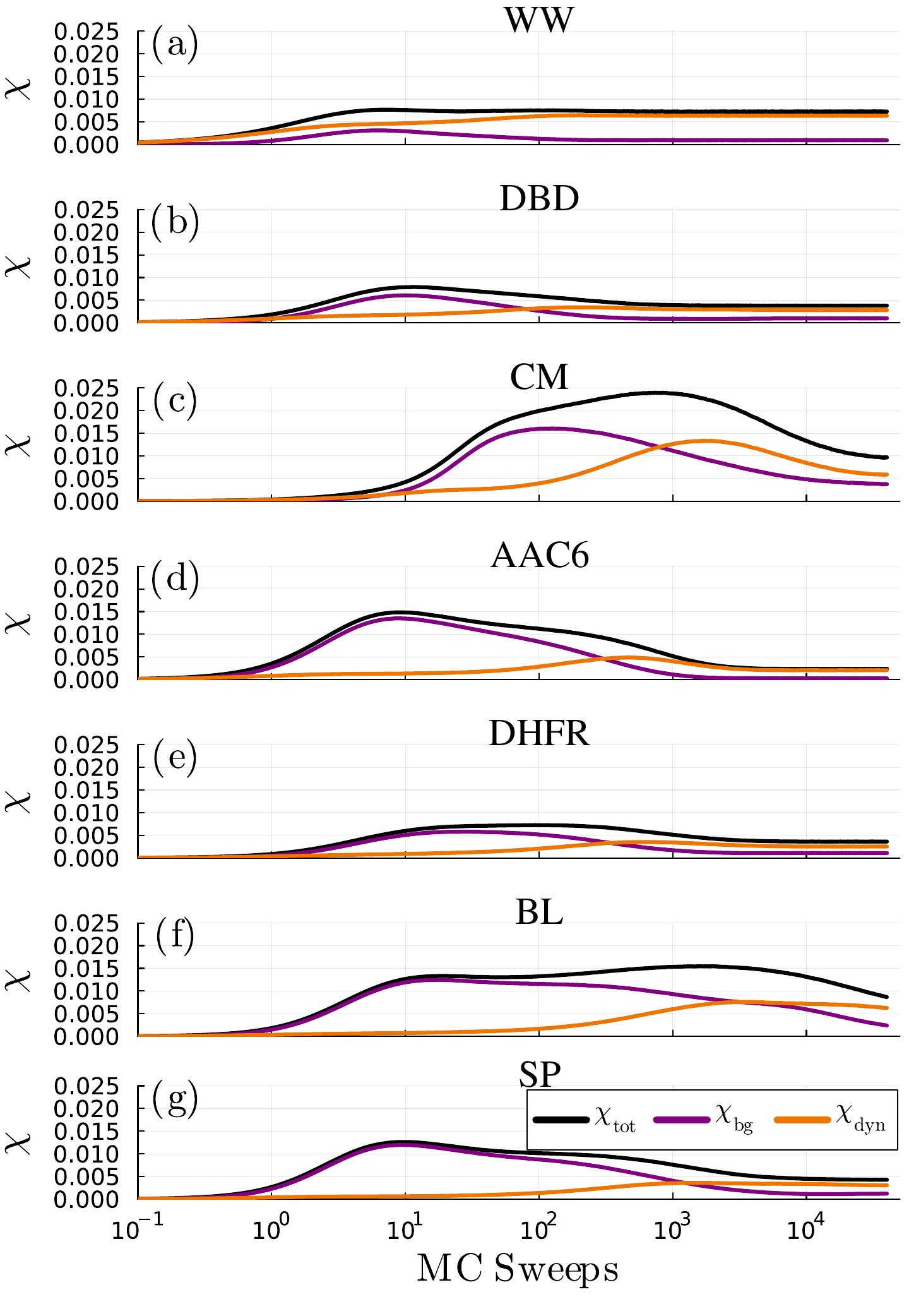}
\caption{Evolution of the different susceptibilities $\chitot$, $\chiback$, and $\chidyn$, for different protein families. (a) WW domain, (b)~DNA-binding domain (DBD), (c) Chorismate Mutase (CM), (d) AAC6, (e) DHFR enzyme, (f) Beta-lactamase (BL), and (g) Serine Protease (SP).
}
\label{fig:SuscEvolution}
\end{figure}

In Fig.~\ref{fig:SuscEvolution} we generalize our results to many protein families, chosen to have different ranges of sequence length, MSA depth (number of natural sequences) and equilibration time scales. These families are also interesting because for some of them, experimental data from Deep Mutational Scans and in vitro evolution are available.

For compactness, we only display the average dynamical fluctuations, with $\chitot$ corresponding to the total variance of $H$, 
$\chidyn$ to the variance due to the stochatisticy of the evolution (averaged over the ancestor), and $\chiback$ to the variance due to the choice of ancestor.
All the $\chitot$ curves display a similar behavior, increasing from zero to a maximum value that is maintained over one or more decades, before finally decreasing to a smaller value corresponding to equilibrium. 
What matters most to us, however, is 
the relative importance of the two terms $\chidyn$ and $\chiback$ in which $\chitot$ can be decomposed.
The key observation is that background-related fluctuations dominate over dynamical ones at short evolutionary times, while the inverse is true at larger times. 
This means that, at least over time scales for which $\chiback \gg \chidyn$, we can hope to reconstruct with good accuracy the ancestral sequence from which a set of evolutionary trajectories started. 
Conversely, at larger time scales, the dynamical noise contribution dominates and the trajectory-to-trajectory fluctuations are large enough
to hide the signal coming from the ancestral sequence, precluding the possibility to reconstruct it.
This behavior is observed in all the protein families that we tested, with the exception of the WW domain in which $\chiback$ never grows to larger values than $\chidyn$. 
We attribute this difference to the small size of the WW domain, which prevents the sequences to accumulate enough epistatic interactions.
However, both the time scales and the relative importance of the two terms contributing to the total susceptibility vary greatly from family to family (Fig.~\ref{fig:SuscEvolution}).
In DBD, AAC6, DHFR, and SP the peak of $\chitot$ is reached quite early in the dynamics and the two contributions $\chidyn$ and $\chiback$ are almost non-overlapping.  
On the other hand, for CM and BL, the peak occurs much later due to the two contributions having a significant overlap. 

For a given family and a typical ancestor, the time scale at which $\chidyn$ and $\chiback$ cross, the former becoming larger than the latter, defines a characteristic 
evolutionary time scale, around which the memory of the ancestor is lost.

\subsection{Epistatic constraints and dynamical fluctuations}
\label{sec:resC}

\begin{figure}
\centering
\includegraphics[width=\linewidth]{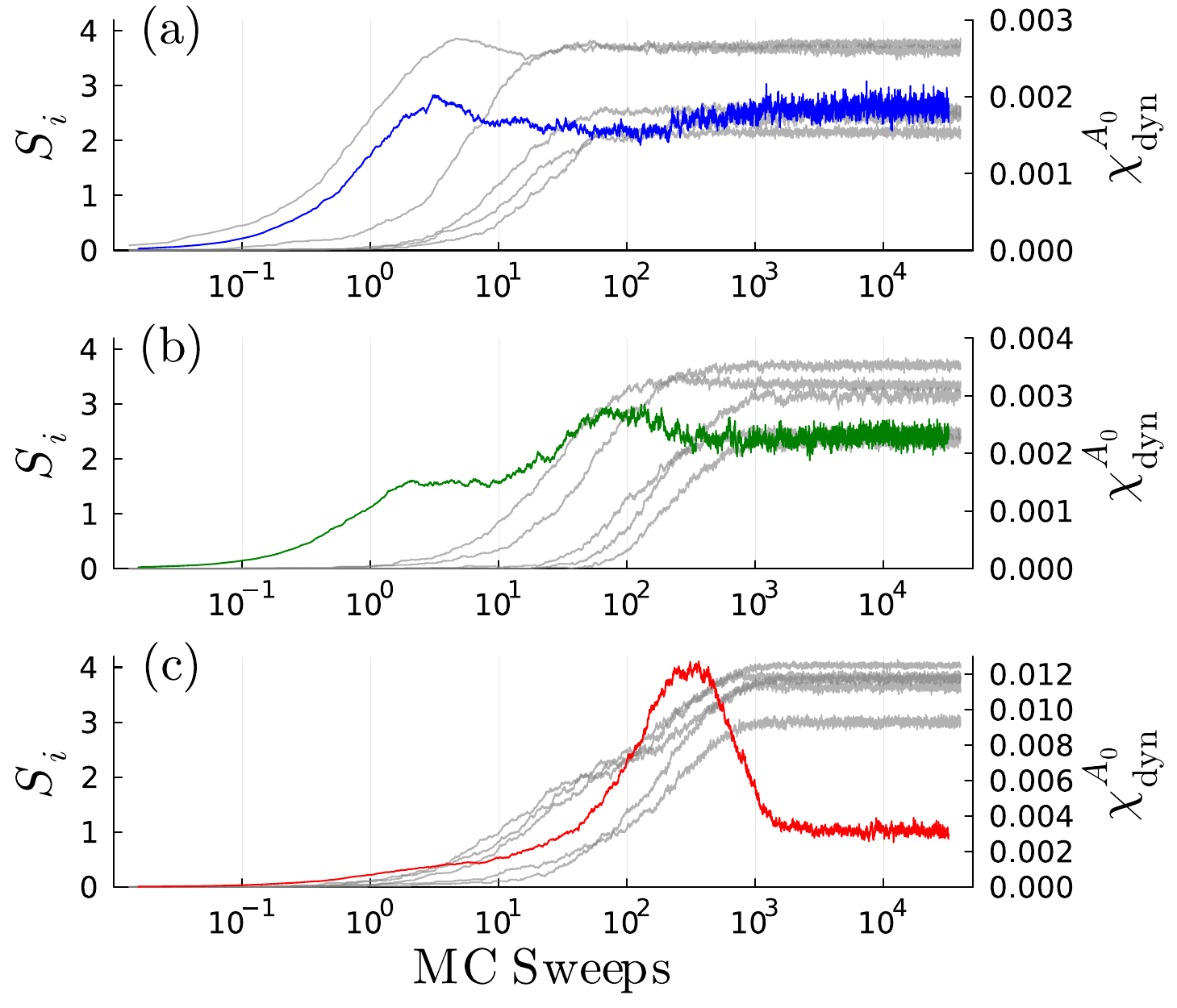}
\caption{Evolution of the entropy of the most epistatically constrained sites for a specific ancestral sequence of DBD plotted together with $\chidynA$ of the same sequence. (a), (b), and (c) are for the blue, green, and red sequences in Fig.~\ref{fig:HDistanceEvolution}, respectively.}
\label{fig:EpistaticityAndChiDyn_timescales}
\end{figure}

Up to this point, we discussed the time dependence of the different contributions to the total fluctuations. 
We argued that, as long as the dominant contribution to $\chitot$ comes from $\chiback$, the ancestral sequence is strongly related to the evolving ones. 
We now want to better understand the transition to the final regime in which dynamical fluctuations, i.e., $\chidyn$, dominate. 

In Fig.~\ref{fig:EpistaticityAndChiDyn_timescales} we show, for the three ancestral sequences highlighted in Fig.~\ref{fig:HDistanceEvolution}, 
how the dynamical part of the susceptibility is strongly related to the epistatically constrained sites (defined as in Sec.~\ref{sec:CDECIE} and in Ref.~\cite{DiBari2024}).
To make this relation clear, we first checked that the time scale at which the dynamical susceptibility reaches its peak is compatible with the time scale at which epistatically constrained sites evolve. 
For a given ancestor $A_0$,
we consider the five most epistatically constrained sites, i.e. those with the largest ${\rm CIE}_i - {\rm CDE}_i^{A_0}$.
For these sites, we consider at time $t$ the frequency of appearance of amino acid $a$, $f_i^t(a)$, in the set of sequences that evolved from $A_0$,
and from it we compute a time-dependent entropy 
\begin{equation}
S_i(t) = -\sum_{a=1}^{21} f_i^t(a) \log_2 f_i^t(a) \ .
\end{equation}
In Fig.~\ref{fig:EpistaticityAndChiDyn_timescales}, $\chidynA(t)$ (colored curve) is superposed to $S_i(t)$ of the five sites (gray curves), as a function of evolutionary
time $t$.
We observe that the peak of $\chidynA$ is reached just before the equilibration of the epistatically constrained sites, i.e. the time at which $S_i(t)$ approaches
the CIE$_i$.
The time of equilibration of epistatically constrained sites also increases from Fig.~\ref{fig:EpistaticityAndChiDyn_timescales}a to Fig.~\ref{fig:EpistaticityAndChiDyn_timescales}c, similarly to what happens for the equilibration of the Hamming distance in the same three sequences in Fig.~\ref{fig:HDistanceEvolution}. 
Epistasis does not only affect the time scale at which the peak of $\chidynA$ is reached, but also the intensity of the peak,
which means that more epistatic ancestors lead to a more heterogeneous dynamics at intermediate times, when the epistatic
sites mutate.

\begin{figure}
\centering
\includegraphics[width=\linewidth]{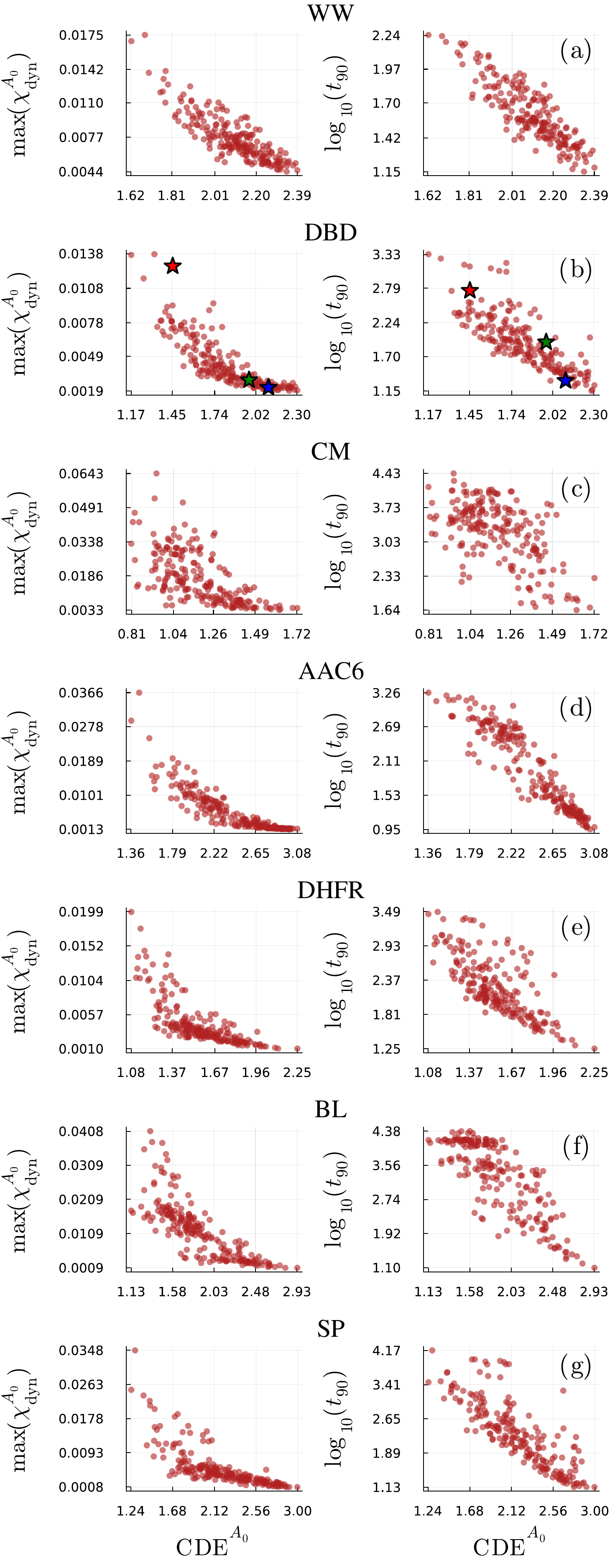}
\caption{Scatter plot of $\textrm{CDE}^{A_0}$ against the maximum of $\chidynA$ (left) and against $\log_{10}(t_{90})$ (right) for each protein family we considered.
For each family, the points correspond to 200 sequences $A_0$ extracted from the natural
MSA with weights (see SM). 
In (b) colored stars are used to highlight the sequences that correspond to the blue, green, and red curves in Fig.~\ref{fig:HDistanceEvolution}. 
}
\label{fig:EpistaticityAndChiDyn_maximum}
\end{figure}

To quantify these observations,
in Fig.~\ref{fig:EpistaticityAndChiDyn_maximum} we consider the same set of families as in Fig.~\ref{fig:SuscEvolution}, namely WW, DBD, CM, AAC6, DHFR, BL, and SP,
spanning a wide range of sequence length (from $L=31$ in WW to $L=220$ in SP).
For each family, we consider a set of ancestors $A_0$
and we report a scatter plot of the following quantities, each $A_0$ being a point:
\begin{itemize}
\item
the value of $\textrm{CDE}^{A_0}$ defined as in Eq.~\eqref{eq:globalCDE},
\item
the maximum value $\max_t \chidynA(t)$, indicated as $\max(\chidynA)$ for simplicity,
\item
 and the time $t_{90}$
at which the average $[H^{A_0}]$ over the trajectories reaches $90 \%$ of the equilibrium ({$t\to\infty$}) value.
\end{itemize}
We observe that both $\max(\chidynA)$ and $t_{90}$ markedly increase as $\textrm{CDE}^{A_0}$ decreases.
Hence, a `highly epistatic' sequence with a small value of $\textrm{CDE}^{A_0}$ (with respect to the typical value of the family) 
has many sites with low context-dependent entropy, which cannot mutate at the beginning of evolution,
leading to a slower and more heterogeneous overall dynamics.
This is for example the case of the red sequence in Fig.~\ref{fig:HDistanceEvolution}, for which we have $\textrm{CDE}^{A_0} = 1.45$, 
while for the blue one $\textrm{CDE}^{A_0} = 2.11$. The three sequences considered in
Fig.~\ref{fig:HDistanceEvolution} are indicated with stars of the corresponding colors in Fig.~\ref{fig:EpistaticityAndChiDyn_maximum}b.

Epistatically constrained sites also carry the specific signature of the ancestral sequence, 
because the conserved sites are roughly the same for every sequence of the family and the variable ones carry little to no information. 
These results thus suggest that 
tracing back an evolutionary trajectory to its ancestral sequence becomes more difficult as one approaches the peak of $\chidyn$, because this is when the epistatic sites, i.e. the sites that carry information about that initial sequence, start to mutate.

\subsection{Cooperative mutational dynamics}
\label{sec:resD}

We have established that, when epistatic sites start to mutate, $\chidynA$ reaches its peak, and that the intensity of this peak is correlated with the amount of 
epistatically correlated sites in the ancestral sequence.
We now want to show that the peak of $\chidynA$ is due to
dynamical correlations between different sites, 
caused by the strong interaction between those sites and the context.
These correlations result in a cooperative mutational process, in which epistatically constrained sites can only mutate because other such sites mutate,
leading to an avalanche of mutations.

\begin{figure*}
\centering
\includegraphics[width=\linewidth]{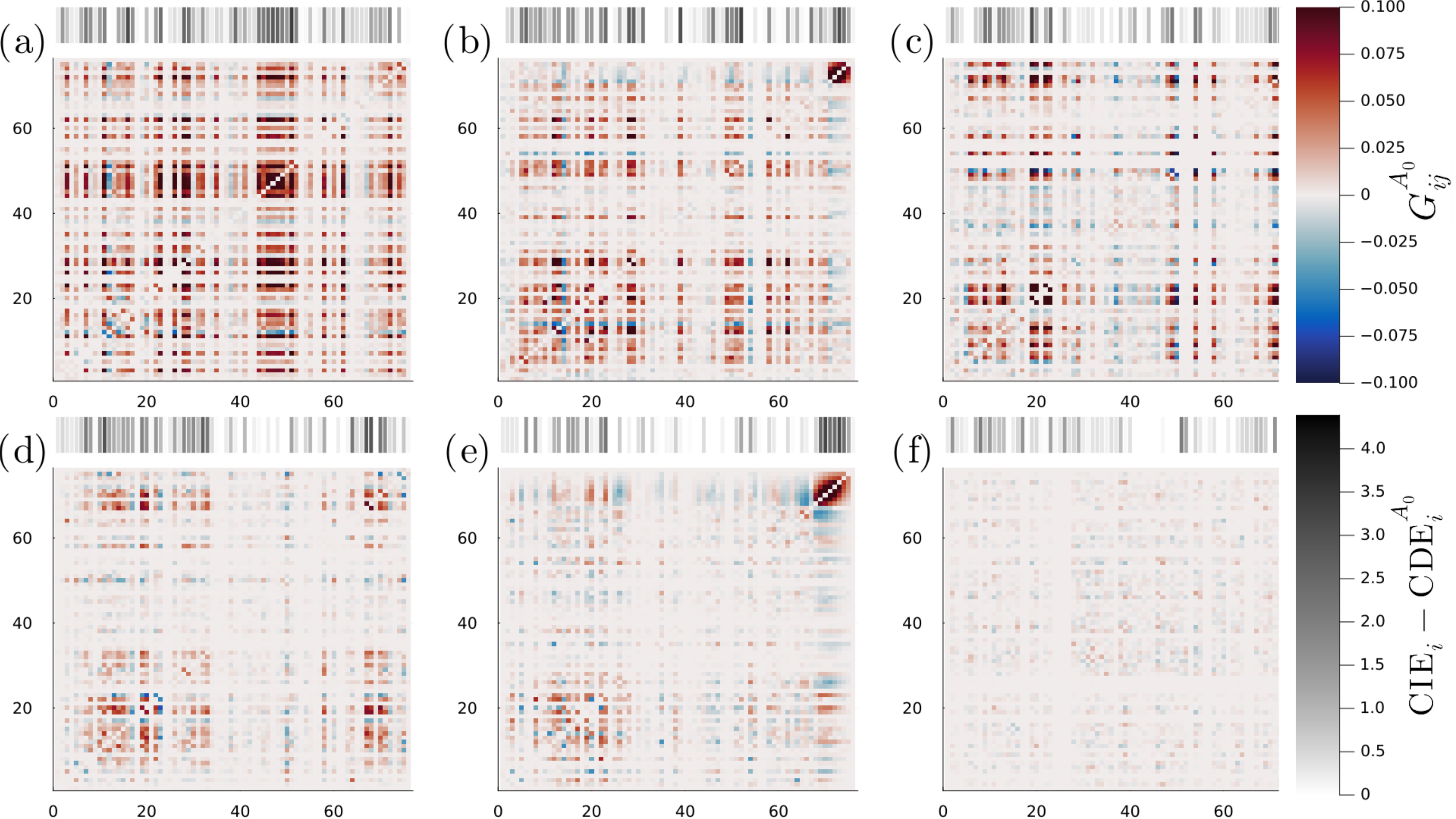}
\caption{Correlations between sites at time $t^*$ at which $\chidynA(t)$ reaches its maximum (with a cutoff at $t_{\rm th} = 1000$ sweeps), 
for different ancestral sequences. The bar above each snapshot encodes the strength of epistatic constraints for the sites in the ancestral sequence.}
\label{fig:Correlations}
\end{figure*}

To precisely quantify this effect,
one can interpret the dynamical susceptibility as the sum of a site-site dynamical correlation function~\cite{berthier2011dynamical}.
The Hamming distance 
is defined in Eq.~\eqref{eq:HA0def}.
Inserting its expression into Eq.~\eqref{eq:chiA0dyndef}, we obtain
\begin{equation}
\label{eq:SuscAndCorr}
\chidynA(t) = \frac{1}{L^2} \sum_{ij} G^{A_0}_{ij}(t) \ ,
\end{equation}
with 
\begin{equation}
\label{eq:CorrDef}
G^{A_0}_{ij}(t) = [\delta_{a^t_i,a^0_i} \delta_{a^t_j, a^0_j}] - [\delta_{a^t_i,a^0_i}] [\delta_{a^t_j, a^0_j}] \ ,
\end{equation}
recalling that the Kronecker delta symbol $\delta_{a^t_i,a^0_i}$
is zero if site $i$ is mutated between the two sequences $A_t$ and $A_0$, 
and one otherwise.
The matrix $G^{A_0}_{ij}$ is a time-dependent correlation matrix between sites $i$ and $j$. 
It is high when the sites are dynamically correlated, i.e. if $a^t_i$ is different from $a^0_i$, then $a^t_j$ is likely different from $a^0_j$ as well, and vice versa. It is low when the sites mutate independently.
Because $\chidynA$ is the sum of $G^{A_0}_{ij}$ over all pairs $(i,j)$, a large value of $\chidynA$  implies that many $(i,j)$ are strongly correlated.
Furthermore, by looking at this correlation matrix one can infer which pairs of sites mutate in a correlated way during evolution.

The value of the off-diagonal part of $G^{A_0}_{ij}(t^*)$, computed at the time $t^*$ at which a peak in the dynamical susceptibility is observed, is shown in Fig.~\ref{fig:Correlations} for six different starting sequences in the DBD family. 

Fig.~\ref{fig:Correlations}a corresponds to an ancestor (red in Fig.~\ref{fig:HDistanceEvolution})
for which a large peak and a longer time to equilibrate are observed in the evolution of $\chidynA$. 
The large value of $\chidynA$, which is just the sum of all matrix elements, implies a large number of positively correlated sites.
Fig.~\ref{fig:Correlations}e corresponds to an ancestor (green in Fig.~\ref{fig:HDistanceEvolution}) for which $\chidynA$ has a smaller peak before saturating. We notice in this case that $G^{A_0}_{ij}(t^*)$ still displays some strong correlations, but the majority of the sites are uncorrelated. 
Finally, Fig.~\ref{fig:Correlations}f corresponds to an ancestor (blue in Fig.~\ref{fig:HDistanceEvolution}) for which $\chidynA$ saturates to a small value. In this case correlations are almost absent at $t^*$, which means that essentially all sites mutate independently.
It is thus evident how different initial sequences give rise to different patterns in the $G^{A_0}_{ij}$ matrix.
At large times the chains reach equilibrium and are almost statistically indistinguishable from natural sequences. This means that $G^{A_0}_{ij}(t \to \infty) \approx G^{A_0}_{ij, {\rm nat}}$, with 
\begin{equation*}
G^{A_0}_{ij, {\rm nat}} =  \langle \delta_{a^{\rm nat}_i,a^0_i} \delta_{a^{\rm nat}_j ,a^0_j} \rangle_{\rm nat}  -  \langle \delta_{a^{\rm nat}_i,a^0_i}\rangle_{\rm nat} \langle \delta_{a^{\rm nat}_j ,a^0_j}\rangle_{\rm nat}, 
\end{equation*}
where $\langle \dots \rangle_{\rm nat}$ is the average computed over the natural sequences.

In section~\ref{sec:resC}, we showed that the presence of a peak in $\chidynA$  is correlated to $\textrm{CDE}^{A_0}$, 
i.e. to the amount of epistatically constrained sites in the ancestral sequence.
In this section, we show that these sites also display large dynamical correlations between themselves.
The bar above each plot in Fig.~\ref{fig:Correlations} is shaded to represent $\textrm{CIE}_i - \textrm{CDE}^{\rm A}_i$, capturing the information about site \(i\) encoded in the background sequence \(A_{\setminus i}\). This coloring reflects the degree to which each site is constrained by epistatic interactions.
The conserved and variable sites are indicated in white, while the epistatically constrained ones are shown in black.
As expected, we see good agreement between these residues and the ones that give a large contribution to $\chidynA$.
Hence, we conclude that epistatically constrained sites mutate cooperatively around the time scale corresponding to the peak of $\chidynA$.

\subsection{Response to environmental variation}
\label{sec:resE}

In the literature on disordered physical systems, it has been established that the dynamical correlations discussed in sections~\ref{sec:resC} and \ref{sec:resD}
are related to the linear response of the dynamics to a change in temperature through a kind of fluctuation-dissipation relation~\cite{Berthier2005}. 

In the present context, the `temperature' is a parameter that controls the probability with which mutations are accepted.
High temperature corresponds to all mutations being accepted (very low selection), 
while low temperature corresponds to only the most beneficial mutations being accepted (very strong selection, i.e. directed evolution).
The value of $T=1$ corresponds to the conditions at which the DCA model is trained on natural sequences, hence $T\sim 1$ corresponds to a neutral drift
dynamics during which the neutral space of sequences that have comparable fitness to natural ones is explored~\cite{bisardi2022modeling}.
Indeed, Ref.~\cite{bisardi2022modeling} has shown that two distinct in vitro evolution experiments, realized with selection pressures comparable to the natural one,
can be described by fitting the temperature in a range slightly above $T\sim 1$. 

\begin{figure}
\centering
\includegraphics[width=\linewidth]{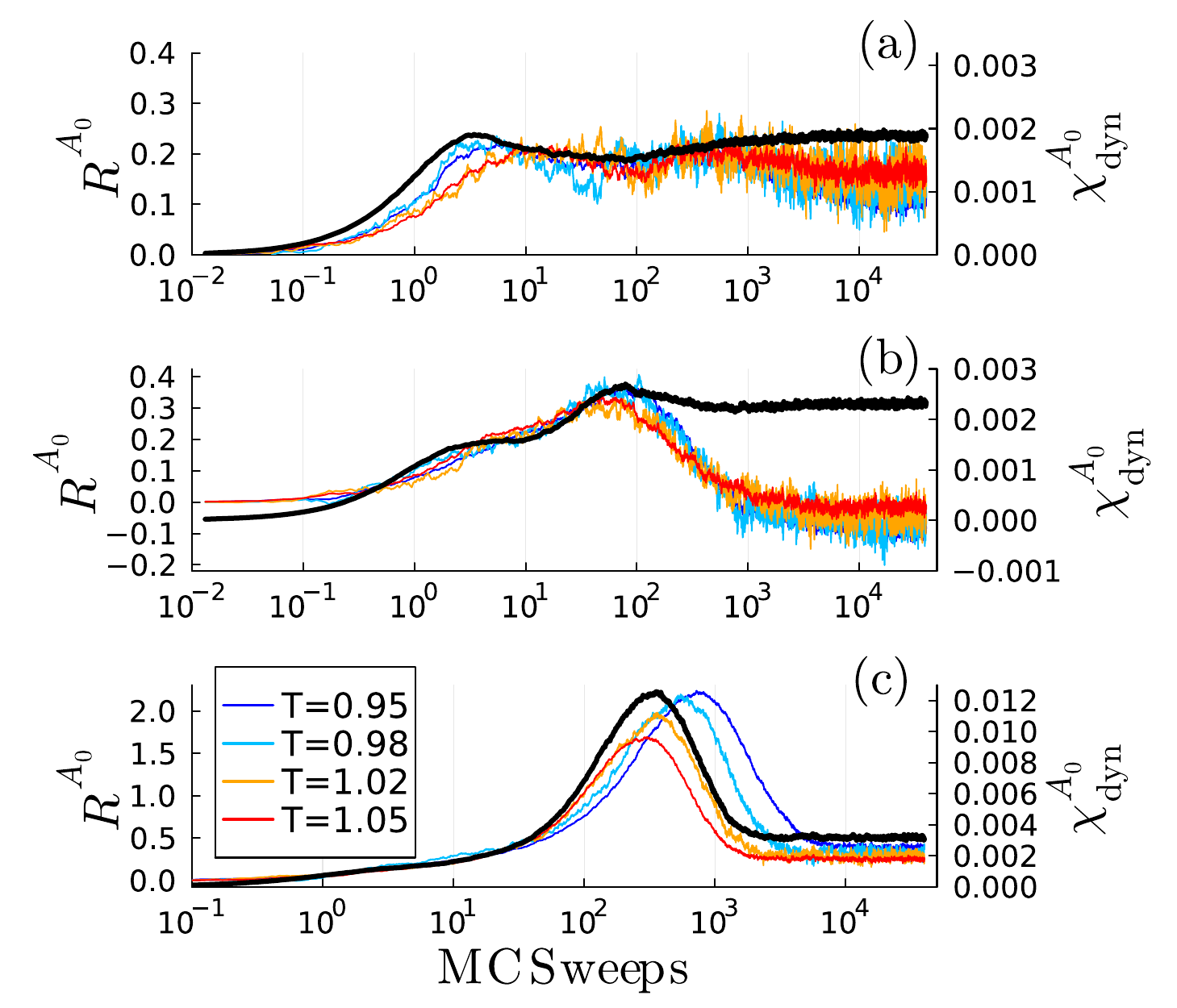}
\caption{Evolution of the dynamical susceptibility $\chidynA(t)$ compared with the linear response of the average Hamming distance to a change of temperature $R^{A_0}(t)$. The three panels correspond to the three ancestral sequences highlighted in Fig.~\ref{fig:HDistanceEvolution},
(a)~blue, (b)~green, (c)~red.}
\label{fig:LinearResponse}
\end{figure}

Hence, in our modeling framework, a change of temperature corresponds to a change of selection strength. 
Following Ref.~\cite{Berthier2005}, 
we check whether the dynamical susceptibilities are related to the response of the average dynamics with respect to the temperature variation.
To estimate such linear response, we consider two evolutionary dynamics starting from the same ancestor, one at temperature $T \neq 1$ and the other at temperature $T=1$,
and we compare the average number of accepted mutations $[H^{A_0}(t; T)]$ between the two evolutions. The dynamical linear response is given by
\begin{equation}
R^{A_0}(t) = \lim_{T\to 1} \frac{[H^{A_0}(t; T)] - [H^{A_0}(t; T=1)]}{T-1} \ ,
\end{equation}
and we emphasize that this quantity depends on the ancestor $A_0$ and on time $t$.
In Fig.~\ref{fig:LinearResponse}, we compare the time dependence of $\chidynA(t)$ with $R^{A_0}(t)$ for the same three ancestors as 
in Fig.~\ref{fig:HDistanceEvolution}.
Note that we cannot take the limit $T\to 1$ due to statistical noise, and we estimate $R^{A_0}(t)$ by observing that the curves for 
several values of $T$ close to 1 are almost superimposed to each other.
Fig.~\ref{fig:LinearResponse} shows that the fluctuation-dissipation relation seems to hold quite well, 
such that $\chidynA(t) \propto R^{A_0}(t)$, at least for times $t$ smaller than the peak of both quantities.

The results in Fig.~\ref{fig:LinearResponse} 
suggest that more epistatically constrained ancestral sequences, which display a stronger peak in the dynamical susceptibility, will also display a stronger response of the evolutionary dynamics to a small change in selection strength. 
Such response is stronger around the time of the peak in $\chidynA(t)$, which is also the time at which epistatically constrained sites mutate cooperatively,
see section~\ref{sec:resD}.
In other words, {\it sequences with more epistatically constrained sites are more sensitive to environmental changes}. 
In the SM, we give an additional analytical argument based on the short-time dynamics, that supports the same conclusion.

\subsection{Limits of ancestral sequence reconstruction}
\label{sec:resF}

In an evolutionary process, predictions can be made in two directions: either starting from the ancestral sequence and predicting its future evolution over a given time; or starting from a set of evolved sequences and attempting to infer the ancestral sequence from which the evolution began.
An attempt at predicting future evolution has been made in Ref.~\cite{bisardi2022modeling} (see Refs.~\cite{widmer2008irreversible,schoenholz2016structural,bapst2020unveiling,jung2025roadmap,jung2024dynamic} for related studies in disordered systems), where however the prediction was limited to small evolutionary time scales.
Because we have repeatedly hinted that our analysis can provide insight into the time scale over which reconstructing the ancestral sequence of an evolutionary process is possible, before concluding, we consider here the ancestral sequence reconstruction problem more explicitly.

\begin{figure}
\centering
\includegraphics[width=\linewidth]{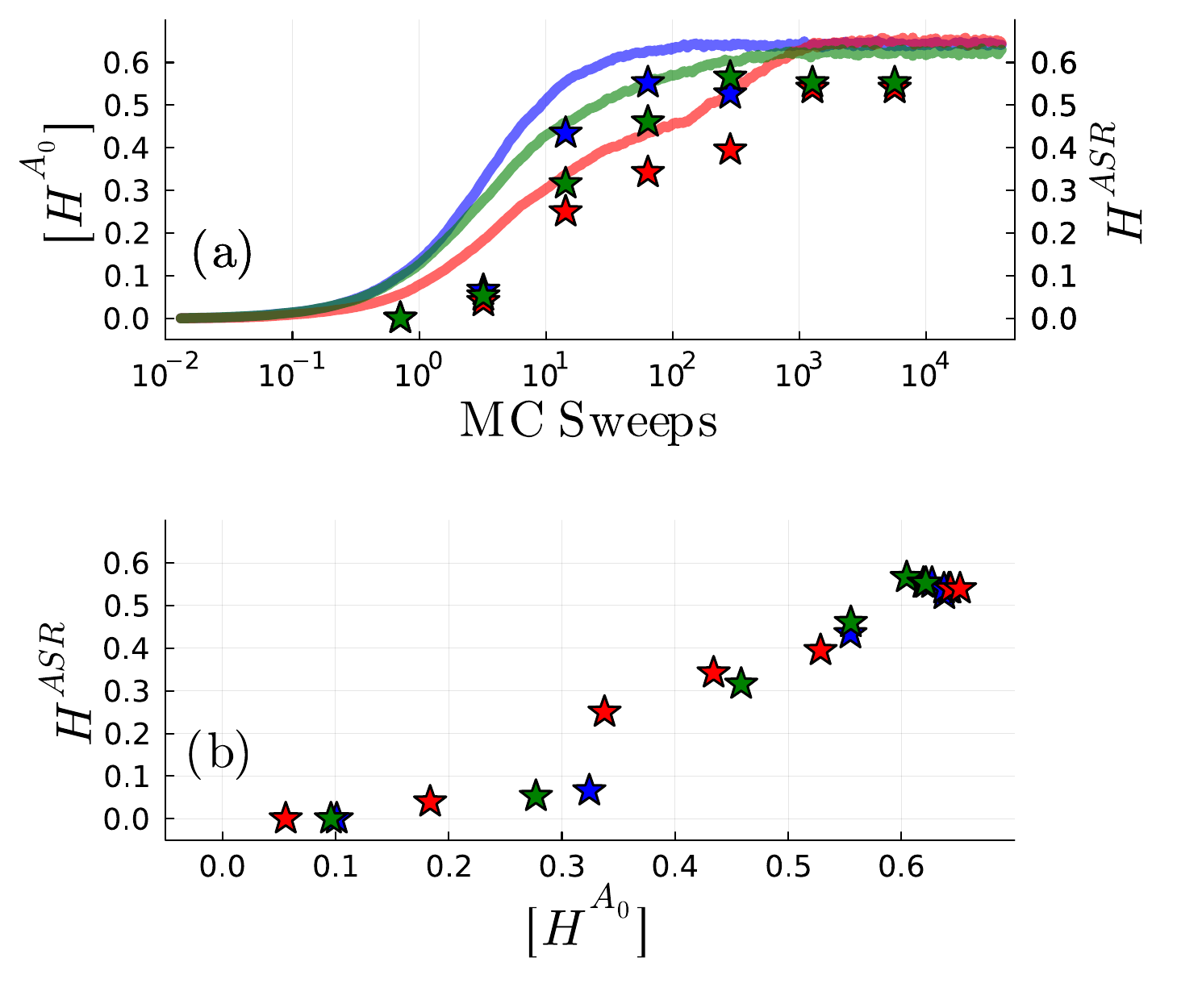}
\caption{(a)~The full lines represent the evolution of the average Hamming distance for the same three ancestors as in Fig.~\ref{fig:HDistanceEvolution}. The star symbols represent the Hamming distance $H^{\rm ASR}$ between the ground truth ancestor and the one reconstructed using the simulated MSAs at different times.
(b)~Hamming distance $H^{\rm ASR}$ between the reconstructed ancestor and the ground truth as a function of the average distance $[H^{A_0}(t)]$ between the ancestor and the MSA used for the  reconstruction. The colors refer to the same sequences of panel~(a). }
\label{fig:ASR_vs_Hamming_over_t}
\end{figure}

For each of the three sequences we analyzed in Fig.~\ref{fig:HDistanceEvolution}, we 
simulate 200 independent evolutionary trajectories, thus obtaining
several MSAs of 200 evolved sequences at different evolutionary times $t$. 
Then, we use the FastML online tool~\cite{FastML} to infer the ancestral sequence starting from those MSAs 
and using a star-shaped phylogenetic tree.
We then check how close the reconstructed sequence is to the actual ancestor. 

The results are shown in Fig.~\ref{fig:ASR_vs_Hamming_over_t}a, where we plot the evolution of the average Hamming distance for the same sequences as in Fig.~\ref{fig:HDistanceEvolution}, and we compare it with the Hamming distance (i.e. the percentage of errors) between the ground truth ancestor and the reconstructed one.
At the beginning of the dynamics the FastML tool is able to reconstruct perfectly the original sequence in all of the three cases. 
This is not surprising, as at short times few mutations appear. 
The situation is similar at large times, where the MSAs generated from the three starting sequences are statistically indistinguishable and the FastML tool performs equally bad in all of them~\cite{gascuel2010inferring,evans2000broadcasting}. 
Thus, it is more insightful to look at intermediate times, where a significant difference between the evolving sequences is present. 
After $\sim 10$ sweeps the variable sites begin to mutate and the ancestral sequence reconstruction tool cannot reproduce the initial sequence perfectly. 
The leap in the error percentage with respect to the previous point is however larger for the blue sequence, which is the one with fewer epistastic sites, while the red one is still recovered reasonably well. This difference is present for the subsequent points as well, until the error saturates. 
In Fig.~\ref{fig:ASR_vs_Hamming_over_t}b we show that the Hamming distance (percentage of errors) of the sequence reconstruction results grows 
roughly proportionally to the average sequence divergence of the evolutionary trajectories
for all of the three studied sequences, but more so for the more epistatic red sequence.
We conclude that, at least with the FastML tool, the possibility of reconstructing ancestral sequences is mostly determined by the Hamming distance of the ancestor from the evolving sequences, i.e. $[H^{A_0}(t)]$,
with more errors accumulating when using datasets of more diverged sequences. 

Our analysis shows that, for a given evolutionary time,  
the amount of diversity $[H^{A_0}(t)]$ depends strongly on the ancestor, 
due to highly non-trivial epistatic dynamical correlations. More epistatically constrained ancestors give rise to less diversity, thus allowing for reconstruction over longer evolutionary times (Fig.~\ref{fig:ASR_vs_Hamming_over_t}a). 
Yet, at comparable amount of diversity, more epistatic ancestors are more difficult to reconstruct (Fig.~\ref{fig:ASR_vs_Hamming_over_t}b), at least using the FastML algorithm
that neglects correlations between sites~\cite{felsenstein}.
These results confirm once more that the evolutionary time limit for predictability is linked to the presence of epistatic sites that carry information about the starting sequence only up to a certain sequence divergence. We further observe that more advanced tools (yet to be fully developed) for ancestral sequence reconstruction might exploit
the epistatic correlations to achieve a better reconstruction~\cite{deleonardis2024reconstruction}, which is not done in FastML and similar algorithms.
We expect that such tools would perform better on more epistatic ancestors such as the red one in Fig.~\ref{fig:ASR_vs_Hamming_over_t}.

\section{Conclusions}

In this study, we explore the role of disorder and stochasticity in protein evolution, focusing on the interplay between epistasis, site correlations, and sequence-dependent fluctuations. 
We simulated evolution in silico using DCA to model the fitness landscape, and an MCMC algorithm to mimic the process of mutation and selection, a methodology that has been validated in previous work~\cite{de2020epistatic,bisardi2022modeling,DiBari2024}.
By employing tools from statistical physics~\cite{kirkpatrick1988comparison,franz1999dynamical,franz2000non,bouchaud2005nonlinear,Berthier2005,franz2011field,berthier2011dynamical,franz2013static,Seoane2018,Berthier2007}, we quantified how the ancestral amino acid sequence significantly influences early evolutionary dynamics, as measured by the dynamical susceptibility.
Our main results are the following.
\begin{itemize}
\item 
Our analysis shows that, during the initial phase of evolution, there is very significant heterogeneity in the dynamics, depending on the choice of ancestral sequence. 
Some ancestral sequences lead to a smooth evolution where each site mutates almost independently of the others, while others lead to a more complex dynamics characterized by intermediate plateaus and significant fluctuations (Sec.~\ref{sec:resA}).

\item
We have shown that, for a variety of protein families,
the noise arising from the starting sequence dominates over stochastic evolutionary fluctuations, implying that one can, in principle, trace the evolutionary trajectory back to its origin (Sec.~\ref{sec:resB}). This effect is more or less pronounced, depending on the family, and our tools allow one to quantify it.
However, as time progresses and epistatically constrained sites evolve, this traceability is lost due to the growing influence of the mutational stochasticity.

\item 
The heterogeneity of the initial evolutionary dynamics can be traced back to the amount
of epistatic constraints in the ancestral sequence. We introduced a quantity, the
average over sites of the context-dependent entropy, and we have shown that this quantity is
strongly correlated with the time at which dynamical heterogeneity reaches its peak and with the strength of the fluctuations at the peak (Sec.~\ref{sec:resC}).
This method for assessing sequence evolvability could guide experimentalists in selecting an appropriate starting sequence for experiments, based on the desired functional behavior. 

\item
The amplitude of the global fluctuations can be expressed in terms of a sum of pairwise dynamical correlations between pairs of residues. More epistatically constrained ancestral sequences show groups of residues that mutate collectively, and those patterns can be identified from the correlation matrix $G_{ij}^{A_0}(t)$ we introduced in Sec.~\ref{sec:resD}.
The observed correlations between sites thus reflect a complex epistatic landscape, where certain residues evolve in a highly context-dependent manner. 
These correlations could also be leveraged in experimental settings; for instance, one may imagine to control the evolvability of a specific residue by targeting sites that exhibit strong correlations with it.

\item
We demonstrated that these epistatic sites, which evolve over long time scales, significantly affect the response of the dynamics to a change in environmental conditions. More epistatically constrained sequences lead to a larger response to a change in environment,
suggesting potential implications on sequence evolvability imposed by environmental pressures, such as antibiotic concentration (Sec.~\ref{sec:resE}).
\item
Finally, we presented a preliminary study of the performance of an Ancestral Sequence Reconstruction (ASR) algorithm (here, FastML), in light of our previous findings. We found that more epistatically constrained ancestors lead to less diversity at comparable time scales, which facilitates their reconstruction. Yet, at comparable diversity, they are more difficult to reconstruct. We believe that this analysis will be instrumental in improving the efficiency of ASR algorithms, which could in principle exploit the correlations identified in this work.
\end{itemize}
More generally, our findings extend the analogy between protein evolution and disordered physical systems, reinforcing the idea that protein dynamics exhibit characteristics of complex and strongly correlated systems. 
However, when comparing these results with statistical physics models exhibiting glassy dynamics, we also found qualitative differences, emphasizing the unique constraints imposed by natural evolution on proteins. 
Future work should aim to further elucidate the connection between epistasis, evolvability, and environmental selection, which may offer insights into evolutionary strategies across diverse biological systems. 
Moreover, more realistic evolutionary dynamics could be considered. 
In this paper we focused on independent Monte Carlo chains, which corresponds to evolution on a star tree.
Introducing a more complex tree structure could affect the results presented here, as the varying distance between the leaves of the tree and its root and the correlations coming from a common ancestor may affect the computation of $\chidynA$.
Furthermore, we believe that most of the ideas presented in this work will soon be amenable to experimental testing, thanks to the increased power of in vitro evolution platforms.

\acknowledgments

We thank Maria Chiara Angelini, Ludovic Berthier, Pierre Barrat-Charlaix, Simona Cocco, Dongkyu Lee, Arvind Murugan, Cl\'ement Nizak, Misaki Ozawa, Andrea Pagnani, Olivier Rivoire, Joe Thornton, Nobuhiko Tokuriki, and Alya Zeinaty for fruitful discussions. 
This research has been supported by first FIS (Italian Science Fund) 2021 funding scheme (FIS783 - SMaC - Statistical Mechanics and Complexity) from MUR, Italian Ministry of University and Research and from the PRIN funding scheme (2022LMHTET - Complexity, disorder and fluctuations: spin glass physics and beyond) from MUR, Italian Ministry of University and Research.

\bibliography{references_nourl} 

\begin{thebibliography}{86}%
\makeatletter
\providecommand \@ifxundefined [1]{%
 \@ifx{#1\undefined}
}%
\providecommand \@ifnum [1]{%
 \ifnum #1\expandafter \@firstoftwo
 \else \expandafter \@secondoftwo
 \fi
}%
\providecommand \@ifx [1]{%
 \ifx #1\expandafter \@firstoftwo
 \else \expandafter \@secondoftwo
 \fi
}%
\providecommand \natexlab [1]{#1}%
\providecommand \enquote  [1]{``#1''}%
\providecommand \bibnamefont  [1]{#1}%
\providecommand \bibfnamefont [1]{#1}%
\providecommand \citenamefont [1]{#1}%
\providecommand \href@noop [0]{\@secondoftwo}%
\providecommand \href [0]{\begingroup \@sanitize@url \@href}%
\providecommand \@href[1]{\@@startlink{#1}\@@href}%
\providecommand \@@href[1]{\endgroup#1\@@endlink}%
\providecommand \@sanitize@url [0]{\catcode `\\12\catcode `\$12\catcode `\&12\catcode `\#12\catcode `\^12\catcode `\_12\catcode `\%12\relax}%
\providecommand \@@startlink[1]{}%
\providecommand \@@endlink[0]{}%
\providecommand \url  [0]{\begingroup\@sanitize@url \@url }%
\providecommand \@url [1]{\endgroup\@href {#1}{\urlprefix }}%
\providecommand \urlprefix  [0]{URL }%
\providecommand \Eprint [0]{\href }%
\providecommand \doibase [0]{http://dx.doi.org/}%
\providecommand \selectlanguage [0]{\@gobble}%
\providecommand \bibinfo  [0]{\@secondoftwo}%
\providecommand \bibfield  [0]{\@secondoftwo}%
\providecommand \translation [1]{[#1]}%
\providecommand \BibitemOpen [0]{}%
\providecommand \bibitemStop [0]{}%
\providecommand \bibitemNoStop [0]{.\EOS\space}%
\providecommand \EOS [0]{\spacefactor3000\relax}%
\providecommand \BibitemShut  [1]{\csname bibitem#1\endcsname}%
\let\auto@bib@innerbib\@empty
\bibitem [{\citenamefont {Blum}\ \emph {et~al.}(2024)\citenamefont {Blum}, \citenamefont {Andreeva}, \citenamefont {Florentino}, \citenamefont {Chuguransky}, \citenamefont {Grego}, \citenamefont {Hobbs}, \citenamefont {Pinto}, \citenamefont {Orr}, \citenamefont {Paysan-Lafosse}, \citenamefont {Ponamareva}, \citenamefont {Salazar}, \citenamefont {Bordin}, \citenamefont {Bork}, \citenamefont {Bridge}, \citenamefont {Colwell}, \citenamefont {Gough}, \citenamefont {Haft}, \citenamefont {Letunic}, \citenamefont {Llinares-López}, \citenamefont {Marchler-Bauer}, \citenamefont {Meng-Papaxanthos}, \citenamefont {Mi}, \citenamefont {Natale}, \citenamefont {Orengo}, \citenamefont {Pandurangan}, \citenamefont {Piovesan}, \citenamefont {Rivoire}, \citenamefont {Sigrist}, \citenamefont {Thanki}, \citenamefont {Thibaud-Nissen}, \citenamefont {Thomas}, \citenamefont {Tosatto}, \citenamefont {Wu},\ and\ \citenamefont {Bateman}}]{blum2024interpro}%
  \BibitemOpen
  \bibfield  {author} {\bibinfo {author} {\bibfnamefont {Matthias}\ \bibnamefont {Blum}}, \bibinfo {author} {\bibfnamefont {Antonina}\ \bibnamefont {Andreeva}}, \bibinfo {author} {\bibfnamefont {Laise Cavalcanti}\ \bibnamefont {Florentino}}, \bibinfo {author} {\bibfnamefont {Sara Rocio}\ \bibnamefont {Chuguransky}}, \bibinfo {author} {\bibfnamefont {Tiago}\ \bibnamefont {Grego}}, \bibinfo {author} {\bibfnamefont {Emma}\ \bibnamefont {Hobbs}}, \bibinfo {author} {\bibfnamefont {Beatriz Lazaro}\ \bibnamefont {Pinto}}, \bibinfo {author} {\bibfnamefont {Ailsa}\ \bibnamefont {Orr}}, \bibinfo {author} {\bibfnamefont {Typhaine}\ \bibnamefont {Paysan-Lafosse}}, \bibinfo {author} {\bibfnamefont {Irina}\ \bibnamefont {Ponamareva}}, \bibinfo {author} {\bibfnamefont {Gustavo A}\ \bibnamefont {Salazar}}, \bibinfo {author} {\bibfnamefont {Nicola}\ \bibnamefont {Bordin}}, \bibinfo {author} {\bibfnamefont {Peer}\ \bibnamefont {Bork}}, \bibinfo {author} {\bibfnamefont {Alan}\ \bibnamefont {Bridge}}, \bibinfo {author}
  {\bibfnamefont {Lucy}\ \bibnamefont {Colwell}}, \bibinfo {author} {\bibfnamefont {Julian}\ \bibnamefont {Gough}}, \bibinfo {author} {\bibfnamefont {Daniel H}\ \bibnamefont {Haft}}, \bibinfo {author} {\bibfnamefont {Ivica}\ \bibnamefont {Letunic}}, \bibinfo {author} {\bibfnamefont {Felipe}\ \bibnamefont {Llinares-López}}, \bibinfo {author} {\bibfnamefont {Aron}\ \bibnamefont {Marchler-Bauer}}, \bibinfo {author} {\bibfnamefont {Laetitia}\ \bibnamefont {Meng-Papaxanthos}}, \bibinfo {author} {\bibfnamefont {Huaiyu}\ \bibnamefont {Mi}}, \bibinfo {author} {\bibfnamefont {Darren A}\ \bibnamefont {Natale}}, \bibinfo {author} {\bibfnamefont {Christine A}\ \bibnamefont {Orengo}}, \bibinfo {author} {\bibfnamefont {Arun P}\ \bibnamefont {Pandurangan}}, \bibinfo {author} {\bibfnamefont {Damiano}\ \bibnamefont {Piovesan}}, \bibinfo {author} {\bibfnamefont {Catherine}\ \bibnamefont {Rivoire}}, \bibinfo {author} {\bibfnamefont {Christian J~A}\ \bibnamefont {Sigrist}}, \bibinfo {author} {\bibfnamefont {Narmada}\
  \bibnamefont {Thanki}}, \bibinfo {author} {\bibfnamefont {Françoise}\ \bibnamefont {Thibaud-Nissen}}, \bibinfo {author} {\bibfnamefont {Paul D}\ \bibnamefont {Thomas}}, \bibinfo {author} {\bibfnamefont {Silvio C~E}\ \bibnamefont {Tosatto}}, \bibinfo {author} {\bibfnamefont {Cathy H}\ \bibnamefont {Wu}}, \ and\ \bibinfo {author} {\bibfnamefont {Alex}\ \bibnamefont {Bateman}},\ }\bibfield  {title} {\enquote {\bibinfo {title} {{I}nter{P}ro: the protein sequence classification resource in 2025},}\ }\href {\doibase 10.1093/nar/gkae1082} {\bibfield  {journal} {\bibinfo  {journal} {Nucleic Acids Research}\ }\textbf {\bibinfo {volume} {53}},\ \bibinfo {pages} {D444--D456} (\bibinfo {year} {2024})}\BibitemShut {NoStop}%
\bibitem [{\citenamefont {Paysan-Lafosse}\ \emph {et~al.}(2024)\citenamefont {Paysan-Lafosse}, \citenamefont {Andreeva}, \citenamefont {Blum}, \citenamefont {Chuguransky}, \citenamefont {Grego}, \citenamefont {Pinto}, \citenamefont {Salazar}, \citenamefont {Bileschi}, \citenamefont {Llinares-López}, \citenamefont {Meng-Papaxanthos}, \citenamefont {Colwell}, \citenamefont {Grishin}, \citenamefont {Schaeffer}, \citenamefont {Clementel}, \citenamefont {Tosatto}, \citenamefont {Sonnhammer}, \citenamefont {Wood},\ and\ \citenamefont {Bateman}}]{paysan2024pfam}%
  \BibitemOpen
  \bibfield  {author} {\bibinfo {author} {\bibfnamefont {Typhaine}\ \bibnamefont {Paysan-Lafosse}}, \bibinfo {author} {\bibfnamefont {Antonina}\ \bibnamefont {Andreeva}}, \bibinfo {author} {\bibfnamefont {Matthias}\ \bibnamefont {Blum}}, \bibinfo {author} {\bibfnamefont {Sara Rocio}\ \bibnamefont {Chuguransky}}, \bibinfo {author} {\bibfnamefont {Tiago}\ \bibnamefont {Grego}}, \bibinfo {author} {\bibfnamefont {Beatriz Lazaro}\ \bibnamefont {Pinto}}, \bibinfo {author} {\bibfnamefont {Gustavo A}\ \bibnamefont {Salazar}}, \bibinfo {author} {\bibfnamefont {Maxwell L}\ \bibnamefont {Bileschi}}, \bibinfo {author} {\bibfnamefont {Felipe}\ \bibnamefont {Llinares-López}}, \bibinfo {author} {\bibfnamefont {Laetitia}\ \bibnamefont {Meng-Papaxanthos}}, \bibinfo {author} {\bibfnamefont {Lucy J}\ \bibnamefont {Colwell}}, \bibinfo {author} {\bibfnamefont {Nick V}\ \bibnamefont {Grishin}}, \bibinfo {author} {\bibfnamefont {R~Dustin}\ \bibnamefont {Schaeffer}}, \bibinfo {author} {\bibfnamefont {Damiano}\ \bibnamefont
  {Clementel}}, \bibinfo {author} {\bibfnamefont {Silvio C~E}\ \bibnamefont {Tosatto}}, \bibinfo {author} {\bibfnamefont {Erik}\ \bibnamefont {Sonnhammer}}, \bibinfo {author} {\bibfnamefont {Valerie}\ \bibnamefont {Wood}}, \ and\ \bibinfo {author} {\bibfnamefont {Alex}\ \bibnamefont {Bateman}},\ }\bibfield  {title} {\enquote {\bibinfo {title} {The {P}fam protein families database: embracing {AI}/{ML}},}\ }\href {\doibase 10.1093/nar/gkae997} {\bibfield  {journal} {\bibinfo  {journal} {Nucleic Acids Research}\ }\textbf {\bibinfo {volume} {53}},\ \bibinfo {pages} {D523--D534} (\bibinfo {year} {2024})}\BibitemShut {NoStop}%
\bibitem [{\citenamefont {Consortium}(2022)}]{Uniprot2023}%
  \BibitemOpen
  \bibfield  {author} {\bibinfo {author} {\bibfnamefont {The~UniProt}\ \bibnamefont {Consortium}},\ }\bibfield  {title} {\enquote {\bibinfo {title} {{U}ni{P}rot: the {U}niversal {P}rotein {K}nowledgebase in 2023},}\ }\href {\doibase 10.1093/nar/gkac1052} {\bibfield  {journal} {\bibinfo  {journal} {Nucleic Acids Research}\ }\textbf {\bibinfo {volume} {51}},\ \bibinfo {pages} {D523--D531} (\bibinfo {year} {2022})}\BibitemShut {NoStop}%
\bibitem [{\citenamefont {Burley}\ \emph {et~al.}(2022)\citenamefont {Burley}, \citenamefont {Bhikadiya}, \citenamefont {Bi}, \citenamefont {Bittrich}, \citenamefont {Chao}, \citenamefont {Chen}, \citenamefont {Craig}, \citenamefont {Crichlow}, \citenamefont {Dalenberg}, \citenamefont {Duarte}, \citenamefont {Dutta}, \citenamefont {Fayazi}, \citenamefont {Feng}, \citenamefont {Flatt}, \citenamefont {Ganesan}, \citenamefont {Ghosh}, \citenamefont {Goodsell}, \citenamefont {Green}, \citenamefont {Guranovic}, \citenamefont {Henry}, \citenamefont {Hudson}, \citenamefont {Khokhriakov}, \citenamefont {Lawson}, \citenamefont {Liang}, \citenamefont {Lowe}, \citenamefont {Peisach}, \citenamefont {Persikova}, \citenamefont {Piehl}, \citenamefont {Rose}, \citenamefont {Sali}, \citenamefont {Segura}, \citenamefont {Sekharan}, \citenamefont {Shao}, \citenamefont {Vallat}, \citenamefont {Voigt}, \citenamefont {Webb}, \citenamefont {Westbrook}, \citenamefont {Whetstone}, \citenamefont {Young}, \citenamefont {Zalevsky},\
  and\ \citenamefont {Zardecki}}]{PDB2022_2}%
  \BibitemOpen
  \bibfield  {author} {\bibinfo {author} {\bibfnamefont {Stephen~K}\ \bibnamefont {Burley}}, \bibinfo {author} {\bibfnamefont {Charmi}\ \bibnamefont {Bhikadiya}}, \bibinfo {author} {\bibfnamefont {Chunxiao}\ \bibnamefont {Bi}}, \bibinfo {author} {\bibfnamefont {Sebastian}\ \bibnamefont {Bittrich}}, \bibinfo {author} {\bibfnamefont {Henry}\ \bibnamefont {Chao}}, \bibinfo {author} {\bibfnamefont {Li}~\bibnamefont {Chen}}, \bibinfo {author} {\bibfnamefont {Paul~A}\ \bibnamefont {Craig}}, \bibinfo {author} {\bibfnamefont {Gregg~V}\ \bibnamefont {Crichlow}}, \bibinfo {author} {\bibfnamefont {Kenneth}\ \bibnamefont {Dalenberg}}, \bibinfo {author} {\bibfnamefont {Jose~M}\ \bibnamefont {Duarte}}, \bibinfo {author} {\bibfnamefont {Shuchismita}\ \bibnamefont {Dutta}}, \bibinfo {author} {\bibfnamefont {Maryam}\ \bibnamefont {Fayazi}}, \bibinfo {author} {\bibfnamefont {Zukang}\ \bibnamefont {Feng}}, \bibinfo {author} {\bibfnamefont {Justin~W}\ \bibnamefont {Flatt}}, \bibinfo {author} {\bibfnamefont {Sai}\ \bibnamefont
  {Ganesan}}, \bibinfo {author} {\bibfnamefont {Sutapa}\ \bibnamefont {Ghosh}}, \bibinfo {author} {\bibfnamefont {David~S}\ \bibnamefont {Goodsell}}, \bibinfo {author} {\bibfnamefont {Rachel~Kramer}\ \bibnamefont {Green}}, \bibinfo {author} {\bibfnamefont {Vladimir}\ \bibnamefont {Guranovic}}, \bibinfo {author} {\bibfnamefont {Jeremy}\ \bibnamefont {Henry}}, \bibinfo {author} {\bibfnamefont {Brian~P}\ \bibnamefont {Hudson}}, \bibinfo {author} {\bibfnamefont {Igor}\ \bibnamefont {Khokhriakov}}, \bibinfo {author} {\bibfnamefont {Catherine~L}\ \bibnamefont {Lawson}}, \bibinfo {author} {\bibfnamefont {Yuhe}\ \bibnamefont {Liang}}, \bibinfo {author} {\bibfnamefont {Robert}\ \bibnamefont {Lowe}}, \bibinfo {author} {\bibfnamefont {Ezra}\ \bibnamefont {Peisach}}, \bibinfo {author} {\bibfnamefont {Irina}\ \bibnamefont {Persikova}}, \bibinfo {author} {\bibfnamefont {Dennis~W}\ \bibnamefont {Piehl}}, \bibinfo {author} {\bibfnamefont {Yana}\ \bibnamefont {Rose}}, \bibinfo {author} {\bibfnamefont {Andrej}\ \bibnamefont
  {Sali}}, \bibinfo {author} {\bibfnamefont {Joan}\ \bibnamefont {Segura}}, \bibinfo {author} {\bibfnamefont {Monica}\ \bibnamefont {Sekharan}}, \bibinfo {author} {\bibfnamefont {Chenghua}\ \bibnamefont {Shao}}, \bibinfo {author} {\bibfnamefont {Brinda}\ \bibnamefont {Vallat}}, \bibinfo {author} {\bibfnamefont {Maria}\ \bibnamefont {Voigt}}, \bibinfo {author} {\bibfnamefont {Ben}\ \bibnamefont {Webb}}, \bibinfo {author} {\bibfnamefont {John~D}\ \bibnamefont {Westbrook}}, \bibinfo {author} {\bibfnamefont {Shamara}\ \bibnamefont {Whetstone}}, \bibinfo {author} {\bibfnamefont {Jasmine~Y}\ \bibnamefont {Young}}, \bibinfo {author} {\bibfnamefont {Arthur}\ \bibnamefont {Zalevsky}}, \ and\ \bibinfo {author} {\bibfnamefont {Christine}\ \bibnamefont {Zardecki}},\ }\bibfield  {title} {\enquote {\bibinfo {title} {{RCSB} protein data bank ({RCSB}.org): delivery of experimentally-determined {PDB} structures alongside one million computed structure models of proteins from artificial intelligence/machine learning},}\ }\href
  {\doibase 10.1093/nar/gkac1077} {\bibfield  {journal} {\bibinfo  {journal} {Nucleic Acids Research}\ }\textbf {\bibinfo {volume} {51}},\ \bibinfo {pages} {D488--D508} (\bibinfo {year} {2022})}\BibitemShut {NoStop}%
\bibitem [{\citenamefont {Kimura}(1983)}]{kimura1985neutral}%
  \BibitemOpen
  \bibfield  {author} {\bibinfo {author} {\bibfnamefont {Motoo}\ \bibnamefont {Kimura}},\ }\href {\doibase https://doi.org/10.1017/CBO9780511623486} {\emph {\bibinfo {title} {The neutral theory of molecular evolution}}}\ (\bibinfo  {publisher} {Cambridge University Press},\ \bibinfo {year} {1983})\BibitemShut {NoStop}%
\bibitem [{\citenamefont {Fowler}\ and\ \citenamefont {Fields}(2014)}]{fowler2014deep}%
  \BibitemOpen
  \bibfield  {author} {\bibinfo {author} {\bibfnamefont {Douglas~M}\ \bibnamefont {Fowler}}\ and\ \bibinfo {author} {\bibfnamefont {Stanley}\ \bibnamefont {Fields}},\ }\bibfield  {title} {\enquote {\bibinfo {title} {Deep mutational scanning: a new style of protein science},}\ }\href {\doibase https://doi.org/10.1038/nmeth.3027} {\bibfield  {journal} {\bibinfo  {journal} {Nature methods}\ }\textbf {\bibinfo {volume} {11}},\ \bibinfo {pages} {801--807} (\bibinfo {year} {2014})}\BibitemShut {NoStop}%
\bibitem [{\citenamefont {Sarkisyan}\ \emph {et~al.}(2016)\citenamefont {Sarkisyan}, \citenamefont {Bolotin}, \citenamefont {Meer}, \citenamefont {Usmanova}, \citenamefont {Mishin}, \citenamefont {Sharonov}, \citenamefont {Ivankov}, \citenamefont {Bozhanova}, \citenamefont {Baranov}, \citenamefont {Soylemez} \emph {et~al.}}]{sarkisyan2016local}%
  \BibitemOpen
  \bibfield  {author} {\bibinfo {author} {\bibfnamefont {Karen~S}\ \bibnamefont {Sarkisyan}}, \bibinfo {author} {\bibfnamefont {Dmitry~A}\ \bibnamefont {Bolotin}}, \bibinfo {author} {\bibfnamefont {Margarita~V}\ \bibnamefont {Meer}}, \bibinfo {author} {\bibfnamefont {Dinara~R}\ \bibnamefont {Usmanova}}, \bibinfo {author} {\bibfnamefont {Alexander~S}\ \bibnamefont {Mishin}}, \bibinfo {author} {\bibfnamefont {George~V}\ \bibnamefont {Sharonov}}, \bibinfo {author} {\bibfnamefont {Dmitry~N}\ \bibnamefont {Ivankov}}, \bibinfo {author} {\bibfnamefont {Nina~G}\ \bibnamefont {Bozhanova}}, \bibinfo {author} {\bibfnamefont {Mikhail~S}\ \bibnamefont {Baranov}}, \bibinfo {author} {\bibfnamefont {Onuralp}\ \bibnamefont {Soylemez}},  \emph {et~al.},\ }\bibfield  {title} {\enquote {\bibinfo {title} {Local fitness landscape of the green fluorescent protein},}\ }\href {\doibase https://doi.org/10.1038/nature17995} {\bibfield  {journal} {\bibinfo  {journal} {Nature}\ }\textbf {\bibinfo {volume} {533}},\ \bibinfo {pages}
  {397--401} (\bibinfo {year} {2016})}\BibitemShut {NoStop}%
\bibitem [{\citenamefont {Notin}\ \emph {et~al.}(2024)\citenamefont {Notin}, \citenamefont {Kollasch}, \citenamefont {Ritter}, \citenamefont {Van~Niekerk}, \citenamefont {Paul}, \citenamefont {Spinner}, \citenamefont {Rollins}, \citenamefont {Shaw}, \citenamefont {Orenbuch}, \citenamefont {Weitzman}, \citenamefont {Frazer}, \citenamefont {Dias}, \citenamefont {Franceschi}, \citenamefont {Orenbuch}, \citenamefont {Gal},\ and\ \citenamefont {Marks}}]{notin2024proteingym}%
  \BibitemOpen
  \bibfield  {author} {\bibinfo {author} {\bibfnamefont {Pascal}\ \bibnamefont {Notin}}, \bibinfo {author} {\bibfnamefont {Aaron}\ \bibnamefont {Kollasch}}, \bibinfo {author} {\bibfnamefont {Daniel}\ \bibnamefont {Ritter}}, \bibinfo {author} {\bibfnamefont {Lood}\ \bibnamefont {Van~Niekerk}}, \bibinfo {author} {\bibfnamefont {Steffanie}\ \bibnamefont {Paul}}, \bibinfo {author} {\bibfnamefont {Han}\ \bibnamefont {Spinner}}, \bibinfo {author} {\bibfnamefont {Nathan}\ \bibnamefont {Rollins}}, \bibinfo {author} {\bibfnamefont {Ada}\ \bibnamefont {Shaw}}, \bibinfo {author} {\bibfnamefont {Rose}\ \bibnamefont {Orenbuch}}, \bibinfo {author} {\bibfnamefont {Ruben}\ \bibnamefont {Weitzman}}, \bibinfo {author} {\bibfnamefont {Jonathan}\ \bibnamefont {Frazer}}, \bibinfo {author} {\bibfnamefont {Mafalda}\ \bibnamefont {Dias}}, \bibinfo {author} {\bibfnamefont {Dinko}\ \bibnamefont {Franceschi}}, \bibinfo {author} {\bibfnamefont {Rose}\ \bibnamefont {Orenbuch}}, \bibinfo {author} {\bibfnamefont {Yarin}\ \bibnamefont
  {Gal}}, \ and\ \bibinfo {author} {\bibfnamefont {Debora~S.}\ \bibnamefont {Marks}},\ }\bibfield  {title} {\enquote {\bibinfo {title} {Proteingym: Large-scale benchmarks for protein fitness prediction and design},}\ }\href {https://doi.org/10.1101/2023.12.07.570727} {\bibfield  {journal} {\bibinfo  {journal} {Advances in Neural Information Processing Systems}\ }\textbf {\bibinfo {volume} {36}} (\bibinfo {year} {2024})}\BibitemShut {NoStop}%
\bibitem [{\citenamefont {Hartl}\ \emph {et~al.}(1997)\citenamefont {Hartl}, \citenamefont {Clark},\ and\ \citenamefont {Clark}}]{hartl1997principles}%
  \BibitemOpen
  \bibfield  {author} {\bibinfo {author} {\bibfnamefont {Daniel~L}\ \bibnamefont {Hartl}}, \bibinfo {author} {\bibfnamefont {Andrew~G}\ \bibnamefont {Clark}}, \ and\ \bibinfo {author} {\bibfnamefont {Andrew~G}\ \bibnamefont {Clark}},\ }\href {\doibase https://doi.org/10.1093/jhered/esm035} {\emph {\bibinfo {title} {Principles of population genetics}}},\ Vol.\ \bibinfo {volume} {116}\ (\bibinfo  {publisher} {Sinauer associates Sunderland, MA},\ \bibinfo {year} {1997})\BibitemShut {NoStop}%
\bibitem [{\citenamefont {Harms}\ and\ \citenamefont {Thornton}(2013)}]{harms2013evolutionary}%
  \BibitemOpen
  \bibfield  {author} {\bibinfo {author} {\bibfnamefont {Michael~J}\ \bibnamefont {Harms}}\ and\ \bibinfo {author} {\bibfnamefont {Joseph~W}\ \bibnamefont {Thornton}},\ }\bibfield  {title} {\enquote {\bibinfo {title} {Evolutionary biochemistry: revealing the historical and physical causes of protein properties},}\ }\href {\doibase https://doi.org/10.1038/nrg3540} {\bibfield  {journal} {\bibinfo  {journal} {Nature Reviews Genetics}\ }\textbf {\bibinfo {volume} {14}},\ \bibinfo {pages} {559--571} (\bibinfo {year} {2013})}\BibitemShut {NoStop}%
\bibitem [{\citenamefont {Weinreich}\ \emph {et~al.}(2013)\citenamefont {Weinreich}, \citenamefont {Lan}, \citenamefont {Wylie},\ and\ \citenamefont {Heckendorn}}]{weinreich2013should}%
  \BibitemOpen
  \bibfield  {author} {\bibinfo {author} {\bibfnamefont {Daniel~M}\ \bibnamefont {Weinreich}}, \bibinfo {author} {\bibfnamefont {Yinghong}\ \bibnamefont {Lan}}, \bibinfo {author} {\bibfnamefont {C~Scott}\ \bibnamefont {Wylie}}, \ and\ \bibinfo {author} {\bibfnamefont {Robert~B}\ \bibnamefont {Heckendorn}},\ }\bibfield  {title} {\enquote {\bibinfo {title} {Should evolutionary geneticists worry about higher-order epistasis?}}\ }\href {\doibase https://doi.org/10.1016/j.gde.2013.10.007} {\bibfield  {journal} {\bibinfo  {journal} {Current opinion in genetics \& development}\ }\textbf {\bibinfo {volume} {23}},\ \bibinfo {pages} {700--707} (\bibinfo {year} {2013})}\BibitemShut {NoStop}%
\bibitem [{\citenamefont {De~Visser}\ and\ \citenamefont {Krug}(2014)}]{de2014empirical}%
  \BibitemOpen
  \bibfield  {author} {\bibinfo {author} {\bibfnamefont {J~Arjan~GM}\ \bibnamefont {De~Visser}}\ and\ \bibinfo {author} {\bibfnamefont {Joachim}\ \bibnamefont {Krug}},\ }\bibfield  {title} {\enquote {\bibinfo {title} {Empirical fitness landscapes and the predictability of evolution},}\ }\href {\doibase https://doi.org/10.1038/nrg3744} {\bibfield  {journal} {\bibinfo  {journal} {Nature Reviews Genetics}\ }\textbf {\bibinfo {volume} {15}},\ \bibinfo {pages} {480--490} (\bibinfo {year} {2014})}\BibitemShut {NoStop}%
\bibitem [{\citenamefont {Starr}\ and\ \citenamefont {Thornton}(2016)}]{starr2016epistasis}%
  \BibitemOpen
  \bibfield  {author} {\bibinfo {author} {\bibfnamefont {Tyler~N}\ \bibnamefont {Starr}}\ and\ \bibinfo {author} {\bibfnamefont {Joseph~W}\ \bibnamefont {Thornton}},\ }\bibfield  {title} {\enquote {\bibinfo {title} {Epistasis in protein evolution},}\ }\href {\doibase https://doi.org/10.1002/pro.2897} {\bibfield  {journal} {\bibinfo  {journal} {Protein science}\ }\textbf {\bibinfo {volume} {25}},\ \bibinfo {pages} {1204--1218} (\bibinfo {year} {2016})}\BibitemShut {NoStop}%
\bibitem [{\citenamefont {Poelwijk}\ \emph {et~al.}(2016)\citenamefont {Poelwijk}, \citenamefont {Krishna},\ and\ \citenamefont {Ranganathan}}]{poelwijk2016context}%
  \BibitemOpen
  \bibfield  {author} {\bibinfo {author} {\bibfnamefont {Frank~J}\ \bibnamefont {Poelwijk}}, \bibinfo {author} {\bibfnamefont {Vinod}\ \bibnamefont {Krishna}}, \ and\ \bibinfo {author} {\bibfnamefont {Rama}\ \bibnamefont {Ranganathan}},\ }\bibfield  {title} {\enquote {\bibinfo {title} {The context-dependence of mutations: a linkage of formalisms},}\ }\href {\doibase https://doi.org/10.1371/journal.pcbi.1004771} {\bibfield  {journal} {\bibinfo  {journal} {PLoS computational biology}\ }\textbf {\bibinfo {volume} {12}},\ \bibinfo {pages} {e1004771} (\bibinfo {year} {2016})}\BibitemShut {NoStop}%
\bibitem [{\citenamefont {Cocco}\ \emph {et~al.}(2018)\citenamefont {Cocco}, \citenamefont {Feinauer}, \citenamefont {Figliuzzi}, \citenamefont {Monasson},\ and\ \citenamefont {Weigt}}]{cocco2018inverse}%
  \BibitemOpen
  \bibfield  {author} {\bibinfo {author} {\bibfnamefont {Simona}\ \bibnamefont {Cocco}}, \bibinfo {author} {\bibfnamefont {Christoph}\ \bibnamefont {Feinauer}}, \bibinfo {author} {\bibfnamefont {Matteo}\ \bibnamefont {Figliuzzi}}, \bibinfo {author} {\bibfnamefont {R{\'e}mi}\ \bibnamefont {Monasson}}, \ and\ \bibinfo {author} {\bibfnamefont {Martin}\ \bibnamefont {Weigt}},\ }\bibfield  {title} {\enquote {\bibinfo {title} {Inverse statistical physics of protein sequences: a key issues review},}\ }\href {\doibase 10.1088/1361-6633/aa9965} {\bibfield  {journal} {\bibinfo  {journal} {Reports on Progress in Physics}\ }\textbf {\bibinfo {volume} {81}},\ \bibinfo {pages} {032601} (\bibinfo {year} {2018})}\BibitemShut {NoStop}%
\bibitem [{\citenamefont {Domingo}\ \emph {et~al.}(2019)\citenamefont {Domingo}, \citenamefont {Baeza-Centurion},\ and\ \citenamefont {Lehner}}]{domingo2019causes}%
  \BibitemOpen
  \bibfield  {author} {\bibinfo {author} {\bibfnamefont {J{\'u}lia}\ \bibnamefont {Domingo}}, \bibinfo {author} {\bibfnamefont {Pablo}\ \bibnamefont {Baeza-Centurion}}, \ and\ \bibinfo {author} {\bibfnamefont {Ben}\ \bibnamefont {Lehner}},\ }\bibfield  {title} {\enquote {\bibinfo {title} {The causes and consequences of genetic interactions (epistasis)},}\ }\href {\doibase 10.1146/annurev-genom-083118-014857} {\bibfield  {journal} {\bibinfo  {journal} {Annual review of genomics and human genetics}\ }\textbf {\bibinfo {volume} {20}},\ \bibinfo {pages} {433--460} (\bibinfo {year} {2019})}\BibitemShut {NoStop}%
\bibitem [{\citenamefont {Johnson}\ \emph {et~al.}(2023)\citenamefont {Johnson}, \citenamefont {Reddy},\ and\ \citenamefont {Desai}}]{johnson2023epistasis}%
  \BibitemOpen
  \bibfield  {author} {\bibinfo {author} {\bibfnamefont {Milo~S}\ \bibnamefont {Johnson}}, \bibinfo {author} {\bibfnamefont {Gautam}\ \bibnamefont {Reddy}}, \ and\ \bibinfo {author} {\bibfnamefont {Michael~M}\ \bibnamefont {Desai}},\ }\bibfield  {title} {\enquote {\bibinfo {title} {Epistasis and evolution: recent advances and an outlook for prediction},}\ }\href {\doibase https://doi.org/10.1186/s12915-023-01585-3} {\bibfield  {journal} {\bibinfo  {journal} {BMC biology}\ }\textbf {\bibinfo {volume} {21}},\ \bibinfo {pages} {120} (\bibinfo {year} {2023})}\BibitemShut {NoStop}%
\bibitem [{\citenamefont {Buda}\ \emph {et~al.}(2023)\citenamefont {Buda}, \citenamefont {Miton},\ and\ \citenamefont {Tokuriki}}]{buda2023pervasive}%
  \BibitemOpen
  \bibfield  {author} {\bibinfo {author} {\bibfnamefont {Karol}\ \bibnamefont {Buda}}, \bibinfo {author} {\bibfnamefont {Charlotte~M}\ \bibnamefont {Miton}}, \ and\ \bibinfo {author} {\bibfnamefont {Nobuhiko}\ \bibnamefont {Tokuriki}},\ }\bibfield  {title} {\enquote {\bibinfo {title} {Pervasive epistasis exposes intramolecular networks in adaptive enzyme evolution},}\ }\href {\doibase https://doi.org/10.1038/s41467-023-44333-5} {\bibfield  {journal} {\bibinfo  {journal} {Nature Communications}\ }\textbf {\bibinfo {volume} {14}},\ \bibinfo {pages} {8508} (\bibinfo {year} {2023})}\BibitemShut {NoStop}%
\bibitem [{\citenamefont {Fantini}\ \emph {et~al.}(2020)\citenamefont {Fantini}, \citenamefont {Lisi}, \citenamefont {De~Los~Rios}, \citenamefont {Cattaneo},\ and\ \citenamefont {Pastore}}]{fantini2020protein}%
  \BibitemOpen
  \bibfield  {author} {\bibinfo {author} {\bibfnamefont {Marco}\ \bibnamefont {Fantini}}, \bibinfo {author} {\bibfnamefont {Simonetta}\ \bibnamefont {Lisi}}, \bibinfo {author} {\bibfnamefont {Paolo}\ \bibnamefont {De~Los~Rios}}, \bibinfo {author} {\bibfnamefont {Antonino}\ \bibnamefont {Cattaneo}}, \ and\ \bibinfo {author} {\bibfnamefont {Annalisa}\ \bibnamefont {Pastore}},\ }\bibfield  {title} {\enquote {\bibinfo {title} {Protein structural information and evolutionary landscape by in vitro evolution},}\ }\href {\doibase https://doi.org/10.1093/molbev/msz256} {\bibfield  {journal} {\bibinfo  {journal} {Molecular biology and evolution}\ }\textbf {\bibinfo {volume} {37}},\ \bibinfo {pages} {1179--1192} (\bibinfo {year} {2020})}\BibitemShut {NoStop}%
\bibitem [{\citenamefont {Stiffler}\ \emph {et~al.}(2020)\citenamefont {Stiffler}, \citenamefont {Poelwijk}, \citenamefont {Brock}, \citenamefont {Stein}, \citenamefont {Riesselman}, \citenamefont {Teyra}, \citenamefont {Sidhu}, \citenamefont {Marks}, \citenamefont {Gauthier},\ and\ \citenamefont {Sander}}]{stiffler2020protein}%
  \BibitemOpen
  \bibfield  {author} {\bibinfo {author} {\bibfnamefont {Michael~A}\ \bibnamefont {Stiffler}}, \bibinfo {author} {\bibfnamefont {Frank~J}\ \bibnamefont {Poelwijk}}, \bibinfo {author} {\bibfnamefont {Kelly~P}\ \bibnamefont {Brock}}, \bibinfo {author} {\bibfnamefont {Richard~R}\ \bibnamefont {Stein}}, \bibinfo {author} {\bibfnamefont {Adam}\ \bibnamefont {Riesselman}}, \bibinfo {author} {\bibfnamefont {Joan}\ \bibnamefont {Teyra}}, \bibinfo {author} {\bibfnamefont {Sachdev~S}\ \bibnamefont {Sidhu}}, \bibinfo {author} {\bibfnamefont {Debora~S}\ \bibnamefont {Marks}}, \bibinfo {author} {\bibfnamefont {Nicholas~P}\ \bibnamefont {Gauthier}}, \ and\ \bibinfo {author} {\bibfnamefont {Chris}\ \bibnamefont {Sander}},\ }\bibfield  {title} {\enquote {\bibinfo {title} {Protein structure from experimental evolution},}\ }\href {\doibase https://doi.org/10.1016/j.cels.2019.11.008} {\bibfield  {journal} {\bibinfo  {journal} {Cell Systems}\ }\textbf {\bibinfo {volume} {10}},\ \bibinfo {pages} {15--24} (\bibinfo {year}
  {2020})}\BibitemShut {NoStop}%
\bibitem [{\citenamefont {Erdo{\u{g}}an}\ \emph {et~al.}(2024)\citenamefont {Erdo{\u{g}}an}, \citenamefont {Dasmeh}, \citenamefont {Socha}, \citenamefont {Chen}, \citenamefont {Life}, \citenamefont {Jun}, \citenamefont {Kiritchkov}, \citenamefont {Kehila}, \citenamefont {Serohijos},\ and\ \citenamefont {Tokuriki}}]{erdougan2023neutral}%
  \BibitemOpen
  \bibfield  {author} {\bibinfo {author} {\bibfnamefont {Ay{\c{s}}e~Nisan}\ \bibnamefont {Erdo{\u{g}}an}}, \bibinfo {author} {\bibfnamefont {Pouria}\ \bibnamefont {Dasmeh}}, \bibinfo {author} {\bibfnamefont {Raymond~Daniel}\ \bibnamefont {Socha}}, \bibinfo {author} {\bibfnamefont {John~Z}\ \bibnamefont {Chen}}, \bibinfo {author} {\bibfnamefont {Benjamin~E}\ \bibnamefont {Life}}, \bibinfo {author} {\bibfnamefont {Rachel}\ \bibnamefont {Jun}}, \bibinfo {author} {\bibfnamefont {Linda}\ \bibnamefont {Kiritchkov}}, \bibinfo {author} {\bibfnamefont {Dan}\ \bibnamefont {Kehila}}, \bibinfo {author} {\bibfnamefont {Adrian~WR}\ \bibnamefont {Serohijos}}, \ and\ \bibinfo {author} {\bibfnamefont {Nobuhiko}\ \bibnamefont {Tokuriki}},\ }\bibfield  {title} {\enquote {\bibinfo {title} {Neutral drift upon threshold-like selection promotes variation in antibiotic resistance phenotype},}\ }\href {\doibase https://doi.org/10.1038/s41467-024-55012-4} {\bibfield  {journal} {\bibinfo  {journal} {Nature Communications}\ }\textbf
  {\bibinfo {volume} {15}},\ \bibinfo {pages} {10813} (\bibinfo {year} {2024})}\BibitemShut {NoStop}%
\bibitem [{\citenamefont {Park}\ \emph {et~al.}(2022)\citenamefont {Park}, \citenamefont {Metzger},\ and\ \citenamefont {Thornton}}]{park2022epistatic}%
  \BibitemOpen
  \bibfield  {author} {\bibinfo {author} {\bibfnamefont {Yeonwoo}\ \bibnamefont {Park}}, \bibinfo {author} {\bibfnamefont {Brian~PH}\ \bibnamefont {Metzger}}, \ and\ \bibinfo {author} {\bibfnamefont {Joseph~W}\ \bibnamefont {Thornton}},\ }\bibfield  {title} {\enquote {\bibinfo {title} {Epistatic drift causes gradual decay of predictability in protein evolution},}\ }\href {\doibase 10.1126/science.abn6895} {\bibfield  {journal} {\bibinfo  {journal} {Science}\ }\textbf {\bibinfo {volume} {376}},\ \bibinfo {pages} {823--830} (\bibinfo {year} {2022})}\BibitemShut {NoStop}%
\bibitem [{\citenamefont {Rix}\ \emph {et~al.}(2024)\citenamefont {Rix}, \citenamefont {Williams}, \citenamefont {Hu}, \citenamefont {Spinner}, \citenamefont {Pisera}, \citenamefont {Marks},\ and\ \citenamefont {Liu}}]{rix2023continuous}%
  \BibitemOpen
  \bibfield  {author} {\bibinfo {author} {\bibfnamefont {Gordon}\ \bibnamefont {Rix}}, \bibinfo {author} {\bibfnamefont {Rory~L.}\ \bibnamefont {Williams}}, \bibinfo {author} {\bibfnamefont {Vincent~J.}\ \bibnamefont {Hu}}, \bibinfo {author} {\bibfnamefont {Aviv}\ \bibnamefont {Spinner}}, \bibinfo {author} {\bibfnamefont {Alexander~(Olek)}\ \bibnamefont {Pisera}}, \bibinfo {author} {\bibfnamefont {Debora~S.}\ \bibnamefont {Marks}}, \ and\ \bibinfo {author} {\bibfnamefont {Chang~C.}\ \bibnamefont {Liu}},\ }\bibfield  {title} {\enquote {\bibinfo {title} {Continuous evolution of user-defined genes at 1 million times the genomic mutation rate},}\ }\href {\doibase 10.1126/science.adm9073} {\bibfield  {journal} {\bibinfo  {journal} {Science}\ }\textbf {\bibinfo {volume} {386}},\ \bibinfo {pages} {9073} (\bibinfo {year} {2024})}\BibitemShut {NoStop}%
\bibitem [{\citenamefont {de~la Paz}\ \emph {et~al.}(2020)\citenamefont {de~la Paz}, \citenamefont {Nartey}, \citenamefont {Yuvaraj},\ and\ \citenamefont {Morcos}}]{de2020epistatic}%
  \BibitemOpen
  \bibfield  {author} {\bibinfo {author} {\bibfnamefont {Jose~Alberto}\ \bibnamefont {de~la Paz}}, \bibinfo {author} {\bibfnamefont {Charisse~M}\ \bibnamefont {Nartey}}, \bibinfo {author} {\bibfnamefont {Monisha}\ \bibnamefont {Yuvaraj}}, \ and\ \bibinfo {author} {\bibfnamefont {Faruck}\ \bibnamefont {Morcos}},\ }\bibfield  {title} {\enquote {\bibinfo {title} {Epistatic contributions promote the unification of incompatible models of neutral molecular evolution},}\ }\href {\doibase https://doi.org/10.1073/pnas.1913071117} {\bibfield  {journal} {\bibinfo  {journal} {Proceedings of the National Academy of Sciences}\ }\textbf {\bibinfo {volume} {117}},\ \bibinfo {pages} {5873--5882} (\bibinfo {year} {2020})}\BibitemShut {NoStop}%
\bibitem [{\citenamefont {Bisardi}\ \emph {et~al.}(2022)\citenamefont {Bisardi}, \citenamefont {Rodriguez-Rivas}, \citenamefont {Zamponi},\ and\ \citenamefont {Weigt}}]{bisardi2022modeling}%
  \BibitemOpen
  \bibfield  {author} {\bibinfo {author} {\bibfnamefont {Matteo}\ \bibnamefont {Bisardi}}, \bibinfo {author} {\bibfnamefont {Juan}\ \bibnamefont {Rodriguez-Rivas}}, \bibinfo {author} {\bibfnamefont {Francesco}\ \bibnamefont {Zamponi}}, \ and\ \bibinfo {author} {\bibfnamefont {Martin}\ \bibnamefont {Weigt}},\ }\bibfield  {title} {\enquote {\bibinfo {title} {Modeling sequence-space exploration and emergence of epistatic signals in protein evolution},}\ }\href {\doibase https://doi.org/10.1093/molbev/msab321} {\bibfield  {journal} {\bibinfo  {journal} {Molecular biology and evolution}\ }\textbf {\bibinfo {volume} {39}},\ \bibinfo {pages} {msab321} (\bibinfo {year} {2022})}\BibitemShut {NoStop}%
\bibitem [{\citenamefont {Alvarez}\ \emph {et~al.}(2022)\citenamefont {Alvarez}, \citenamefont {Nartey}, \citenamefont {Mercado},\ and\ \citenamefont {Morcos}}]{alvarez2022novel}%
  \BibitemOpen
  \bibfield  {author} {\bibinfo {author} {\bibfnamefont {Sophia}\ \bibnamefont {Alvarez}}, \bibinfo {author} {\bibfnamefont {Charisse}\ \bibnamefont {Nartey}}, \bibinfo {author} {\bibfnamefont {Nicholas}\ \bibnamefont {Mercado}}, \ and\ \bibinfo {author} {\bibfnamefont {Faruck}\ \bibnamefont {Morcos}},\ }\bibfield  {title} {\enquote {\bibinfo {title} {Novel sequence space explored by functional proteins generated through computational evolution-based design},}\ }\href {\doibase 10.1016/j.bpj.2021.11.2476} {\bibfield  {journal} {\bibinfo  {journal} {Biophysical Journal}\ }\textbf {\bibinfo {volume} {121}},\ \bibinfo {pages} {45a} (\bibinfo {year} {2022})}\BibitemShut {NoStop}%
\bibitem [{\citenamefont {Alvarez}\ \emph {et~al.}(2024)\citenamefont {Alvarez}, \citenamefont {Nartey}, \citenamefont {Mercado}, \citenamefont {de~la Paz}, \citenamefont {Huseinbegovic},\ and\ \citenamefont {Morcos}}]{alvarez2024vivo}%
  \BibitemOpen
  \bibfield  {author} {\bibinfo {author} {\bibfnamefont {Sophia}\ \bibnamefont {Alvarez}}, \bibinfo {author} {\bibfnamefont {Charisse~M}\ \bibnamefont {Nartey}}, \bibinfo {author} {\bibfnamefont {Nicholas}\ \bibnamefont {Mercado}}, \bibinfo {author} {\bibfnamefont {Jose~Alberto}\ \bibnamefont {de~la Paz}}, \bibinfo {author} {\bibfnamefont {Tea}\ \bibnamefont {Huseinbegovic}}, \ and\ \bibinfo {author} {\bibfnamefont {Faruck}\ \bibnamefont {Morcos}},\ }\bibfield  {title} {\enquote {\bibinfo {title} {In vivo functional phenotypes from a computational epistatic model of evolution},}\ }\href {\doibase https://doi.org/10.1073/pnas.2308895121} {\bibfield  {journal} {\bibinfo  {journal} {Proceedings of the National Academy of Sciences}\ }\textbf {\bibinfo {volume} {121}},\ \bibinfo {pages} {e2308895121} (\bibinfo {year} {2024})}\BibitemShut {NoStop}%
\bibitem [{\citenamefont {Biswas}\ \emph {et~al.}(2024)\citenamefont {Biswas}, \citenamefont {Choudhuri}, \citenamefont {Arnold}, \citenamefont {Lyumkis}, \citenamefont {Haldane},\ and\ \citenamefont {Levy}}]{biswas2024kinetic}%
  \BibitemOpen
  \bibfield  {author} {\bibinfo {author} {\bibfnamefont {Avik}\ \bibnamefont {Biswas}}, \bibinfo {author} {\bibfnamefont {Indrani}\ \bibnamefont {Choudhuri}}, \bibinfo {author} {\bibfnamefont {Eddy}\ \bibnamefont {Arnold}}, \bibinfo {author} {\bibfnamefont {Dmitry}\ \bibnamefont {Lyumkis}}, \bibinfo {author} {\bibfnamefont {Allan}\ \bibnamefont {Haldane}}, \ and\ \bibinfo {author} {\bibfnamefont {Ronald~M}\ \bibnamefont {Levy}},\ }\bibfield  {title} {\enquote {\bibinfo {title} {Kinetic coevolutionary models predict the temporal emergence of hiv-1 resistance mutations under drug selection pressure},}\ }\href {\doibase https://doi.org/10.1073/pnas.2316662121} {\bibfield  {journal} {\bibinfo  {journal} {Proceedings of the National Academy of Sciences}\ }\textbf {\bibinfo {volume} {121}},\ \bibinfo {pages} {e2316662121} (\bibinfo {year} {2024})}\BibitemShut {NoStop}%
\bibitem [{\citenamefont {Bari}\ \emph {et~al.}(2024)\citenamefont {Bari}, \citenamefont {Bisardi}, \citenamefont {Cotogno}, \citenamefont {Weigt},\ and\ \citenamefont {Zamponi}}]{DiBari2024}%
  \BibitemOpen
  \bibfield  {author} {\bibinfo {author} {\bibfnamefont {Leonardo~Di}\ \bibnamefont {Bari}}, \bibinfo {author} {\bibfnamefont {Matteo}\ \bibnamefont {Bisardi}}, \bibinfo {author} {\bibfnamefont {Sabrina}\ \bibnamefont {Cotogno}}, \bibinfo {author} {\bibfnamefont {Martin}\ \bibnamefont {Weigt}}, \ and\ \bibinfo {author} {\bibfnamefont {Francesco}\ \bibnamefont {Zamponi}},\ }\bibfield  {title} {\enquote {\bibinfo {title} {Emergent time scales of epistasis in protein evolution},}\ }\href {\doibase 10.1073/pnas.2406807121} {\bibfield  {journal} {\bibinfo  {journal} {Proceedings of the National Academy of Sciences}\ }\textbf {\bibinfo {volume} {121}},\ \bibinfo {pages} {e2406807121} (\bibinfo {year} {2024})}\BibitemShut {NoStop}%
\bibitem [{\citenamefont {Morcos}\ \emph {et~al.}(2011)\citenamefont {Morcos}, \citenamefont {Pagnani}, \citenamefont {Lunt}, \citenamefont {Bertolino}, \citenamefont {Marks}, \citenamefont {Sander}, \citenamefont {Zecchina}, \citenamefont {Onuchic}, \citenamefont {Hwa},\ and\ \citenamefont {Weigt}}]{Morcos2011}%
  \BibitemOpen
  \bibfield  {author} {\bibinfo {author} {\bibfnamefont {Faruck}\ \bibnamefont {Morcos}}, \bibinfo {author} {\bibfnamefont {Andrea}\ \bibnamefont {Pagnani}}, \bibinfo {author} {\bibfnamefont {Bryan}\ \bibnamefont {Lunt}}, \bibinfo {author} {\bibfnamefont {Arianna}\ \bibnamefont {Bertolino}}, \bibinfo {author} {\bibfnamefont {Debora~S.}\ \bibnamefont {Marks}}, \bibinfo {author} {\bibfnamefont {Chris}\ \bibnamefont {Sander}}, \bibinfo {author} {\bibfnamefont {Riccardo}\ \bibnamefont {Zecchina}}, \bibinfo {author} {\bibfnamefont {José~N.}\ \bibnamefont {Onuchic}}, \bibinfo {author} {\bibfnamefont {Terence}\ \bibnamefont {Hwa}}, \ and\ \bibinfo {author} {\bibfnamefont {Martin}\ \bibnamefont {Weigt}},\ }\bibfield  {title} {\enquote {\bibinfo {title} {Direct-coupling analysis of residue coevolution captures native contacts across many protein families},}\ }\href {\doibase 10.1073/pnas.1111471108} {\bibfield  {journal} {\bibinfo  {journal} {Proceedings of the National Academy of Sciences}\ }\textbf {\bibinfo
  {volume} {108}},\ \bibinfo {pages} {E1293--E1301} (\bibinfo {year} {2011})}\BibitemShut {NoStop}%
\bibitem [{\citenamefont {Ferguson}\ \emph {et~al.}(2013)\citenamefont {Ferguson}, \citenamefont {Mann}, \citenamefont {Omarjee}, \citenamefont {Ndung’u}, \citenamefont {Walker},\ and\ \citenamefont {Chakraborty}}]{ferguson2013translating}%
  \BibitemOpen
  \bibfield  {author} {\bibinfo {author} {\bibfnamefont {Andrew~L}\ \bibnamefont {Ferguson}}, \bibinfo {author} {\bibfnamefont {Jaclyn~K}\ \bibnamefont {Mann}}, \bibinfo {author} {\bibfnamefont {Saleha}\ \bibnamefont {Omarjee}}, \bibinfo {author} {\bibfnamefont {Thumbi}\ \bibnamefont {Ndung’u}}, \bibinfo {author} {\bibfnamefont {Bruce~D}\ \bibnamefont {Walker}}, \ and\ \bibinfo {author} {\bibfnamefont {Arup~K}\ \bibnamefont {Chakraborty}},\ }\bibfield  {title} {\enquote {\bibinfo {title} {Translating {HIV} sequences into quantitative fitness landscapes predicts viral vulnerabilities for rational immunogen design},}\ }\href {\doibase 10.1016/j.immuni.2012.11.022} {\bibfield  {journal} {\bibinfo  {journal} {Immunity}\ }\textbf {\bibinfo {volume} {38}},\ \bibinfo {pages} {606--617} (\bibinfo {year} {2013})}\BibitemShut {NoStop}%
\bibitem [{\citenamefont {Figliuzzi}\ \emph {et~al.}(2016)\citenamefont {Figliuzzi}, \citenamefont {Jacquier}, \citenamefont {Schug}, \citenamefont {Tenaillon},\ and\ \citenamefont {Weigt}}]{figliuzzi2016coevolutionary}%
  \BibitemOpen
  \bibfield  {author} {\bibinfo {author} {\bibfnamefont {Matteo}\ \bibnamefont {Figliuzzi}}, \bibinfo {author} {\bibfnamefont {Herv{\'e}}\ \bibnamefont {Jacquier}}, \bibinfo {author} {\bibfnamefont {Alexander}\ \bibnamefont {Schug}}, \bibinfo {author} {\bibfnamefont {Oliver}\ \bibnamefont {Tenaillon}}, \ and\ \bibinfo {author} {\bibfnamefont {Martin}\ \bibnamefont {Weigt}},\ }\bibfield  {title} {\enquote {\bibinfo {title} {Coevolutionary landscape inference and the context-dependence of mutations in beta-lactamase {TEM}-1},}\ }\href {\doibase https://doi.org/10.1093/molbev/msv211} {\bibfield  {journal} {\bibinfo  {journal} {Molecular biology and evolution}\ }\textbf {\bibinfo {volume} {33}},\ \bibinfo {pages} {268--280} (\bibinfo {year} {2016})}\BibitemShut {NoStop}%
\bibitem [{\citenamefont {Levy}\ \emph {et~al.}(2017)\citenamefont {Levy}, \citenamefont {Haldane},\ and\ \citenamefont {Flynn}}]{levy2017potts}%
  \BibitemOpen
  \bibfield  {author} {\bibinfo {author} {\bibfnamefont {Ronald~M}\ \bibnamefont {Levy}}, \bibinfo {author} {\bibfnamefont {Allan}\ \bibnamefont {Haldane}}, \ and\ \bibinfo {author} {\bibfnamefont {William~F}\ \bibnamefont {Flynn}},\ }\bibfield  {title} {\enquote {\bibinfo {title} {Potts hamiltonian models of protein co-variation, free energy landscapes, and evolutionary fitness},}\ }\href {\doibase https://doi.org/10.1016/j.sbi.2016.11.004} {\bibfield  {journal} {\bibinfo  {journal} {Current opinion in structural biology}\ }\textbf {\bibinfo {volume} {43}},\ \bibinfo {pages} {55--62} (\bibinfo {year} {2017})}\BibitemShut {NoStop}%
\bibitem [{\citenamefont {Couce}\ \emph {et~al.}(2017)\citenamefont {Couce}, \citenamefont {Caudwell}, \citenamefont {Feinauer}, \citenamefont {Hindr{\'e}}, \citenamefont {Feugeas}, \citenamefont {Weigt}, \citenamefont {Lenski}, \citenamefont {Schneider},\ and\ \citenamefont {Tenaillon}}]{couce2017mutator}%
  \BibitemOpen
  \bibfield  {author} {\bibinfo {author} {\bibfnamefont {Alejandro}\ \bibnamefont {Couce}}, \bibinfo {author} {\bibfnamefont {Larissa~Viraphong}\ \bibnamefont {Caudwell}}, \bibinfo {author} {\bibfnamefont {Christoph}\ \bibnamefont {Feinauer}}, \bibinfo {author} {\bibfnamefont {Thomas}\ \bibnamefont {Hindr{\'e}}}, \bibinfo {author} {\bibfnamefont {Jean-Paul}\ \bibnamefont {Feugeas}}, \bibinfo {author} {\bibfnamefont {Martin}\ \bibnamefont {Weigt}}, \bibinfo {author} {\bibfnamefont {Richard~E}\ \bibnamefont {Lenski}}, \bibinfo {author} {\bibfnamefont {Dominique}\ \bibnamefont {Schneider}}, \ and\ \bibinfo {author} {\bibfnamefont {Olivier}\ \bibnamefont {Tenaillon}},\ }\bibfield  {title} {\enquote {\bibinfo {title} {Mutator genomes decay, despite sustained fitness gains, in a long-term experiment with bacteria},}\ }\href {\doibase 10.1073/pnas.1705887114} {\bibfield  {journal} {\bibinfo  {journal} {Proceedings of the National Academy of Sciences}\ }\textbf {\bibinfo {volume} {114}},\ \bibinfo {pages}
  {E9026--E9035} (\bibinfo {year} {2017})}\BibitemShut {NoStop}%
\bibitem [{\citenamefont {Vigu{\'e}}\ and\ \citenamefont {Tenaillon}(2023)}]{vigue2023predicting}%
  \BibitemOpen
  \bibfield  {author} {\bibinfo {author} {\bibfnamefont {Lucile}\ \bibnamefont {Vigu{\'e}}}\ and\ \bibinfo {author} {\bibfnamefont {Olivier}\ \bibnamefont {Tenaillon}},\ }\bibfield  {title} {\enquote {\bibinfo {title} {Predicting the effect of mutations to investigate recent events of selection across 60,472 escherichia coli strains},}\ }\href {\doibase 10.1073/pnas.2304177120} {\bibfield  {journal} {\bibinfo  {journal} {Proceedings of the National Academy of Sciences}\ }\textbf {\bibinfo {volume} {120}},\ \bibinfo {pages} {e2304177120} (\bibinfo {year} {2023})}\BibitemShut {NoStop}%
\bibitem [{\citenamefont {Biswas}\ \emph {et~al.}(2019)\citenamefont {Biswas}, \citenamefont {Haldane}, \citenamefont {Arnold},\ and\ \citenamefont {Levy}}]{Biswas2019}%
  \BibitemOpen
  \bibfield  {author} {\bibinfo {author} {\bibfnamefont {Avik}\ \bibnamefont {Biswas}}, \bibinfo {author} {\bibfnamefont {Allan}\ \bibnamefont {Haldane}}, \bibinfo {author} {\bibfnamefont {Eddy}\ \bibnamefont {Arnold}}, \ and\ \bibinfo {author} {\bibfnamefont {Ronald~M}\ \bibnamefont {Levy}},\ }\bibfield  {title} {\enquote {\bibinfo {title} {Epistasis and entrenchment of drug resistance in hiv-1 subtype b},}\ }\href {\doibase 10.7554/eLife.50524} {\bibfield  {journal} {\bibinfo  {journal} {eLife}\ }\textbf {\bibinfo {volume} {8}},\ \bibinfo {pages} {e50524} (\bibinfo {year} {2019})}\BibitemShut {NoStop}%
\bibitem [{\citenamefont {Lyons}\ \emph {et~al.}(2020)\citenamefont {Lyons}, \citenamefont {Zou}, \citenamefont {Xu},\ and\ \citenamefont {Zhang}}]{lyons2020idiosyncratic}%
  \BibitemOpen
  \bibfield  {author} {\bibinfo {author} {\bibfnamefont {Daniel~M}\ \bibnamefont {Lyons}}, \bibinfo {author} {\bibfnamefont {Zhengting}\ \bibnamefont {Zou}}, \bibinfo {author} {\bibfnamefont {Haiqing}\ \bibnamefont {Xu}}, \ and\ \bibinfo {author} {\bibfnamefont {Jianzhi}\ \bibnamefont {Zhang}},\ }\bibfield  {title} {\enquote {\bibinfo {title} {Idiosyncratic epistasis creates universals in mutational effects and evolutionary trajectories},}\ }\href {\doibase https://doi.org/10.1038/s41559-020-01286-y} {\bibfield  {journal} {\bibinfo  {journal} {Nature Ecology \& Evolution}\ }\textbf {\bibinfo {volume} {4}},\ \bibinfo {pages} {1685--1693} (\bibinfo {year} {2020})}\BibitemShut {NoStop}%
\bibitem [{\citenamefont {Vigu{\'e}}\ \emph {et~al.}(2022)\citenamefont {Vigu{\'e}}, \citenamefont {Croce}, \citenamefont {Petitjean}, \citenamefont {Rupp{\'e}}, \citenamefont {Tenaillon},\ and\ \citenamefont {Weigt}}]{vigue2022deciphering}%
  \BibitemOpen
  \bibfield  {author} {\bibinfo {author} {\bibfnamefont {Lucile}\ \bibnamefont {Vigu{\'e}}}, \bibinfo {author} {\bibfnamefont {Giancarlo}\ \bibnamefont {Croce}}, \bibinfo {author} {\bibfnamefont {Marie}\ \bibnamefont {Petitjean}}, \bibinfo {author} {\bibfnamefont {Etienne}\ \bibnamefont {Rupp{\'e}}}, \bibinfo {author} {\bibfnamefont {Olivier}\ \bibnamefont {Tenaillon}}, \ and\ \bibinfo {author} {\bibfnamefont {Martin}\ \bibnamefont {Weigt}},\ }\bibfield  {title} {\enquote {\bibinfo {title} {Deciphering polymorphism in 61,157 {E}scherichia coli genomes via epistatic sequence landscapes},}\ }\href {\doibase https://doi.org/10.1038/s41467-022-31643-3} {\bibfield  {journal} {\bibinfo  {journal} {Nature Communications}\ }\textbf {\bibinfo {volume} {13}},\ \bibinfo {pages} {4030} (\bibinfo {year} {2022})}\BibitemShut {NoStop}%
\bibitem [{\citenamefont {Chen}\ \emph {et~al.}(2024)\citenamefont {Chen}, \citenamefont {Bisardi}, \citenamefont {Lee}, \citenamefont {Cotogno}, \citenamefont {Zamponi}, \citenamefont {Weigt},\ and\ \citenamefont {Tokuriki}}]{chen2023understanding}%
  \BibitemOpen
  \bibfield  {author} {\bibinfo {author} {\bibfnamefont {John~Z}\ \bibnamefont {Chen}}, \bibinfo {author} {\bibfnamefont {Matteo}\ \bibnamefont {Bisardi}}, \bibinfo {author} {\bibfnamefont {Dongkyu}\ \bibnamefont {Lee}}, \bibinfo {author} {\bibfnamefont {Sabrina}\ \bibnamefont {Cotogno}}, \bibinfo {author} {\bibfnamefont {Francesco}\ \bibnamefont {Zamponi}}, \bibinfo {author} {\bibfnamefont {Martin}\ \bibnamefont {Weigt}}, \ and\ \bibinfo {author} {\bibfnamefont {Nobuhiko}\ \bibnamefont {Tokuriki}},\ }\bibfield  {title} {\enquote {\bibinfo {title} {Understanding epistatic networks in the {B}1 beta-lactamases through coevolutionary statistical modeling and deep mutational scanning},}\ }\href {\doibase https://doi.org/10.1038/s41467-024-52614-w} {\bibfield  {journal} {\bibinfo  {journal} {Nature Communications}\ }\textbf {\bibinfo {volume} {15}},\ \bibinfo {pages} {8441} (\bibinfo {year} {2024})}\BibitemShut {NoStop}%
\bibitem [{\citenamefont {Kirkpatrick}\ and\ \citenamefont {Thirumalai}(1988)}]{kirkpatrick1988comparison}%
  \BibitemOpen
  \bibfield  {author} {\bibinfo {author} {\bibfnamefont {Theodore~R}\ \bibnamefont {Kirkpatrick}}\ and\ \bibinfo {author} {\bibfnamefont {Devarajan}\ \bibnamefont {Thirumalai}},\ }\bibfield  {title} {\enquote {\bibinfo {title} {Comparison between dynamical theories and metastable states in regular and glassy mean-field spin models with underlying first-order-like phase transitions},}\ }\href {\doibase 10.1103/physreva.37.4439} {\bibfield  {journal} {\bibinfo  {journal} {Physical Review A}\ }\textbf {\bibinfo {volume} {37}},\ \bibinfo {pages} {4439} (\bibinfo {year} {1988})}\BibitemShut {NoStop}%
\bibitem [{\citenamefont {Franz}\ \emph {et~al.}(1999)\citenamefont {Franz}, \citenamefont {Donati}, \citenamefont {Parisi},\ and\ \citenamefont {Glotzer}}]{franz1999dynamical}%
  \BibitemOpen
  \bibfield  {author} {\bibinfo {author} {\bibfnamefont {Silvio}\ \bibnamefont {Franz}}, \bibinfo {author} {\bibfnamefont {Claudio}\ \bibnamefont {Donati}}, \bibinfo {author} {\bibfnamefont {Giorgio}\ \bibnamefont {Parisi}}, \ and\ \bibinfo {author} {\bibfnamefont {Sharon~C}\ \bibnamefont {Glotzer}},\ }\bibfield  {title} {\enquote {\bibinfo {title} {On dynamical correlations in supercooled liquids},}\ }\href {\doibase 10.1080/13642819908223066} {\bibfield  {journal} {\bibinfo  {journal} {Philosophical Magazine B}\ }\textbf {\bibinfo {volume} {79}},\ \bibinfo {pages} {1827--1831} (\bibinfo {year} {1999})}\BibitemShut {NoStop}%
\bibitem [{\citenamefont {Franz}\ and\ \citenamefont {Parisi}(2000)}]{franz2000non}%
  \BibitemOpen
  \bibfield  {author} {\bibinfo {author} {\bibfnamefont {Silvio}\ \bibnamefont {Franz}}\ and\ \bibinfo {author} {\bibfnamefont {Giorgio}\ \bibnamefont {Parisi}},\ }\bibfield  {title} {\enquote {\bibinfo {title} {On non-linear susceptibility in supercooled liquids},}\ }\href {\doibase 10.1088/0953-8984/12/29/305} {\bibfield  {journal} {\bibinfo  {journal} {Journal of Physics: Condensed Matter}\ }\textbf {\bibinfo {volume} {12}},\ \bibinfo {pages} {6335} (\bibinfo {year} {2000})}\BibitemShut {NoStop}%
\bibitem [{\citenamefont {Bouchaud}\ and\ \citenamefont {Biroli}(2005)}]{bouchaud2005nonlinear}%
  \BibitemOpen
  \bibfield  {author} {\bibinfo {author} {\bibfnamefont {Jean-Philippe}\ \bibnamefont {Bouchaud}}\ and\ \bibinfo {author} {\bibfnamefont {Giulio}\ \bibnamefont {Biroli}},\ }\bibfield  {title} {\enquote {\bibinfo {title} {Nonlinear susceptibility in glassy systems: A probe for cooperative dynamical length scales},}\ }\href {\doibase https://doi.org/10.1103/PhysRevB.72.064204} {\bibfield  {journal} {\bibinfo  {journal} {Physical Review B—Condensed Matter and Materials Physics}\ }\textbf {\bibinfo {volume} {72}},\ \bibinfo {pages} {064204} (\bibinfo {year} {2005})}\BibitemShut {NoStop}%
\bibitem [{\citenamefont {Berthier}\ \emph {et~al.}(2005)\citenamefont {Berthier}, \citenamefont {Biroli}, \citenamefont {Bouchaud}, \citenamefont {Cipelletti}, \citenamefont {Masri}, \citenamefont {L'Hôte}, \citenamefont {Ladieu},\ and\ \citenamefont {Pierno}}]{Berthier2005}%
  \BibitemOpen
  \bibfield  {author} {\bibinfo {author} {\bibfnamefont {L.}~\bibnamefont {Berthier}}, \bibinfo {author} {\bibfnamefont {G.}~\bibnamefont {Biroli}}, \bibinfo {author} {\bibfnamefont {J.-P.}\ \bibnamefont {Bouchaud}}, \bibinfo {author} {\bibfnamefont {L.}~\bibnamefont {Cipelletti}}, \bibinfo {author} {\bibfnamefont {D.~El}\ \bibnamefont {Masri}}, \bibinfo {author} {\bibfnamefont {D.}~\bibnamefont {L'Hôte}}, \bibinfo {author} {\bibfnamefont {F.}~\bibnamefont {Ladieu}}, \ and\ \bibinfo {author} {\bibfnamefont {M.}~\bibnamefont {Pierno}},\ }\bibfield  {title} {\enquote {\bibinfo {title} {Direct experimental evidence of a growing length scale accompanying the glass transition},}\ }\href {\doibase 10.1126/science.1120714} {\bibfield  {journal} {\bibinfo  {journal} {Science}\ }\textbf {\bibinfo {volume} {310}},\ \bibinfo {pages} {1797--1800} (\bibinfo {year} {2005})}\BibitemShut {NoStop}%
\bibitem [{\citenamefont {Berthier}\ and\ \citenamefont {Jack}(2007)}]{Berthier2007}%
  \BibitemOpen
  \bibfield  {author} {\bibinfo {author} {\bibfnamefont {Ludovic}\ \bibnamefont {Berthier}}\ and\ \bibinfo {author} {\bibfnamefont {Robert~L.}\ \bibnamefont {Jack}},\ }\bibfield  {title} {\enquote {\bibinfo {title} {Structure and dynamics of glass formers: Predictability at large length scales},}\ }\href {\doibase 10.1103/PhysRevE.76.041509} {\bibfield  {journal} {\bibinfo  {journal} {Phys. Rev. E}\ }\textbf {\bibinfo {volume} {76}},\ \bibinfo {pages} {041509} (\bibinfo {year} {2007})}\BibitemShut {NoStop}%
\bibitem [{\citenamefont {Franz}\ \emph {et~al.}(2011)\citenamefont {Franz}, \citenamefont {Parisi}, \citenamefont {Ricci-Tersenghi},\ and\ \citenamefont {Rizzo}}]{franz2011field}%
  \BibitemOpen
  \bibfield  {author} {\bibinfo {author} {\bibfnamefont {Silvio}\ \bibnamefont {Franz}}, \bibinfo {author} {\bibfnamefont {Giorgio}\ \bibnamefont {Parisi}}, \bibinfo {author} {\bibfnamefont {Federico}\ \bibnamefont {Ricci-Tersenghi}}, \ and\ \bibinfo {author} {\bibfnamefont {Tommaso}\ \bibnamefont {Rizzo}},\ }\bibfield  {title} {\enquote {\bibinfo {title} {Field theory of fluctuations in glasses},}\ }\href {\doibase https://doi.org/10.1140/epje/i2011-11102-0} {\bibfield  {journal} {\bibinfo  {journal} {The European Physical Journal E}\ }\textbf {\bibinfo {volume} {34}},\ \bibinfo {pages} {1--17} (\bibinfo {year} {2011})}\BibitemShut {NoStop}%
\bibitem [{\citenamefont {Berthier}\ \emph {et~al.}(2011)\citenamefont {Berthier}, \citenamefont {Biroli}, \citenamefont {Bouchaud}, \citenamefont {Cipelletti},\ and\ \citenamefont {van Saarloos}}]{berthier2011dynamical}%
  \BibitemOpen
  \bibfield  {author} {\bibinfo {author} {\bibfnamefont {Ludovic}\ \bibnamefont {Berthier}}, \bibinfo {author} {\bibfnamefont {Giulio}\ \bibnamefont {Biroli}}, \bibinfo {author} {\bibfnamefont {Jean-Philippe}\ \bibnamefont {Bouchaud}}, \bibinfo {author} {\bibfnamefont {Luca}\ \bibnamefont {Cipelletti}}, \ and\ \bibinfo {author} {\bibfnamefont {Wim}\ \bibnamefont {van Saarloos}},\ }\href {\doibase https://doi.org/10.1093/acprof:oso/9780199691470.001.0001} {\emph {\bibinfo {title} {Dynamical heterogeneities in glasses, colloids, and granular media}}},\ Vol.\ \bibinfo {volume} {150}\ (\bibinfo  {publisher} {OUP Oxford},\ \bibinfo {year} {2011})\BibitemShut {NoStop}%
\bibitem [{\citenamefont {Franz}\ \emph {et~al.}(2013)\citenamefont {Franz}, \citenamefont {Jacquin}, \citenamefont {Parisi}, \citenamefont {Urbani},\ and\ \citenamefont {Zamponi}}]{franz2013static}%
  \BibitemOpen
  \bibfield  {author} {\bibinfo {author} {\bibfnamefont {Silvio}\ \bibnamefont {Franz}}, \bibinfo {author} {\bibfnamefont {Hugo}\ \bibnamefont {Jacquin}}, \bibinfo {author} {\bibfnamefont {Giorgio}\ \bibnamefont {Parisi}}, \bibinfo {author} {\bibfnamefont {Pierfrancesco}\ \bibnamefont {Urbani}}, \ and\ \bibinfo {author} {\bibfnamefont {Francesco}\ \bibnamefont {Zamponi}},\ }\bibfield  {title} {\enquote {\bibinfo {title} {Static replica approach to critical correlations in glassy systems},}\ }\href {\doibase 10.1063/1.4776213} {\bibfield  {journal} {\bibinfo  {journal} {The Journal of Chemical Physics}\ }\textbf {\bibinfo {volume} {138}},\ \bibinfo {pages} {12A540} (\bibinfo {year} {2013})}\BibitemShut {NoStop}%
\bibitem [{\citenamefont {Seoane}\ and\ \citenamefont {Zamponi}(2018)}]{Seoane2018}%
  \BibitemOpen
  \bibfield  {author} {\bibinfo {author} {\bibfnamefont {Beatriz}\ \bibnamefont {Seoane}}\ and\ \bibinfo {author} {\bibfnamefont {Francesco}\ \bibnamefont {Zamponi}},\ }\bibfield  {title} {\enquote {\bibinfo {title} {Spin-glass-like aging in colloidal and granular glasses},}\ }\href {\doibase 10.1039/C8SM00859K} {\bibfield  {journal} {\bibinfo  {journal} {Soft Matter}\ }\textbf {\bibinfo {volume} {14}},\ \bibinfo {pages} {5222--5234} (\bibinfo {year} {2018})}\BibitemShut {NoStop}%
\bibitem [{\citenamefont {Folena}\ \emph {et~al.}(2022)\citenamefont {Folena}, \citenamefont {Biroli}, \citenamefont {Charbonneau}, \citenamefont {Hu},\ and\ \citenamefont {Zamponi}}]{Folena2022}%
  \BibitemOpen
  \bibfield  {author} {\bibinfo {author} {\bibfnamefont {Giampaolo}\ \bibnamefont {Folena}}, \bibinfo {author} {\bibfnamefont {Giulio}\ \bibnamefont {Biroli}}, \bibinfo {author} {\bibfnamefont {Patrick}\ \bibnamefont {Charbonneau}}, \bibinfo {author} {\bibfnamefont {Yi}~\bibnamefont {Hu}}, \ and\ \bibinfo {author} {\bibfnamefont {Francesco}\ \bibnamefont {Zamponi}},\ }\bibfield  {title} {\enquote {\bibinfo {title} {Equilibrium fluctuations in mean-field disordered models},}\ }\href {\doibase 10.1103/PhysRevE.106.024605} {\bibfield  {journal} {\bibinfo  {journal} {Phys. Rev. E}\ }\textbf {\bibinfo {volume} {106}},\ \bibinfo {pages} {024605} (\bibinfo {year} {2022})}\BibitemShut {NoStop}%
\bibitem [{\citenamefont {Widmer-Cooper}\ \emph {et~al.}(2008)\citenamefont {Widmer-Cooper}, \citenamefont {Perry}, \citenamefont {Harrowell},\ and\ \citenamefont {Reichman}}]{widmer2008irreversible}%
  \BibitemOpen
  \bibfield  {author} {\bibinfo {author} {\bibfnamefont {Asaph}\ \bibnamefont {Widmer-Cooper}}, \bibinfo {author} {\bibfnamefont {Heidi}\ \bibnamefont {Perry}}, \bibinfo {author} {\bibfnamefont {Peter}\ \bibnamefont {Harrowell}}, \ and\ \bibinfo {author} {\bibfnamefont {David~R}\ \bibnamefont {Reichman}},\ }\bibfield  {title} {\enquote {\bibinfo {title} {Irreversible reorganization in a supercooled liquid originates from localized soft modes},}\ }\href {\doibase https://doi.org/10.1038/nphys1025} {\bibfield  {journal} {\bibinfo  {journal} {Nature Physics}\ }\textbf {\bibinfo {volume} {4}},\ \bibinfo {pages} {711--715} (\bibinfo {year} {2008})}\BibitemShut {NoStop}%
\bibitem [{\citenamefont {Schoenholz}\ \emph {et~al.}(2016)\citenamefont {Schoenholz}, \citenamefont {Cubuk}, \citenamefont {Sussman}, \citenamefont {Kaxiras},\ and\ \citenamefont {Liu}}]{schoenholz2016structural}%
  \BibitemOpen
  \bibfield  {author} {\bibinfo {author} {\bibfnamefont {Samuel~S}\ \bibnamefont {Schoenholz}}, \bibinfo {author} {\bibfnamefont {Ekin~D}\ \bibnamefont {Cubuk}}, \bibinfo {author} {\bibfnamefont {Daniel~M}\ \bibnamefont {Sussman}}, \bibinfo {author} {\bibfnamefont {Efthimios}\ \bibnamefont {Kaxiras}}, \ and\ \bibinfo {author} {\bibfnamefont {Andrea~J}\ \bibnamefont {Liu}},\ }\bibfield  {title} {\enquote {\bibinfo {title} {A structural approach to relaxation in glassy liquids},}\ }\href {\doibase https://doi.org/10.1038/nphys3644} {\bibfield  {journal} {\bibinfo  {journal} {Nature Physics}\ }\textbf {\bibinfo {volume} {12}},\ \bibinfo {pages} {469--471} (\bibinfo {year} {2016})}\BibitemShut {NoStop}%
\bibitem [{\citenamefont {Bapst}\ \emph {et~al.}(2020)\citenamefont {Bapst}, \citenamefont {Keck}, \citenamefont {Grabska-Barwi{\'n}ska}, \citenamefont {Donner}, \citenamefont {Cubuk}, \citenamefont {Schoenholz}, \citenamefont {Obika}, \citenamefont {Nelson}, \citenamefont {Back}, \citenamefont {Hassabis} \emph {et~al.}}]{bapst2020unveiling}%
  \BibitemOpen
  \bibfield  {author} {\bibinfo {author} {\bibfnamefont {Victor}\ \bibnamefont {Bapst}}, \bibinfo {author} {\bibfnamefont {Thomas}\ \bibnamefont {Keck}}, \bibinfo {author} {\bibfnamefont {A}~\bibnamefont {Grabska-Barwi{\'n}ska}}, \bibinfo {author} {\bibfnamefont {Craig}\ \bibnamefont {Donner}}, \bibinfo {author} {\bibfnamefont {Ekin~Dogus}\ \bibnamefont {Cubuk}}, \bibinfo {author} {\bibfnamefont {Samuel~S}\ \bibnamefont {Schoenholz}}, \bibinfo {author} {\bibfnamefont {Annette}\ \bibnamefont {Obika}}, \bibinfo {author} {\bibfnamefont {Alexander~WR}\ \bibnamefont {Nelson}}, \bibinfo {author} {\bibfnamefont {Trevor}\ \bibnamefont {Back}}, \bibinfo {author} {\bibfnamefont {Demis}\ \bibnamefont {Hassabis}},  \emph {et~al.},\ }\bibfield  {title} {\enquote {\bibinfo {title} {Unveiling the predictive power of static structure in glassy systems},}\ }\href {\doibase https://doi.org/10.1038/s41567-020-0842-8} {\bibfield  {journal} {\bibinfo  {journal} {Nature physics}\ }\textbf {\bibinfo {volume} {16}},\ \bibinfo
  {pages} {448--454} (\bibinfo {year} {2020})}\BibitemShut {NoStop}%
\bibitem [{\citenamefont {Jung}\ \emph {et~al.}(2024)\citenamefont {Jung}, \citenamefont {Biroli},\ and\ \citenamefont {Berthier}}]{jung2024dynamic}%
  \BibitemOpen
  \bibfield  {author} {\bibinfo {author} {\bibfnamefont {Gerhard}\ \bibnamefont {Jung}}, \bibinfo {author} {\bibfnamefont {Giulio}\ \bibnamefont {Biroli}}, \ and\ \bibinfo {author} {\bibfnamefont {Ludovic}\ \bibnamefont {Berthier}},\ }\bibfield  {title} {\enquote {\bibinfo {title} {Dynamic heterogeneity at the experimental glass transition predicted by transferable machine learning},}\ }\href {\doibase https://doi.org/10.1103/PhysRevB.109.064205} {\bibfield  {journal} {\bibinfo  {journal} {Physical Review B}\ }\textbf {\bibinfo {volume} {109}},\ \bibinfo {pages} {064205} (\bibinfo {year} {2024})}\BibitemShut {NoStop}%
\bibitem [{\citenamefont {Jung}\ \emph {et~al.}(2025)\citenamefont {Jung}, \citenamefont {Alkemade}, \citenamefont {Bapst}, \citenamefont {Coslovich}, \citenamefont {Filion}, \citenamefont {Landes}, \citenamefont {Liu}, \citenamefont {Pezzicoli}, \citenamefont {Shiba}, \citenamefont {Volpe}, \citenamefont {Zamponi}, \citenamefont {Berthier},\ and\ \citenamefont {Biroli}}]{jung2025roadmap}%
  \BibitemOpen
  \bibfield  {author} {\bibinfo {author} {\bibfnamefont {Gerhard}\ \bibnamefont {Jung}}, \bibinfo {author} {\bibfnamefont {Rinske~M}\ \bibnamefont {Alkemade}}, \bibinfo {author} {\bibfnamefont {Victor}\ \bibnamefont {Bapst}}, \bibinfo {author} {\bibfnamefont {Daniele}\ \bibnamefont {Coslovich}}, \bibinfo {author} {\bibfnamefont {Laura}\ \bibnamefont {Filion}}, \bibinfo {author} {\bibfnamefont {Fran{\c{c}}ois~P}\ \bibnamefont {Landes}}, \bibinfo {author} {\bibfnamefont {Andrea~J}\ \bibnamefont {Liu}}, \bibinfo {author} {\bibfnamefont {Francesco~Saverio}\ \bibnamefont {Pezzicoli}}, \bibinfo {author} {\bibfnamefont {Hayato}\ \bibnamefont {Shiba}}, \bibinfo {author} {\bibfnamefont {Giovanni}\ \bibnamefont {Volpe}}, \bibinfo {author} {\bibfnamefont {Francesco}\ \bibnamefont {Zamponi}}, \bibinfo {author} {\bibfnamefont {Ludovic}\ \bibnamefont {Berthier}}, \ and\ \bibinfo {author} {\bibfnamefont {Giulio}\ \bibnamefont {Biroli}},\ }\bibfield  {title} {\enquote {\bibinfo {title} {Roadmap on machine learning glassy
  dynamics},}\ }\href {\doibase https://doi.org/10.1038/s42254-024-00791-4} {\bibfield  {journal} {\bibinfo  {journal} {Nature Reviews Physics}\ ,\ \bibinfo {pages} {1--14}} (\bibinfo {year} {2025})}\BibitemShut {NoStop}%
\bibitem [{\citenamefont {Otwinowski}(2018)}]{otwinowski2018biophysical}%
  \BibitemOpen
  \bibfield  {author} {\bibinfo {author} {\bibfnamefont {Jakub}\ \bibnamefont {Otwinowski}},\ }\bibfield  {title} {\enquote {\bibinfo {title} {Biophysical inference of epistasis and the effects of mutations on protein stability and function},}\ }\href {\doibase https://doi.org/10.1093/molbev/msy141} {\bibfield  {journal} {\bibinfo  {journal} {Molecular biology and evolution}\ }\textbf {\bibinfo {volume} {35}},\ \bibinfo {pages} {2345--2354} (\bibinfo {year} {2018})}\BibitemShut {NoStop}%
\bibitem [{\citenamefont {Otwinowski}\ \emph {et~al.}(2018)\citenamefont {Otwinowski}, \citenamefont {McCandlish},\ and\ \citenamefont {Plotkin}}]{otwinowski2018inferring}%
  \BibitemOpen
  \bibfield  {author} {\bibinfo {author} {\bibfnamefont {Jakub}\ \bibnamefont {Otwinowski}}, \bibinfo {author} {\bibfnamefont {David~M}\ \bibnamefont {McCandlish}}, \ and\ \bibinfo {author} {\bibfnamefont {Joshua~B}\ \bibnamefont {Plotkin}},\ }\bibfield  {title} {\enquote {\bibinfo {title} {Inferring the shape of global epistasis},}\ }\href {\doibase https://doi.org/10.1073/pnas.1804015115} {\bibfield  {journal} {\bibinfo  {journal} {Proceedings of the National Academy of Sciences}\ }\textbf {\bibinfo {volume} {115}},\ \bibinfo {pages} {E7550--E7558} (\bibinfo {year} {2018})}\BibitemShut {NoStop}%
\bibitem [{\citenamefont {Reddy}\ and\ \citenamefont {Desai}(2021)}]{reddy2021global}%
  \BibitemOpen
  \bibfield  {author} {\bibinfo {author} {\bibfnamefont {Gautam}\ \bibnamefont {Reddy}}\ and\ \bibinfo {author} {\bibfnamefont {Michael~M}\ \bibnamefont {Desai}},\ }\bibfield  {title} {\enquote {\bibinfo {title} {Global epistasis emerges from a generic model of a complex trait},}\ }\href {\doibase https://doi.org/10.7554/eLife.64740} {\bibfield  {journal} {\bibinfo  {journal} {Elife}\ }\textbf {\bibinfo {volume} {10}},\ \bibinfo {pages} {e64740} (\bibinfo {year} {2021})}\BibitemShut {NoStop}%
\bibitem [{\citenamefont {Schulte}\ \emph {et~al.}(2025)\citenamefont {Schulte}, \citenamefont {Alqatari}, \citenamefont {Rossi},\ and\ \citenamefont {Zamponi}}]{schulte2025functional}%
  \BibitemOpen
  \bibfield  {author} {\bibinfo {author} {\bibfnamefont {Anna~Ottavia}\ \bibnamefont {Schulte}}, \bibinfo {author} {\bibfnamefont {Samar}\ \bibnamefont {Alqatari}}, \bibinfo {author} {\bibfnamefont {Saverio}\ \bibnamefont {Rossi}}, \ and\ \bibinfo {author} {\bibfnamefont {Francesco}\ \bibnamefont {Zamponi}},\ }\bibfield  {title} {\enquote {\bibinfo {title} {Functional bottlenecks can emerge from non-epistatic underlying traits},}\ }\href {10.1101/2025.05.20.655048} {\bibfield  {journal} {\bibinfo  {journal} {bioRxiv:2025.05.20.655048}\ } (\bibinfo {year} {2025})}\BibitemShut {NoStop}%
\bibitem [{\citenamefont {Sailer}\ and\ \citenamefont {Harms}(2017{\natexlab{a}})}]{sailer2017detecting}%
  \BibitemOpen
  \bibfield  {author} {\bibinfo {author} {\bibfnamefont {Zachary~R}\ \bibnamefont {Sailer}}\ and\ \bibinfo {author} {\bibfnamefont {Michael~J}\ \bibnamefont {Harms}},\ }\bibfield  {title} {\enquote {\bibinfo {title} {Detecting high-order epistasis in nonlinear genotype-phenotype maps},}\ }\href {\doibase https://doi.org/10.1534/genetics.116.195214} {\bibfield  {journal} {\bibinfo  {journal} {Genetics}\ }\textbf {\bibinfo {volume} {205}},\ \bibinfo {pages} {1079--1088} (\bibinfo {year} {2017}{\natexlab{a}})}\BibitemShut {NoStop}%
\bibitem [{\citenamefont {Sailer}\ and\ \citenamefont {Harms}(2017{\natexlab{b}})}]{sailer2017high}%
  \BibitemOpen
  \bibfield  {author} {\bibinfo {author} {\bibfnamefont {Zachary~R}\ \bibnamefont {Sailer}}\ and\ \bibinfo {author} {\bibfnamefont {Michael~J}\ \bibnamefont {Harms}},\ }\bibfield  {title} {\enquote {\bibinfo {title} {High-order epistasis shapes evolutionary trajectories},}\ }\href {\doibase https://doi.org/10.1371/journal.pcbi.1005541} {\bibfield  {journal} {\bibinfo  {journal} {PLoS computational biology}\ }\textbf {\bibinfo {volume} {13}},\ \bibinfo {pages} {e1005541} (\bibinfo {year} {2017}{\natexlab{b}})}\BibitemShut {NoStop}%
\bibitem [{\citenamefont {Domingo}\ \emph {et~al.}(2018)\citenamefont {Domingo}, \citenamefont {Diss},\ and\ \citenamefont {Lehner}}]{domingo2018pairwise}%
  \BibitemOpen
  \bibfield  {author} {\bibinfo {author} {\bibfnamefont {J{\'u}lia}\ \bibnamefont {Domingo}}, \bibinfo {author} {\bibfnamefont {Guillaume}\ \bibnamefont {Diss}}, \ and\ \bibinfo {author} {\bibfnamefont {Ben}\ \bibnamefont {Lehner}},\ }\bibfield  {title} {\enquote {\bibinfo {title} {Pairwise and higher-order genetic interactions during the evolution of a t{RNA}},}\ }\href {\doibase https://doi.org/10.1038/s41586-018-0170-7} {\bibfield  {journal} {\bibinfo  {journal} {Nature}\ }\textbf {\bibinfo {volume} {558}},\ \bibinfo {pages} {117--121} (\bibinfo {year} {2018})}\BibitemShut {NoStop}%
\bibitem [{\citenamefont {Poelwijk}\ \emph {et~al.}(2019)\citenamefont {Poelwijk}, \citenamefont {Socolich},\ and\ \citenamefont {Ranganathan}}]{poelwijk2019learning}%
  \BibitemOpen
  \bibfield  {author} {\bibinfo {author} {\bibfnamefont {Frank~J}\ \bibnamefont {Poelwijk}}, \bibinfo {author} {\bibfnamefont {Michael}\ \bibnamefont {Socolich}}, \ and\ \bibinfo {author} {\bibfnamefont {Rama}\ \bibnamefont {Ranganathan}},\ }\bibfield  {title} {\enquote {\bibinfo {title} {Learning the pattern of epistasis linking genotype and phenotype in a protein},}\ }\href {\doibase https://doi.org/10.1038/s41467-019-12130-8} {\bibfield  {journal} {\bibinfo  {journal} {Nature communications}\ }\textbf {\bibinfo {volume} {10}},\ \bibinfo {pages} {4213} (\bibinfo {year} {2019})}\BibitemShut {NoStop}%
\bibitem [{\citenamefont {Ballal}\ \emph {et~al.}(2020)\citenamefont {Ballal}, \citenamefont {Laurendon}, \citenamefont {Salmon}, \citenamefont {Vardakou}, \citenamefont {Cheema}, \citenamefont {Defernez}, \citenamefont {O’Maille},\ and\ \citenamefont {Morozov}}]{ballal2020sparse}%
  \BibitemOpen
  \bibfield  {author} {\bibinfo {author} {\bibfnamefont {Aditya}\ \bibnamefont {Ballal}}, \bibinfo {author} {\bibfnamefont {Caroline}\ \bibnamefont {Laurendon}}, \bibinfo {author} {\bibfnamefont {Melissa}\ \bibnamefont {Salmon}}, \bibinfo {author} {\bibfnamefont {Maria}\ \bibnamefont {Vardakou}}, \bibinfo {author} {\bibfnamefont {Jitender}\ \bibnamefont {Cheema}}, \bibinfo {author} {\bibfnamefont {Marianne}\ \bibnamefont {Defernez}}, \bibinfo {author} {\bibfnamefont {Paul~E}\ \bibnamefont {O’Maille}}, \ and\ \bibinfo {author} {\bibfnamefont {Alexandre~V}\ \bibnamefont {Morozov}},\ }\bibfield  {title} {\enquote {\bibinfo {title} {Sparse epistatic patterns in the evolution of terpene synthases},}\ }\href {\doibase 10.1093/molbev/msaa052} {\bibfield  {journal} {\bibinfo  {journal} {Molecular biology and evolution}\ }\textbf {\bibinfo {volume} {37}},\ \bibinfo {pages} {1907--1924} (\bibinfo {year} {2020})}\BibitemShut {NoStop}%
\bibitem [{\citenamefont {Phillips}\ \emph {et~al.}(2021)\citenamefont {Phillips}, \citenamefont {Lawrence}, \citenamefont {Moulana}, \citenamefont {Dupic}, \citenamefont {Chang}, \citenamefont {Johnson}, \citenamefont {Cvijovic}, \citenamefont {Mora}, \citenamefont {Walczak},\ and\ \citenamefont {Desai}}]{phillips2021binding}%
  \BibitemOpen
  \bibfield  {author} {\bibinfo {author} {\bibfnamefont {Angela~M}\ \bibnamefont {Phillips}}, \bibinfo {author} {\bibfnamefont {Katherine~R}\ \bibnamefont {Lawrence}}, \bibinfo {author} {\bibfnamefont {Alief}\ \bibnamefont {Moulana}}, \bibinfo {author} {\bibfnamefont {Thomas}\ \bibnamefont {Dupic}}, \bibinfo {author} {\bibfnamefont {Jeffrey}\ \bibnamefont {Chang}}, \bibinfo {author} {\bibfnamefont {Milo~S}\ \bibnamefont {Johnson}}, \bibinfo {author} {\bibfnamefont {Ivana}\ \bibnamefont {Cvijovic}}, \bibinfo {author} {\bibfnamefont {Thierry}\ \bibnamefont {Mora}}, \bibinfo {author} {\bibfnamefont {Aleksandra~M}\ \bibnamefont {Walczak}}, \ and\ \bibinfo {author} {\bibfnamefont {Michael~M}\ \bibnamefont {Desai}},\ }\bibfield  {title} {\enquote {\bibinfo {title} {Binding affinity landscapes constrain the evolution of broadly neutralizing anti-influenza antibodies},}\ }\href {\doibase https://doi.org/10.7554/eLife.71393} {\bibfield  {journal} {\bibinfo  {journal} {Elife}\ }\textbf {\bibinfo {volume} {10}},\
  \bibinfo {pages} {e71393} (\bibinfo {year} {2021})}\BibitemShut {NoStop}%
\bibitem [{\citenamefont {Miton}\ \emph {et~al.}(2021)\citenamefont {Miton}, \citenamefont {Buda},\ and\ \citenamefont {Tokuriki}}]{miton2021epistasis}%
  \BibitemOpen
  \bibfield  {author} {\bibinfo {author} {\bibfnamefont {Charlotte~M}\ \bibnamefont {Miton}}, \bibinfo {author} {\bibfnamefont {Karol}\ \bibnamefont {Buda}}, \ and\ \bibinfo {author} {\bibfnamefont {Nobuhiko}\ \bibnamefont {Tokuriki}},\ }\bibfield  {title} {\enquote {\bibinfo {title} {Epistasis and intramolecular networks in protein evolution},}\ }\href {\doibase https://doi.org/10.1016/j.sbi.2021.04.007} {\bibfield  {journal} {\bibinfo  {journal} {Current opinion in structural biology}\ }\textbf {\bibinfo {volume} {69}},\ \bibinfo {pages} {160--168} (\bibinfo {year} {2021})}\BibitemShut {NoStop}%
\bibitem [{\citenamefont {Lunzer}\ \emph {et~al.}(2010)\citenamefont {Lunzer}, \citenamefont {Golding},\ and\ \citenamefont {Dean}}]{lunzer2010pervasive}%
  \BibitemOpen
  \bibfield  {author} {\bibinfo {author} {\bibfnamefont {Mark}\ \bibnamefont {Lunzer}}, \bibinfo {author} {\bibfnamefont {G~Brian}\ \bibnamefont {Golding}}, \ and\ \bibinfo {author} {\bibfnamefont {Antony~M}\ \bibnamefont {Dean}},\ }\bibfield  {title} {\enquote {\bibinfo {title} {Pervasive cryptic epistasis in molecular evolution},}\ }\href {\doibase https://doi.org/10.1371/journal.pgen.1001162} {\bibfield  {journal} {\bibinfo  {journal} {PLoS genetics}\ }\textbf {\bibinfo {volume} {6}},\ \bibinfo {pages} {e1001162} (\bibinfo {year} {2010})}\BibitemShut {NoStop}%
\bibitem [{\citenamefont {Miton}\ and\ \citenamefont {Tokuriki}(2016)}]{miton2016mutational}%
  \BibitemOpen
  \bibfield  {author} {\bibinfo {author} {\bibfnamefont {Charlotte~M}\ \bibnamefont {Miton}}\ and\ \bibinfo {author} {\bibfnamefont {Nobuhiko}\ \bibnamefont {Tokuriki}},\ }\bibfield  {title} {\enquote {\bibinfo {title} {How mutational epistasis impairs predictability in protein evolution and design},}\ }\href {\doibase https://doi.org/10.1002/pro.2876} {\bibfield  {journal} {\bibinfo  {journal} {Protein Science}\ }\textbf {\bibinfo {volume} {25}},\ \bibinfo {pages} {1260--1272} (\bibinfo {year} {2016})}\BibitemShut {NoStop}%
\bibitem [{\citenamefont {Rivoire}\ \emph {et~al.}(2016)\citenamefont {Rivoire}, \citenamefont {Reynolds},\ and\ \citenamefont {Ranganathan}}]{rivoire2016evolution}%
  \BibitemOpen
  \bibfield  {author} {\bibinfo {author} {\bibfnamefont {Olivier}\ \bibnamefont {Rivoire}}, \bibinfo {author} {\bibfnamefont {Kimberly~A}\ \bibnamefont {Reynolds}}, \ and\ \bibinfo {author} {\bibfnamefont {Rama}\ \bibnamefont {Ranganathan}},\ }\bibfield  {title} {\enquote {\bibinfo {title} {Evolution-based functional decomposition of proteins},}\ }\href {\doibase https://doi.org/10.1371/journal.pcbi.1004817} {\bibfield  {journal} {\bibinfo  {journal} {PLoS computational biology}\ }\textbf {\bibinfo {volume} {12}},\ \bibinfo {pages} {e1004817} (\bibinfo {year} {2016})}\BibitemShut {NoStop}%
\bibitem [{\citenamefont {Starr}\ \emph {et~al.}(2018)\citenamefont {Starr}, \citenamefont {Flynn}, \citenamefont {Mishra}, \citenamefont {Bolon},\ and\ \citenamefont {Thornton}}]{starr2018pervasive}%
  \BibitemOpen
  \bibfield  {author} {\bibinfo {author} {\bibfnamefont {Tyler~N}\ \bibnamefont {Starr}}, \bibinfo {author} {\bibfnamefont {Julia~M}\ \bibnamefont {Flynn}}, \bibinfo {author} {\bibfnamefont {Parul}\ \bibnamefont {Mishra}}, \bibinfo {author} {\bibfnamefont {Daniel~NA}\ \bibnamefont {Bolon}}, \ and\ \bibinfo {author} {\bibfnamefont {Joseph~W}\ \bibnamefont {Thornton}},\ }\bibfield  {title} {\enquote {\bibinfo {title} {Pervasive contingency and entrenchment in a billion years of {H}sp90 evolution},}\ }\href {\doibase https://doi.org/10.1073/pnas.1718133115} {\bibfield  {journal} {\bibinfo  {journal} {Proceedings of the National Academy of Sciences}\ }\textbf {\bibinfo {volume} {115}},\ \bibinfo {pages} {4453--4458} (\bibinfo {year} {2018})}\BibitemShut {NoStop}%
\bibitem [{\citenamefont {Bakerlee}\ \emph {et~al.}(2022)\citenamefont {Bakerlee}, \citenamefont {Nguyen~Ba}, \citenamefont {Shulgina}, \citenamefont {Rojas~Echenique},\ and\ \citenamefont {Desai}}]{bakerlee2022idiosyncratic}%
  \BibitemOpen
  \bibfield  {author} {\bibinfo {author} {\bibfnamefont {Christopher~W}\ \bibnamefont {Bakerlee}}, \bibinfo {author} {\bibfnamefont {Alex~N}\ \bibnamefont {Nguyen~Ba}}, \bibinfo {author} {\bibfnamefont {Yekaterina}\ \bibnamefont {Shulgina}}, \bibinfo {author} {\bibfnamefont {Jose~I}\ \bibnamefont {Rojas~Echenique}}, \ and\ \bibinfo {author} {\bibfnamefont {Michael~M}\ \bibnamefont {Desai}},\ }\bibfield  {title} {\enquote {\bibinfo {title} {Idiosyncratic epistasis leads to global fitness--correlated trends},}\ }\href {\doibase 10.1126/science.abm4774} {\bibfield  {journal} {\bibinfo  {journal} {Science}\ }\textbf {\bibinfo {volume} {376}},\ \bibinfo {pages} {630--635} (\bibinfo {year} {2022})}\BibitemShut {NoStop}%
\bibitem [{\citenamefont {Papkou}\ \emph {et~al.}(2023)\citenamefont {Papkou}, \citenamefont {Garcia-Pastor}, \citenamefont {Escudero},\ and\ \citenamefont {Wagner}}]{Papkou2023}%
  \BibitemOpen
  \bibfield  {author} {\bibinfo {author} {\bibfnamefont {Andrei}\ \bibnamefont {Papkou}}, \bibinfo {author} {\bibfnamefont {Lucia}\ \bibnamefont {Garcia-Pastor}}, \bibinfo {author} {\bibfnamefont {José~Antonio}\ \bibnamefont {Escudero}}, \ and\ \bibinfo {author} {\bibfnamefont {Andreas}\ \bibnamefont {Wagner}},\ }\bibfield  {title} {\enquote {\bibinfo {title} {A rugged yet easily navigable fitness landscape},}\ }\href {\doibase 10.1126/science.adh3860} {\bibfield  {journal} {\bibinfo  {journal} {Science}\ }\textbf {\bibinfo {volume} {382}},\ \bibinfo {pages} {901} (\bibinfo {year} {2023})}\BibitemShut {NoStop}%
\bibitem [{\citenamefont {Somermeyer}\ \emph {et~al.}(2022)\citenamefont {Somermeyer}, \citenamefont {Fleiss}, \citenamefont {Mishin}, \citenamefont {Bozhanova}, \citenamefont {Igolkina}, \citenamefont {Meiler}, \citenamefont {Pujol}, \citenamefont {Putintseva}, \citenamefont {Sarkisyan},\ and\ \citenamefont {Kondrashov}}]{Somermeyer2022}%
  \BibitemOpen
  \bibfield  {author} {\bibinfo {author} {\bibfnamefont {Louisa~Gonzalez}\ \bibnamefont {Somermeyer}}, \bibinfo {author} {\bibfnamefont {Aubin}\ \bibnamefont {Fleiss}}, \bibinfo {author} {\bibfnamefont {Alexander~S.}\ \bibnamefont {Mishin}}, \bibinfo {author} {\bibfnamefont {Nina~G.}\ \bibnamefont {Bozhanova}}, \bibinfo {author} {\bibfnamefont {Anna~A.}\ \bibnamefont {Igolkina}}, \bibinfo {author} {\bibfnamefont {Jens}\ \bibnamefont {Meiler}}, \bibinfo {author} {\bibfnamefont {Maria Elisenda~Alaball}\ \bibnamefont {Pujol}}, \bibinfo {author} {\bibfnamefont {Ekaterina~V.}\ \bibnamefont {Putintseva}}, \bibinfo {author} {\bibfnamefont {Karen~S.}\ \bibnamefont {Sarkisyan}}, \ and\ \bibinfo {author} {\bibfnamefont {Fyodor~A.}\ \bibnamefont {Kondrashov}},\ }\bibfield  {title} {\enquote {\bibinfo {title} {Heterogeneity of the gfp fitness landscape and data-driven protein design},}\ }\href {\doibase 10.7554/eLife.75842} {\bibfield  {journal} {\bibinfo  {journal} {eLife}\ }\textbf {\bibinfo {volume} {11}},\ \bibinfo
  {pages} {e75842} (\bibinfo {year} {2022})}\BibitemShut {NoStop}%
\bibitem [{\citenamefont {Schulz}\ \emph {et~al.}(2025)\citenamefont {Schulz}, \citenamefont {Tan}, \citenamefont {Wu},\ and\ \citenamefont {Wang}}]{Schulz2025}%
  \BibitemOpen
  \bibfield  {author} {\bibinfo {author} {\bibfnamefont {Steven}\ \bibnamefont {Schulz}}, \bibinfo {author} {\bibfnamefont {Timothy J.~C.}\ \bibnamefont {Tan}}, \bibinfo {author} {\bibfnamefont {Nicholas~C.}\ \bibnamefont {Wu}}, \ and\ \bibinfo {author} {\bibfnamefont {Shenshen}\ \bibnamefont {Wang}},\ }\bibfield  {title} {\enquote {\bibinfo {title} {Epistatic hotspots organize antibody fitness landscape and boost evolvability},}\ }\href {\doibase 10.1073/pnas.2413884122} {\bibfield  {journal} {\bibinfo  {journal} {Proceedings of the National Academy of Sciences}\ }\textbf {\bibinfo {volume} {122}},\ \bibinfo {pages} {e2413884122} (\bibinfo {year} {2025})}\BibitemShut {NoStop}%
\bibitem [{\citenamefont {Olson}\ \emph {et~al.}(2014)\citenamefont {Olson}, \citenamefont {Wu},\ and\ \citenamefont {Sun}}]{olson2014comprehensive}%
  \BibitemOpen
  \bibfield  {author} {\bibinfo {author} {\bibfnamefont {C~Anders}\ \bibnamefont {Olson}}, \bibinfo {author} {\bibfnamefont {Nicholas~C}\ \bibnamefont {Wu}}, \ and\ \bibinfo {author} {\bibfnamefont {Ren}\ \bibnamefont {Sun}},\ }\bibfield  {title} {\enquote {\bibinfo {title} {A comprehensive biophysical description of pairwise epistasis throughout an entire protein domain},}\ }\href {\doibase https://doi.org/10.1016/j.cub.2014.09.072} {\bibfield  {journal} {\bibinfo  {journal} {Current biology}\ }\textbf {\bibinfo {volume} {24}},\ \bibinfo {pages} {2643--2651} (\bibinfo {year} {2014})}\BibitemShut {NoStop}%
\bibitem [{\citenamefont {Romanowicz}\ \emph {et~al.}(2025)\citenamefont {Romanowicz}, \citenamefont {Resnick}, \citenamefont {Hinton},\ and\ \citenamefont {Plesa}}]{Romanowicz2025}%
  \BibitemOpen
  \bibfield  {author} {\bibinfo {author} {\bibfnamefont {Karl~J.}\ \bibnamefont {Romanowicz}}, \bibinfo {author} {\bibfnamefont {Carmen}\ \bibnamefont {Resnick}}, \bibinfo {author} {\bibfnamefont {Samuel~R.}\ \bibnamefont {Hinton}}, \ and\ \bibinfo {author} {\bibfnamefont {Calin}\ \bibnamefont {Plesa}},\ }\bibfield  {title} {\enquote {\bibinfo {title} {Exploring antibiotic resistance in diverse homologs of the dihydrofolate reductase protein family through broad mutational scanning},}\ }\href {https://doi.org/10.1101/2025.01.23.634126} {\bibfield  {journal} {\bibinfo  {journal} {bioRxiv:2025.01.23.634126}\ } (\bibinfo {year} {2025})}\BibitemShut {NoStop}%
\bibitem [{\citenamefont {Figliuzzi}\ \emph {et~al.}(2018)\citenamefont {Figliuzzi}, \citenamefont {Barrat-Charlaix},\ and\ \citenamefont {Weigt}}]{Figliuzzi2018}%
  \BibitemOpen
  \bibfield  {author} {\bibinfo {author} {\bibfnamefont {Matteo}\ \bibnamefont {Figliuzzi}}, \bibinfo {author} {\bibfnamefont {Pierre}\ \bibnamefont {Barrat-Charlaix}}, \ and\ \bibinfo {author} {\bibfnamefont {Martin}\ \bibnamefont {Weigt}},\ }\bibfield  {title} {\enquote {\bibinfo {title} {{How Pairwise Coevolutionary Models Capture the Collective Residue Variability in Proteins?}}}\ }\href {\doibase 10.1093/molbev/msy007} {\bibfield  {journal} {\bibinfo  {journal} {Molecular Biology and Evolution}\ }\textbf {\bibinfo {volume} {35}},\ \bibinfo {pages} {1018--1027} (\bibinfo {year} {2018})}\BibitemShut {NoStop}%
\bibitem [{\citenamefont {Muntoni}\ \emph {et~al.}(2021)\citenamefont {Muntoni}, \citenamefont {Pagnani}, \citenamefont {Weigt},\ and\ \citenamefont {Zamponi}}]{muntoni2021adabmdca}%
  \BibitemOpen
  \bibfield  {author} {\bibinfo {author} {\bibfnamefont {Anna~Paola}\ \bibnamefont {Muntoni}}, \bibinfo {author} {\bibfnamefont {Andrea}\ \bibnamefont {Pagnani}}, \bibinfo {author} {\bibfnamefont {Martin}\ \bibnamefont {Weigt}}, \ and\ \bibinfo {author} {\bibfnamefont {Francesco}\ \bibnamefont {Zamponi}},\ }\bibfield  {title} {\enquote {\bibinfo {title} {adabm{DCA}: adaptive {B}oltzmann machine learning for biological sequences},}\ }\href {\doibase https://doi.org/10.1186/s12859-021-04441-9} {\bibfield  {journal} {\bibinfo  {journal} {BMC bioinformatics}\ }\textbf {\bibinfo {volume} {22}},\ \bibinfo {pages} {1--19} (\bibinfo {year} {2021})}\BibitemShut {NoStop}%
\bibitem [{\citenamefont {Russ}\ \emph {et~al.}(2020)\citenamefont {Russ}, \citenamefont {Figliuzzi}, \citenamefont {Stocker}, \citenamefont {Barrat-Charlaix}, \citenamefont {Socolich}, \citenamefont {Kast}, \citenamefont {Hilvert}, \citenamefont {Monasson}, \citenamefont {Cocco}, \citenamefont {Weigt},\ and\ \citenamefont {Ranganathan}}]{russ2020evolution}%
  \BibitemOpen
  \bibfield  {author} {\bibinfo {author} {\bibfnamefont {William~P}\ \bibnamefont {Russ}}, \bibinfo {author} {\bibfnamefont {Matteo}\ \bibnamefont {Figliuzzi}}, \bibinfo {author} {\bibfnamefont {Christian}\ \bibnamefont {Stocker}}, \bibinfo {author} {\bibfnamefont {Pierre}\ \bibnamefont {Barrat-Charlaix}}, \bibinfo {author} {\bibfnamefont {Michael}\ \bibnamefont {Socolich}}, \bibinfo {author} {\bibfnamefont {Peter}\ \bibnamefont {Kast}}, \bibinfo {author} {\bibfnamefont {Donald}\ \bibnamefont {Hilvert}}, \bibinfo {author} {\bibfnamefont {Remi}\ \bibnamefont {Monasson}}, \bibinfo {author} {\bibfnamefont {Simona}\ \bibnamefont {Cocco}}, \bibinfo {author} {\bibfnamefont {Martin}\ \bibnamefont {Weigt}}, \ and\ \bibinfo {author} {\bibfnamefont {Rama}\ \bibnamefont {Ranganathan}},\ }\bibfield  {title} {\enquote {\bibinfo {title} {An evolution-based model for designing chorismate mutase enzymes},}\ }\href {\doibase 10.1126/science.aba3304} {\bibfield  {journal} {\bibinfo  {journal} {Science}\ }\textbf {\bibinfo
  {volume} {369}},\ \bibinfo {pages} {440--445} (\bibinfo {year} {2020})}\BibitemShut {NoStop}%
\bibitem [{\citenamefont {Ashkenazy}\ \emph {et~al.}(2012)\citenamefont {Ashkenazy}, \citenamefont {Penn}, \citenamefont {Doron-Faigenboim}, \citenamefont {Cohen}, \citenamefont {Cannarozzi}, \citenamefont {Zomer},\ and\ \citenamefont {Pupko}}]{FastML}%
  \BibitemOpen
  \bibfield  {author} {\bibinfo {author} {\bibfnamefont {Haim}\ \bibnamefont {Ashkenazy}}, \bibinfo {author} {\bibfnamefont {Osnat}\ \bibnamefont {Penn}}, \bibinfo {author} {\bibfnamefont {Adi}\ \bibnamefont {Doron-Faigenboim}}, \bibinfo {author} {\bibfnamefont {Ofir}\ \bibnamefont {Cohen}}, \bibinfo {author} {\bibfnamefont {Gina}\ \bibnamefont {Cannarozzi}}, \bibinfo {author} {\bibfnamefont {Oren}\ \bibnamefont {Zomer}}, \ and\ \bibinfo {author} {\bibfnamefont {Tal}\ \bibnamefont {Pupko}},\ }\bibfield  {title} {\enquote {\bibinfo {title} {{FastML: a web server for probabilistic reconstruction of ancestral sequences}},}\ }\href {\doibase 10.1093/nar/gks498} {\bibfield  {journal} {\bibinfo  {journal} {Nucleic Acids Research}\ }\textbf {\bibinfo {volume} {40}},\ \bibinfo {pages} {W580--W584} (\bibinfo {year} {2012})}\BibitemShut {NoStop}%
\bibitem [{\citenamefont {Gascuel}\ and\ \citenamefont {Steel}(2010)}]{gascuel2010inferring}%
  \BibitemOpen
  \bibfield  {author} {\bibinfo {author} {\bibfnamefont {Olivier}\ \bibnamefont {Gascuel}}\ and\ \bibinfo {author} {\bibfnamefont {Mike}\ \bibnamefont {Steel}},\ }\bibfield  {title} {\enquote {\bibinfo {title} {Inferring ancestral sequences in taxon-rich phylogenies},}\ }\href {\doibase https://doi.org/10.1016/j.mbs.2010.07.002} {\bibfield  {journal} {\bibinfo  {journal} {Mathematical biosciences}\ }\textbf {\bibinfo {volume} {227}},\ \bibinfo {pages} {125--135} (\bibinfo {year} {2010})}\BibitemShut {NoStop}%
\bibitem [{\citenamefont {Evans}\ \emph {et~al.}(2000)\citenamefont {Evans}, \citenamefont {Kenyon}, \citenamefont {Peres},\ and\ \citenamefont {Schulman}}]{evans2000broadcasting}%
  \BibitemOpen
  \bibfield  {author} {\bibinfo {author} {\bibfnamefont {William}\ \bibnamefont {Evans}}, \bibinfo {author} {\bibfnamefont {Claire}\ \bibnamefont {Kenyon}}, \bibinfo {author} {\bibfnamefont {Yuval}\ \bibnamefont {Peres}}, \ and\ \bibinfo {author} {\bibfnamefont {Leonard~J}\ \bibnamefont {Schulman}},\ }\bibfield  {title} {\enquote {\bibinfo {title} {Broadcasting on trees and the ising model},}\ }\href {\doibase 10.1214/aoap/1019487349} {\bibfield  {journal} {\bibinfo  {journal} {Annals of Applied Probability}\ ,\ \bibinfo {pages} {410--433}} (\bibinfo {year} {2000})}\BibitemShut {NoStop}%
\bibitem [{\citenamefont {Felsenstein}(2003)}]{felsenstein}%
  \BibitemOpen
  \bibfield  {author} {\bibinfo {author} {\bibfnamefont {Joseph}\ \bibnamefont {Felsenstein}},\ }\href@noop {} {\emph {\bibinfo {title} {Inferring phylogenies}}}\ (\bibinfo  {publisher} {Oxford University Press},\ \bibinfo {year} {2003})\BibitemShut {NoStop}%
\bibitem [{\citenamefont {De~Leonardis}\ \emph {et~al.}(2025)\citenamefont {De~Leonardis}, \citenamefont {Pagnani},\ and\ \citenamefont {Barrat-Charlaix}}]{deleonardis2024reconstruction}%
  \BibitemOpen
  \bibfield  {author} {\bibinfo {author} {\bibfnamefont {Matteo}\ \bibnamefont {De~Leonardis}}, \bibinfo {author} {\bibfnamefont {Andrea}\ \bibnamefont {Pagnani}}, \ and\ \bibinfo {author} {\bibfnamefont {Pierre}\ \bibnamefont {Barrat-Charlaix}},\ }\bibfield  {title} {\enquote {\bibinfo {title} {Reconstruction of ancestral protein sequences using autoregressive generative models},}\ }\href {\doibase 10.1093/molbev/msaf070} {\bibfield  {journal} {\bibinfo  {journal} {Molecular Biology and Evolution}\ }\textbf {\bibinfo {volume} {42}},\ \bibinfo {pages} {msaf070} (\bibinfo {year} {2025})}\BibitemShut {NoStop}%
\bibitem [{\citenamefont {Mistry}\ \emph {et~al.}(2020)\citenamefont {Mistry}, \citenamefont {Chuguransky}, \citenamefont {Williams}, \citenamefont {Qureshi}, \citenamefont {Salazar}, \citenamefont {Sonnhammer}, \citenamefont {Tosatto}, \citenamefont {Paladin}, \citenamefont {Raj}, \citenamefont {Richardson}, \citenamefont {Finn},\ and\ \citenamefont {Bateman}}]{PFAM2021}%
  \BibitemOpen
  \bibfield  {author} {\bibinfo {author} {\bibfnamefont {Jaina}\ \bibnamefont {Mistry}}, \bibinfo {author} {\bibfnamefont {Sara}\ \bibnamefont {Chuguransky}}, \bibinfo {author} {\bibfnamefont {Lowri}\ \bibnamefont {Williams}}, \bibinfo {author} {\bibfnamefont {Matloob}\ \bibnamefont {Qureshi}}, \bibinfo {author} {\bibfnamefont {Gustavo A}\ \bibnamefont {Salazar}}, \bibinfo {author} {\bibfnamefont {Erik L~L}\ \bibnamefont {Sonnhammer}}, \bibinfo {author} {\bibfnamefont {Silvio C~E}\ \bibnamefont {Tosatto}}, \bibinfo {author} {\bibfnamefont {Lisanna}\ \bibnamefont {Paladin}}, \bibinfo {author} {\bibfnamefont {Shriya}\ \bibnamefont {Raj}}, \bibinfo {author} {\bibfnamefont {Lorna~J}\ \bibnamefont {Richardson}}, \bibinfo {author} {\bibfnamefont {Robert~D}\ \bibnamefont {Finn}}, \ and\ \bibinfo {author} {\bibfnamefont {Alex}\ \bibnamefont {Bateman}},\ }\bibfield  {title} {\enquote {\bibinfo {title} {Pfam: The protein families database in 2021},}\ }\href {\doibase 10.1093/nar/gkaa913} {\bibfield  {journal}
  {\bibinfo  {journal} {Nucleic Acids Research}\ }\textbf {\bibinfo {volume} {49}},\ \bibinfo {pages} {D412--D419} (\bibinfo {year} {2020})}\BibitemShut {NoStop}%
\bibitem [{\citenamefont {Calvanese}\ \emph {et~al.}(2024)\citenamefont {Calvanese}, \citenamefont {Lambert}, \citenamefont {Nghe}, \citenamefont {Zamponi},\ and\ \citenamefont {Weigt}}]{Calvanese_2024}%
  \BibitemOpen
  \bibfield  {author} {\bibinfo {author} {\bibfnamefont {Francesco}\ \bibnamefont {Calvanese}}, \bibinfo {author} {\bibfnamefont {Camille~N}\ \bibnamefont {Lambert}}, \bibinfo {author} {\bibfnamefont {Philippe}\ \bibnamefont {Nghe}}, \bibinfo {author} {\bibfnamefont {Francesco}\ \bibnamefont {Zamponi}}, \ and\ \bibinfo {author} {\bibfnamefont {Martin}\ \bibnamefont {Weigt}},\ }\bibfield  {title} {\enquote {\bibinfo {title} {Towards parsimonious generative modeling of rna families},}\ }\href {\doibase 10.1093/nar/gkae289} {\bibfield  {journal} {\bibinfo  {journal} {Nucleic Acids Research}\ }\textbf {\bibinfo {volume} {52}},\ \bibinfo {pages} {5465–5477} (\bibinfo {year} {2024})}\BibitemShut {NoStop}%
\end{thebibliography}%

\clearpage

\beginsupplement

\centerline{\bf SUPPLEMENTAL MATERIAL}

\section{Construction of the natural MSAs}

The alignments of natural sequences used in this work have been constructed as follows. 
We believe that the details of the alignment procedure do not affect the results presented in the paper.
\begin{itemize}
\item
WW: we downloaded the alignment corresponding to the PF00397 family from Pfam~\cite{PFAM2021}
and excluded sequences with more than 20\% gaps.
\item 
DBD: we used as seed the alignment from Ref.~\cite{park2022epistatic} 
(221 sequences)
and ran {\tt hmmsearch} on {\tt uniref90}, excluding sequences with more than 20\% gaps.
\item
Chorismate Mutase (CM): we used the alignment from Ref.~\cite{russ2020evolution}.

\item
AAC6: we used the same alignment as in Ref.~\cite{bisardi2022modeling}.

\item 
DHFR: we downloaded the alignment corresponding to the PF00186 family from Pfam~\cite{PFAM2021}
and excluded sequences with more than 20\% gaps.

\item
BL: we used the same alignment as in Ref.~\cite{bisardi2022modeling}.

\item
SP: we downloaded the alignment corresponding to the PF00089 family from Pfam~\cite{PFAM2021}
and excluded sequences with more than 20\% gaps.

\end{itemize}
To compute the empirical frequencies used to train the bmDCA model, each sequence is assigned a weight, given by the inverse of the number of other sequences at distance smaller than
$20\%$. The effective number of sequences is the sum of such weights. See e.g. Ref.~\cite{muntoni2021adabmdca} for details.

In table~\ref{tab:example} we show for each family the length of the aligned protein sequences $L$, the depth (number of aligned sequences) $M$ of the alignment, and its effective depth $M_{\rm eff}$, which serves as a proxy for estimating the diversity of sequences within the alignment.
\begin{table}[h!]
\centering
\begin{tabular}{|c|c|c|c|}
\hline
\textbf{Family} & \textbf{$L$} & \textbf{$M$} & \textbf{$M_{\rm eff}$} \\ \hline
WW  & 31            &  157809            & 10791            \\ \hline
DBD            & 76            & 24944           & 3129           \\ \hline
CM            &  96            & 1259             &  937           \\ \hline
AAC6            &  117           &  43576           &  21873           \\ \hline
DHFR            &       160      & 36612 & 8785            \\ \hline
BL           & 202           &   18334         & 6875       \\ \hline
SP            &     220    &   53422   & 22892             \\ \hline
\end{tabular}
\caption{Schematic summary of the characteristics of the multiple sequence alignments used for our analysis.}
\label{tab:example}
\end{table}

The three DBD sequences that correspond to the blue, green, and red colors are identified, respectively, with the code \texttt{UniRef90\_E4X5B7/28-107}, \texttt{UniRef90\_O61854/33-105}, and \texttt{UniRef90\_A0A818T1L9/14-89}.
They come, respectively, from the organisms {\it Oikopleura dioica (Tunicate)}, {\it Caenorhabditis elegans}, and {\it Rotaria sp. Silwood1}.

\section{Different models and evolutionary schemes}

\begin{figure}[t]
\centering
\includegraphics[width=1\linewidth]{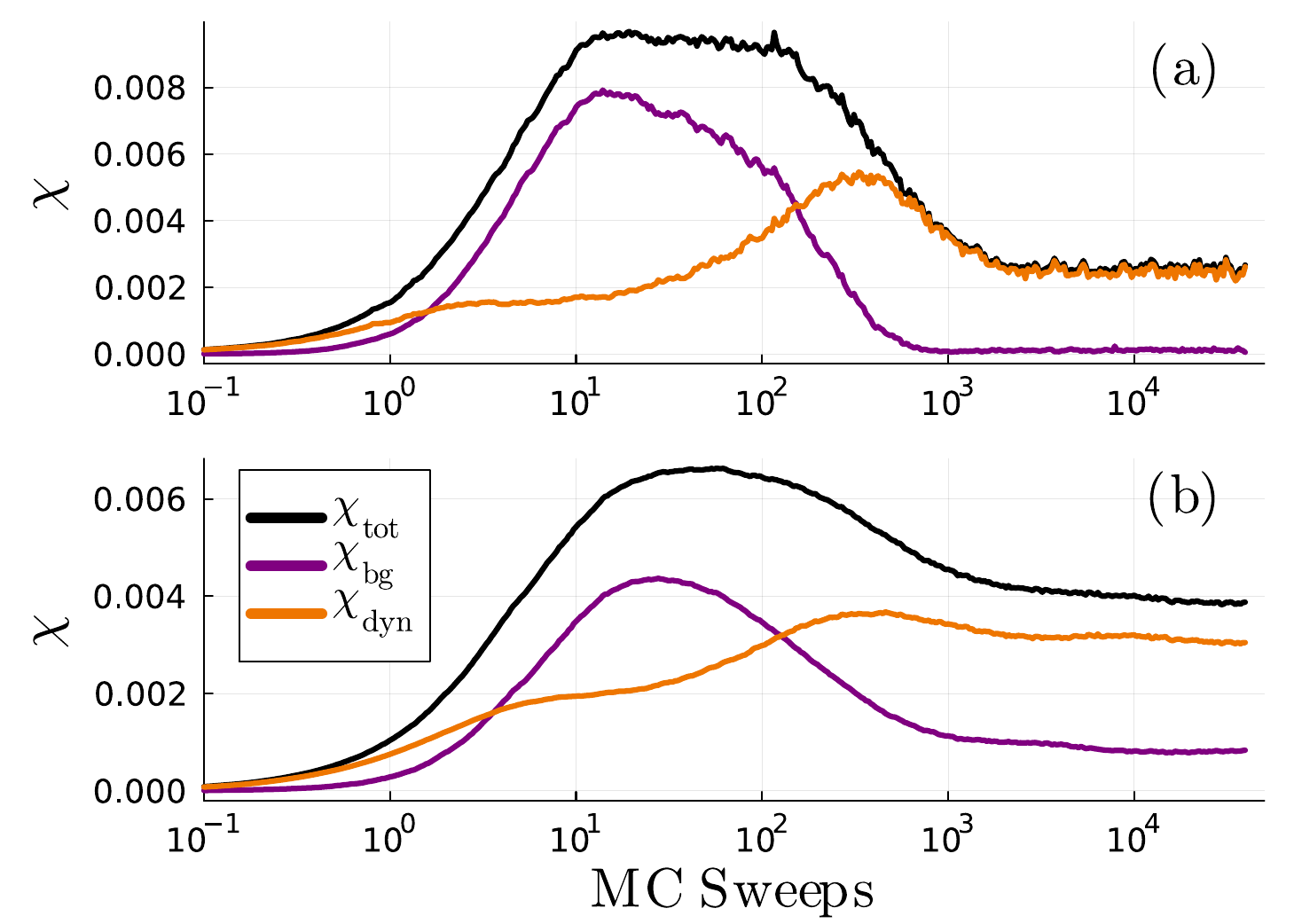}
\caption{Evolution of $\chiback$, $\chidyn$ and $\chitot$ as a function of Monte Carlo sweeps using a sparse edge-activated version~\cite{Calvanese_2024} of the Potts model (a) and using the dynamics of Ref.~\cite{DiBari2024} implementing insertions (b), deletions and single nucleotide substitutions for the DBD family.} 
\label{fig:results_different_models}
\end{figure}

Different variants of the bmDCA model that we employed in the main text have recently been suggested. 
We performed simulations using a sparse edge-activated~\cite{Calvanese_2024} version for the DBD family, without any significant change in the results (Fig.~\ref{fig:results_different_models}a).

As we already pointed out, the evolutionary protocol that we implemented is a simple Metropolis Monte Carlo defined over amino acid sequences, which is schematically described as follows:
\begin{enumerate}
    \item Initialize with a protein sequence $A^{0}$.
    \item At each iteration:
    \begin{enumerate}
        \item Select a random site \( i \in \{1, \dots, L\} \).
        \item Propose a mutation \( a_i' \neq a_i \) by selecting a new amino acid at site \( i \) uniformly from the alphabet.
        \item Compute the energy difference:
        \begin{equation}
            \Delta H = H(A') - H(A)\ ,
        \end{equation}
        where \( A' \) is the sequence with the proposed mutation.
        \item Accept the mutation with probability
        \begin{equation}
            P_{\text{accept}} = \min(1, e^{-\Delta H / T})\ .
        \end{equation}
        If accepted, update \( a_i \leftarrow a_i' \); otherwise, keep~\( a_i \).
    \end{enumerate}
    \item Repeat for a sufficiently large number of iterations to reach equilibrium.
\end{enumerate}
After equilibration, sequences sampled from this procedure approximate the equilibrium distribution of the Potts model.

This dynamics is far from being realistic, especially at short time-scales. As a matter of fact, a more detailed approach was recently described in~\cite{DiBari2024}, where it was shown that an algorithm working with single-nucleotide substitutions (instead of amino acid substitutions), insertions and deletions was able to correctly reproduce the dynamics both at long and short time scales.
We checked that this more refined algorithm, which was shown to over-perform a simple Metropolis at short time scales~\cite{bisardi2022modeling},  produces qualitatively similar results for the quantities of interest for our work (Fig.~\ref{fig:results_different_models}b).

\section{Profile models}

As epistatic sites play an important role in our description, it is interesting to study a dynamics in which they are totally absent. 
This can be done by using a profile model in which only a local field acts on each site, eliminating the coupling between different protein residues. 
Such profile models can be built in two ways, analogously to what has been done in Ref.~\cite{DiBari2024}.  
We call \textit{global profile} the model in which the local field is set equal to
$h^{\rm GP}_i(a) = -\log f_i(a)$, in such a way as to reproduce the one-point frequencies.
In this case, the sites evolve following their local field independently from one another. 
\begin{figure}
\centering
\includegraphics[width=1\linewidth]{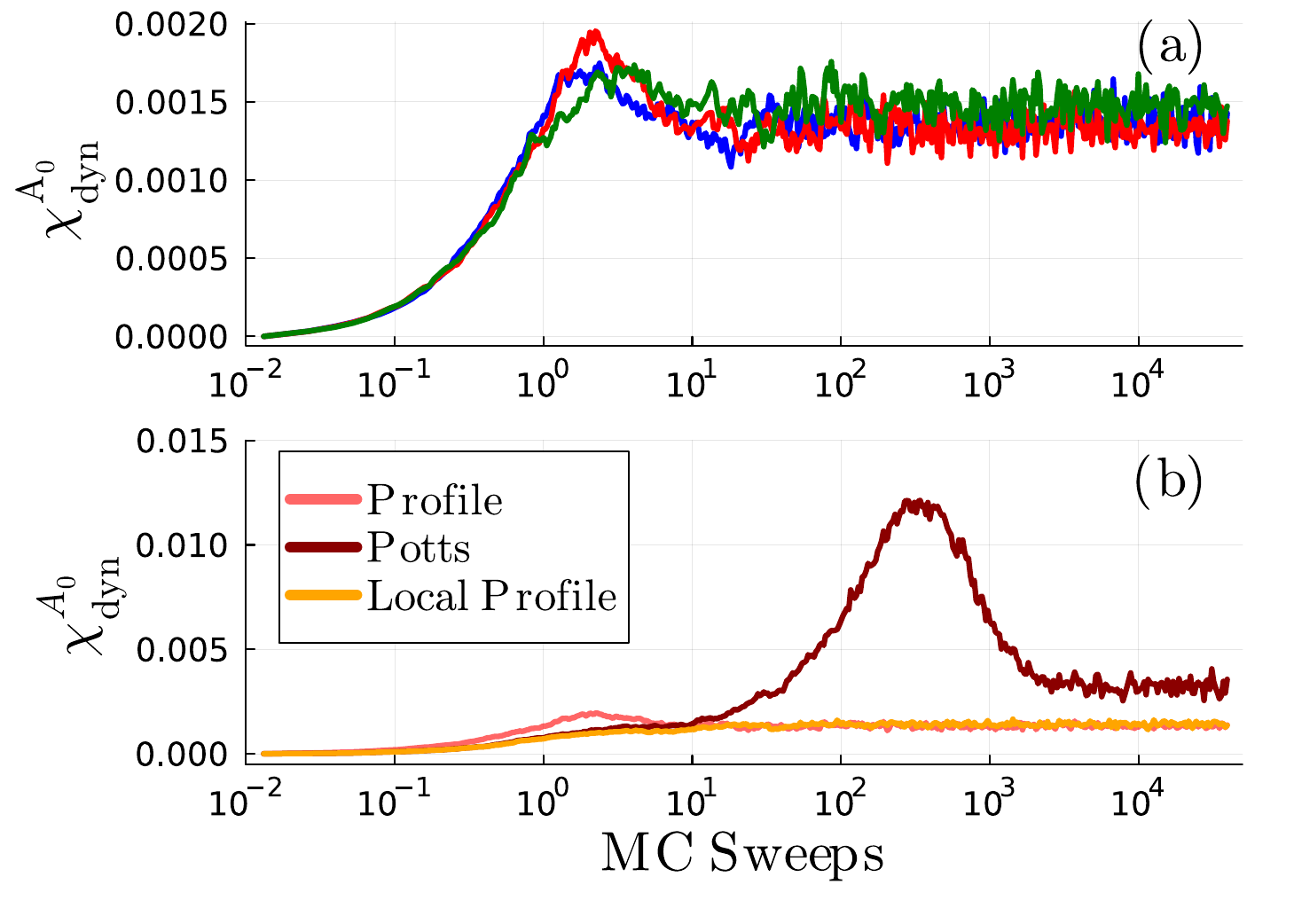}
\caption{Evolution of the dynamical susceptibility for the profile models. In (a) we show the evolution of $\chidynA$ for the same starting sequences used in the main text studied with the global profile model, while in (b) we compare the results for the red sequence obtained using the two profile models and the full one.
} 
\label{fig:chi_dyn_profile}
\end{figure} 
In Fig.~\ref{fig:chi_dyn_profile}a we show how the dynamical susceptibility evolves for the same three sequences we studied in the main text.
We saw above that the blue sequence had a smaller $\chidynA$ and that it converged to the long-time limit quite fast. 
Differently, the red sequence showed a large peak and only converged after many sweeps. 
In this figure we see that when using a global profile model the starting sequence does not affect the $\chidynA$ anymore. 
As the interactions are set to zero, the initial condition plays a much less important role and all the curves seem to behave in a rather similar way, showing a small peak at a homogeneous time scale. 
Couplings between residues and hence epistasis are needed to observe the qualitative behavior presented in the main text.

One can also define a \textit{local profile} model, in which the fields acting on the residues are obtained as 
\begin{equation}\label{eqS:hA}
    h^{A_0}_i(a_i) = h_i(a_i) + \sum_j J_{ij}(a_i,a^0_j) \ ,
\end{equation}
and therefore depend on the initial sequence $A_0$.
We compare both these profile models and the Potts one Fig.~\ref{fig:chi_dyn_profile}b for one sequence $A_0$.
At the very beginning of the dynamics the three models are very similar, as only few mutations occurred. 
Already after few steps, less than a sweep, the global profile model displays the small peak discussed above, due to the uncorrelated mutations occurring at different residues. 
The local profile takes into account the role of the initial condition and hence it follows a dynamics similar to the full Potts model for a longer time ($\sim 10$ sweeps). 
However, for even longer times, in the full Potts model the evolution of the background changes the local field acting on each residue, allowing sites to evolve in a coordinated manner. This results in the presence of the peak, as discussed in the main text.
In the local profile this correlated evolution cannot take place, as the local fields are kept constant. 
Hence, the dynamical susceptibility quickly saturates to its long time limit value.

\section{Different definition of susceptibility}

\begin{figure}[t]
\centering
\includegraphics[width=0.9\linewidth]{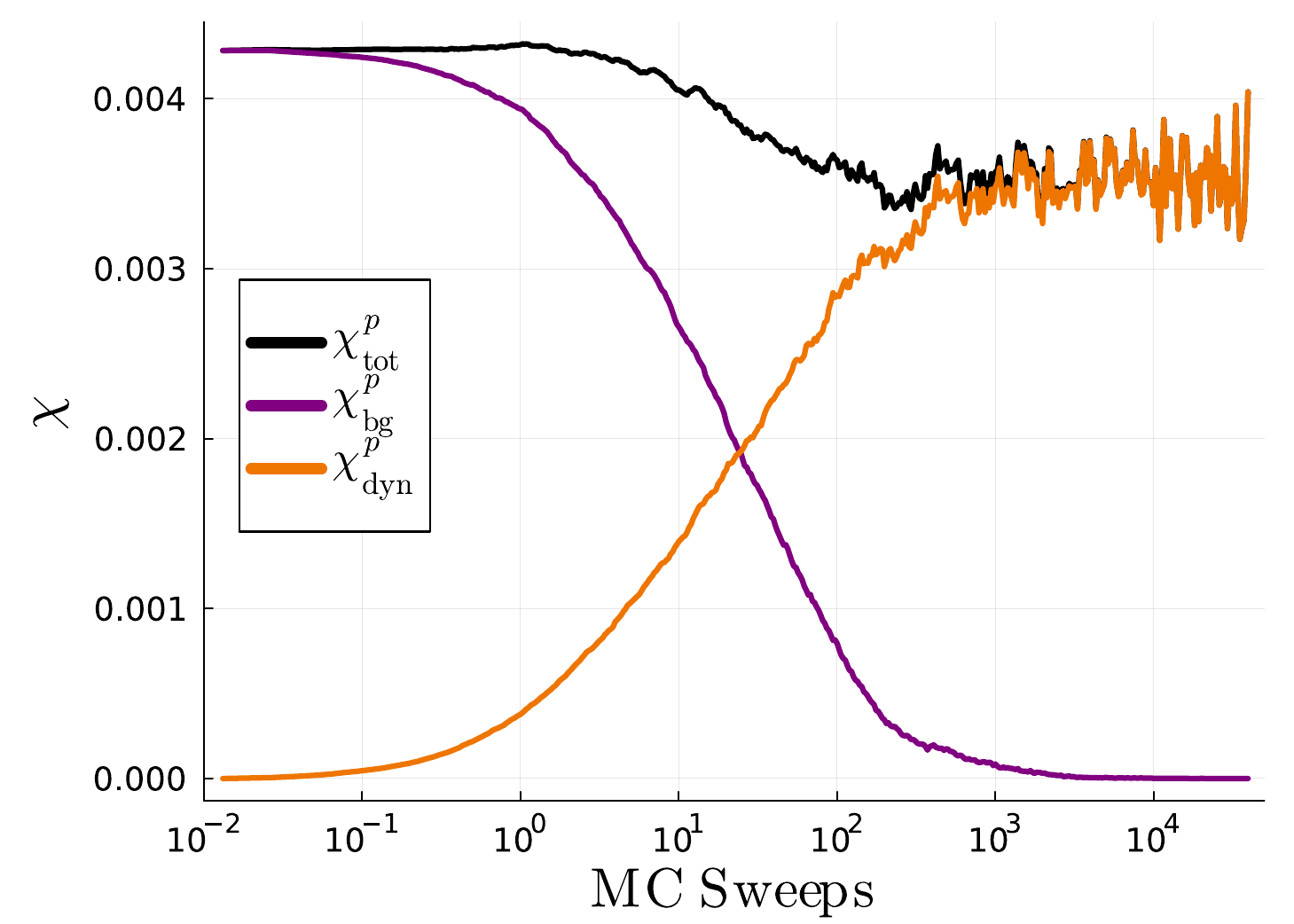}
\caption{Evolution of $\chiback^p$, $\chidyn^p$ and $\chitot^p$ as a function of Monte Carlo sweeps for the DBD family.} 
\label{fig:chi_p}
\end{figure}

It is also interesting to consider a different definition of susceptibility. In particular we already defined $H^{A_0}_r(t,0)$ as the Hamming distance between the sequence that evolved for $t$ steps under the predefined evolutionary noise $r$ starting from the initial sequence $A_0$ and the initial sequence itself.
We can define two supplementary Hamming distances, this time computed at fixed Monte Carlo sweeps between different independent simulations of the system. In particular we define
\begin{equation}
\label{eq:clone_susceptibility}
H^{A,B}_{r,r'}(t) = 1 - \frac{1}{L} \sum_i \delta_{ a^{A}_{r,i}(t) , a^{B}_{r',i} (t)}
\end{equation}
as the Hamming distance between sequences that have independently evolved from different initial sequences $A$ and $B$ for the same time $t$. 
The $[\dots]$ average is then computed by summing over all different realizations of the couples of evolutionary noises $(r,r')$, while the $\langle \dots \rangle$ average is computed by summing over all couples of different initial sequences $(A,B)$.
It is interesting to compare the particular case $H^{A,A}_{r,r'}(t)$ with what we showed before. 
Because our dynamics is time-reversible, we expect that $ \langle [H^{A,A}_{r,r'}(t/2)] \rangle \sim \langle [H^{A}(t)] \rangle $. 
We checked that this is indeed the case.

In the main text we showed that $\chidynA$ can be expressed as the sum of a correlation function between sites (see Eqs.~\ref{eq:SuscAndCorr} and~\ref{eq:CorrDef}). If two sites both mutated after a time $t$ with respect to the initial sequence $A_0$, then they presented a high correlation. 
In this case, $\chi^{A,B}_{\rm dyn}$ can be still seen as the sum of site correlations, but, differently from the previous case, two sites $i$ and $j$ are highly correlated if after time $t$ the realizations starting from the two sequences $A$ and $B$ display the same amino acids in those sites, namely 
\begin{equation*}
\label{eq:g_pairwise}
G^{A,B}_{ij,{\rm dyn}}(t) = [\delta_{a^{t,A}_i,a^{t,B}_i} \delta_{a^{t,A}_j,a^{t,B}_j}] - [\delta_{a^{t,A}_i,a^{t,B}_i}][\delta_{a^{t,A}_j,a^{t,B}_j}]. 
\end{equation*}
Analogously to what is presented in the main text, we can define $\chiback^p$, $\chidyn^p$ and $\chitot^p$ where $p$ apex indicates that we are considering a pairwise metric. Because $H^{A,B}_{r,r'}(t)$ strongly depends on which couples $A,B$ are chosen to initialize the two dynamical evolutions, we can easily expect that the impact of the background initialization $\chiback^p$ will be dominant at the beginning of evolution. Instead, when equilibrium is reached, $H^{A,B}_{r,r'}(t)$ tends to the mean pairwise hamming between the sequences in the training set as the model is generative. Hence, at later times the memory of the initial condition is lost and $\chiback^p$ vanishes. This is confirmed by Fig.~\ref{fig:chi_p}.

\section{Response to environmental variation}

\begin{figure}[t]
\centering
\includegraphics[width=0.9\linewidth]{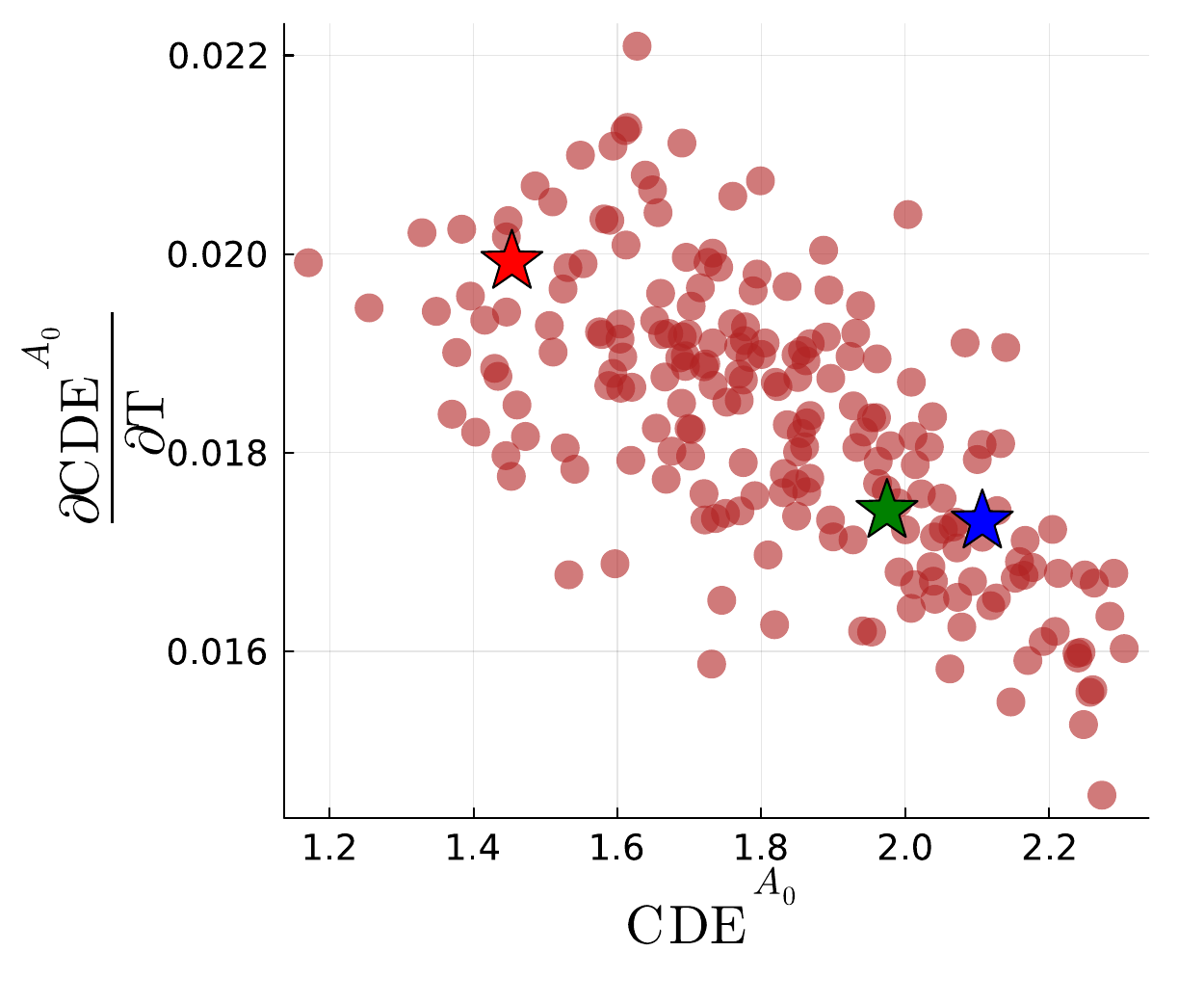}
\caption{Derivative of the context-dependent entropy with respect to temperature (i.e. inverse selection pressure), as a function of the sequence mutability $\textrm{CDE}^{A_0}$ for 200 sequences of the DBD family. The colored stars refer to the three sequences of Fig.2.}
\label{fig:LinearResponseCDE}
\end{figure}

To complement the analysis presented in the main text concerning the response to a change in selection pressure,
we checked how much the mutability of the ancestral sequence $\textrm{CDE}^{A_0}$ changes with respect to a shift in selection pressure.
In Fig.~\ref{fig:LinearResponseCDE} we plot such derivative as a function of $\textrm{CDE}^{A_0}$ for a set of sequences of the DBD family. 
It is clear that epistatically constrained sequences have a higher change in mutability when subject to a change in selection pressure (i.e. a change in the simulation temperature). As a matter of fact, the average context-dependent entropy of a given sequence $A_0$ can be expressed as a function of the local fields $h^{A_0}_i(a_i)$ defined in Eq.~\eqref{eqS:hA} as follows 
\begin{equation}
\textrm{CDE}^{A_0} = - \frac1L \sum_i \sum_{a=1}^{21} \frac{e^{-\beta h^{A_0}_i(a)}}{z^{A_0}_i} \log_2 \frac{e^{-\beta h^{A_0}_i(a)}}{z^{A_0}_i} \ ,
\end{equation}   
with $z^{A_0}_i = \sum_{a=1}^{21} e^{-\beta h^{A_0}_i(a)}$. Hence, the derivative with respect to inverse selective pressure can be analytically computed as follows:
\begin{equation}
\frac{\partial \textrm{CDE}^{A_0}}{\partial T} \propto \frac1L \sum_i \sum_{a=1}^{21} \langle h^{A_0}_i(a)^2 \rangle - \langle h^{A_0}_i(a) \rangle^2   \ ,
\end{equation}
where $\langle \bullet \rangle = \sum_{a=1}^{21} \bullet e^{-\beta h^{A_0}_i(a)} / z^{A_0}_i $.
This expression indicates that sites having a greater variance of local fields are more prone to change their mutability as a sudden change in selective pressure takes place.
This analysis offers interesting insights into the evolvability of protein sequences which should be corroborated by evolution experiments.

\end{document}